%% file: THMCategorysubmit.tex
\documentclass[12pt]{article}
\usepackage{amsmath,amsthm,amssymb}
\usepackage[dvipdfmx]{graphicx} 
\newcommand{\bm}[1]{\mbox{\boldmath$#1$}}
\newcommand{\maprightu}[1]{%
\smash{\mathop{%
\hbox to 1cm{\rightarrowfill}}\limits^{#1}}}
\newcommand{\maprightd}[1]{%
\smash{\mathop{%
\hbox to 1cm{\rightarrowfill}}\limits_{#1}}}
\newcommand{\mapleftu}[1]{%
\smash{\mathop{%
\hbox to 1cm{\leftarrowfill}}\limits^{#1}}}
\newcommand{\mapleftd}[1]{%
\smash{\mathop{%
\hbox to 1cm{\leftarrowfill}}\limits_{#1}}}
\begin{document}
\baselineskip15pt

\centerline{\Large {\bf Singularity Confinement  }}
\bigskip
\centerline{\Large {\bf and}}
\bigskip
\centerline{\Large {\bf   Projective Resolution of Triangulated Category\footnote{A part of this work was reported by one of the authors (S.S) in the conference on ``Nonlinear Mathematical Physics: Twenty Years of JNMP", held at Nordfjordeid, Norway, from June 4 to June 14, 2013. }}}
\bigskip

\centerline{by}

\centerline{ Satoru SAITO$^{1,*}$\, Tsukasa YUMIBAYASHI$^{2,\ddag}$ and Yuki WAKIMOTO$^{2,\ddag}$}
\bigskip

$^1$ {\it {\small Hakusan 4-19-10, Midoriku, Yokohama 226-0006, Japan}}

$^2$ {\it {\small Department of Physics, Tokyo Metropolitan University, Tokyo, 192-0397 Japan}}

$^*$ {\it {\small E-mail: saito$\underbar\ $ru@nifty.com}}

\bigskip

\begin{center}
\begin{minipage}{13cm}
{\bf Abstract}
{\small
We proposed, in our previous paper, to characterize the Hirota-Miwa equation by means of the theory of triangulated category. We extend our argument in this paper to support the idea. In particular we show in detail how the singularity confinement, a phenomenon which was proposed to characterize integrable maps, can be associated with the projective resolution of the triangulated category. }
\end{minipage}
\end{center}

\section{Introduction}

Let us consider a dynamical system in which a rule is fixed to change the state of an object, say $X$, at every step. We call such a system deterministic. If we can predict the state of $X$ at any time, when an initial state of $X$ is given arbitrary, we say the system integrable. The Kepler motion is an integrable system such that an orbit is determined entirely if initial position and initial velocity are given. Such an integrable system is, however,  quite unusual, because the probability we encounter an integrable system is almost zero when a deterministic system is given arbitrary.

We are interested in finding a way to discriminate integrable systems from nonintegrable ones. From the view of category theory in mathematics \cite{MacLane} an integrable system is a category in which objects are the states of $X$ and the morphism is the change of the states. 
In our previous paper \cite{SJNMP} we discussed this problem and proposed that the triangulated category might characterize the Hirota-Miwa equation (HM eq), a completely integrable difference equation from which infinitely many soliton equations of the KP hierarchy can be derived \cite{H, M}. 

We would like to extend our previous argument in this paper. In particular we discuss in detail how the singularity confinement, a phenomenon which was proposed to characterize integrable maps \cite{GRPH, GRPH2}, can be associated with the projective resolution of the triangulated category. \\

The plan of this paper is as follows. In \S 2 we discuss geometrical feature of the HM eq. It is convenient to describe the HM eq by means of the discrete geometry on a lattice space, such that the notion of distinguished triangles becomes manifest. 
This idea was already explained briefly in our previous paper \cite{SJNMP} but we extend the idea much more in detail in this paper. Based on this geometrical structure we fix in \S 3 the deterministic rule of information transfer from one place to another on the lattice space. It will be shown in \S 4 that all these rules can be described in terms of the triangulated category.

We have proposed in \cite{SJNMP} a possible interpretation of the singularity confinement as a projective resolution of the triangulated category. Our main subject of this paper is an investigation of this conjecture in detail, which will be presented in \S 5.  

\section{Geometrical feature of the HM eq.}

Before starting our argument it will be worthwhile to review briefly some features of the HM eq. \cite{H, M}. We consider the following formula throughout this paper:
\begin{equation}
a_{14}a_{23}\tau_{14}(p)\tau_{23}(p)-a_{24}a_{13}\tau_{24}(p)\tau_{13}(p)+a_{34}a_{12}\tau_{34}(p)\tau_{12}(p)=0,
\label{HM}
\end{equation}
\[
\tau(p)\in\mathbb{C},\quad p=(p_1,p_2,p_3,p_4)\in \mathbb{C}^4,\quad  a_{ij}=-a_{ji}\in \mathbb{C}.
\]
In above and hereafter we use the abbreviations, such as
\begin{equation}
\tau_j(p):=D_j\tau(p)=\tau(p+\delta_j),
\quad \tau_{ij}(p):=D_iD_j\tau(p)=\tau(p+\delta_i+\delta_j), 
\label{D}
\end{equation}
\[
\delta_j:=(\delta_{1j},\delta_{2j},\delta_{3j},\delta_{4j})
\]
with the Kronecker symbol $\delta_{ij}$.

\begin{enumerate}
\item[1)] This is the simplest, but nontrivial Pl\"ucker relation identically satisfied by determinants.
\item[2)]  
If we substitute 
\begin{equation}
\tau(p)=\prod_{i,j}\Bigl({E(z_i,z_j)\over z_i-z_j}\Bigr)^{p_ip_j}\theta\Bigl(\zeta+\sum_{j}p_j
w(z_j)\Bigr),\quad a_{ij}=z_i-z_j
\label{Krichever}
\end{equation}
into  (\ref{HM}), we obtain an identity, called Fay's trisecant formula, for the hyperelliptic function $\theta$ and the prime form $E(z_i,z_j)$ defined on a Riemann surface of arbitrary genus \cite{Fay, Krichever}.
\item[3)]
This formula characterizes the Jacobi  varieties among the Abel varieties \cite{Mulase, Shiota}.
\item[4)]
From this single equation all soliton equations in the KP hierarchy can be derived corresponding to various continuous limits of independent variables \cite{H, M}. 
\item[5)]
The solutions of (\ref{HM}) were identified with the points of universal Grassmannian by Sato and known as the $\tau$ functions \cite{Sato}. 
\end{enumerate}

\subsection{Nature of $\tau$ functions}

\begin{figure}[h]
\centering
\input{Riemannsurface.tex}
\caption{Riemann surface}
\label{Riemann}
\end{figure}
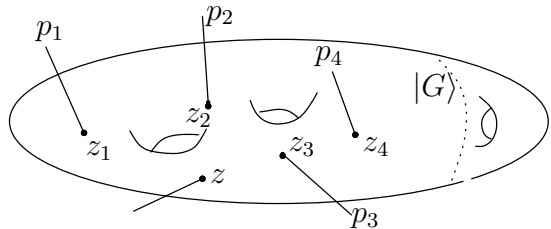

It was shown in \cite{S} that the $\tau$ functions can be represented by means of tachyon correlation functions of the string theory. Since it provides the most convenient formulation in our argument we will use the notion of string theory in what follows. The 4 point string (tachyon) correlation function is given by
\begin{equation}
\Phi(p,z;G)=\langle 0|V(p_1,z_1)V(p_2,z_2)V(p_3,z_3)V(p_4,z_4)|G\rangle,
\label{0VVVVG}
\end{equation}
where $z=(z_1,z_2,z_3,z_4)\in\mathbb{Z}^4$ is a set of parameters determined by $a_{ij}$'s of the equation (\ref{HM}).
Here 
\[
V(p_j,z_j)= :\exp(ip_jX(z_j)):
\]
is the vertex operator of momentum $p_j$ of a string attached at $z_j$ of the string world sheet specified by the state vector $|G\rangle$. The string coordinate $X(z)$ is an operator which acts on the state $|\cdot\rangle$, while the symbol $::$ means the normal order product. It was proved in \cite{S} that the substitution of the ratio
\begin{equation}
\tau(p)={\Phi(p,z;G)\over \Phi(p,z;0)}
\label{string amp}
\end{equation}
into (\ref{HM}) yields exactly Fay's formula associated with the Riemann surface Fig.\ref{Riemann} corresponding to the world sheet $|G\rangle$. Hence the point $z_j$ on the world sheet is a puncture of the Riemann surface.  Notice that, since we do not integrate over $z_j$'s, thus has no problem of divergence, we define the vertex operator $V(p,z)$ with no ghost field $c$.  

Although every solution of (\ref{HM}) is obtained by specifying the state $|G\rangle$ of the general solution (\ref{string amp}), we do not discuss explicit forms of $|G\rangle$ in this paper. Therefore we simply write $\Phi(p,z;G)$ as $\Phi(p,z)$ unless it is necessary. On the other hand the main property of $\tau$ functions is determined by the nature of the vertex operators as we will see now. Since they satisfy
\begin{equation}
V(p,z)V(p',z')=(-1)^{pp'}V(p',z')V(p,z),
\label{VV}
\end{equation}
we see immediately that the fields $\psi_\pm(z):=V(\pm 1, z)$ have the properties
\begin{equation}
\psi_\pm(z)\psi_\pm(z')=-\psi_\pm(z')\psi_\pm(z),\quad
\psi_\pm(z)\psi_\mp(z')=-\psi_\mp(z')\psi_\pm(z),
\label{pp=-pp}
\end{equation}
and, in particular, the following holds:
\begin{equation}
\psi_+(z)\psi_+(z)=0,\quad \psi_-(z)\psi_-(z)=0.
\label{pp=0}
\end{equation}
Hence $\psi_{\pm}(z)$ are Grassmann fields.

By taking this property into account we define operators $\hat D_s^{\pm 1}$ by
\begin{equation}
\hat D_s^{\pm 1}\Phi(p,z):=
\langle 0|V(p_1,z_1)V(p_2,z_2)V(p_3,z_3)V(p_4,z_4)\psi_\pm(z_s)|G\rangle
\label{D tau}
\end{equation}
to describe the insertion of $\psi_\pm(z_s)$ into $\Phi(p,z)$ of (\ref{0VVVVG}). 
When $z_s$ of $\psi_\pm(z_s)$ coincides with one of $z=(z_1,z_2,z_3,z_4)$ in (\ref{D tau}), say $s=j$, the insertion of  $\psi_\pm(z_s)$ is equivalent to change $p_j$ to $p_j\pm 1$, up to a phase factor which comes from the exchange of order of $\psi_\pm(z_j)$ with vertex operators. If we use the notations such as
\[
\Phi_j(p,z)=\hat D_j\Phi(p,z),\quad \Phi_{ij}(p,z)=\hat D_i\hat D_j\Phi(p,z),\qquad i,j=1,2,3,4,
\]
in these particular cases, we obtain
\[
\Phi_{ij}(p,z)=-\Phi_{ji}(p,z),\qquad i,j=1,2,3,4,
\]
hence
\begin{equation}
\Phi_{ii}(p,z)=0, \qquad i=1,2,3,4.
\label{tau(p)=0}
\end{equation}
This is an expression of (\ref{pp=0}). Thus we have found that the zero of $\Phi(p,z)$ is associated with a coincidence of two punctures on the Riemann surface.

In the theory of KP hierarchy \cite{DJKM1, DJKM2, DJKM3} the operators $\ \exp \psi_\pm(z_j)=1+\psi_\pm(z_j)\ $ are known as elements of the symmetry group $GL(\infty)$ which act on the state $|G\rangle$. Correspondingly we call 
\[
\hat D_j=e^{\hat D_j}-1
\]
a `difference' operator, in analogy with the differential operator $\partial_j$.\\

Despite of this odd behavior of the correlation function $\Phi$ under the operation of $\hat D_j^\pm$, the solutions (\ref{string amp}) of the HM eq behave regularly. It owes to the fact that both $\Phi(p,z;G)$ and $\Phi(p,z;0)$ are shifted by $\hat D_j$ simultaneously in $\tau(p)$. The extra phase factors arising from the exchange of order of vertex operators and $\psi_\pm(z_j)$ in (\ref{D tau}) cancel exactly. As a result we find
\[
\hat D_j\tau(p)=D_j\tau(p):=\tau(p+\delta_j)=:\tau_j(p),
\]
in agreement with our previous notation (\ref{D}).  Nevertheless it is important to notice that the zero point of $\Phi$ function (\ref{tau(p)=0}) is an \underline{indeterminate point} of $\tau(p)$. This fact will play a central role in our analysis of the singularity confinement in \S 5.

\subsection{Difference geometry on lattice spaces}

Although the variable $p$ of the $\tau$ function is on $\mathbb{C}^4$, the solutions of the HM eq are on a lattice space $\mathbb{Z}^4$ embedded in $\mathbb{C}^4$, which is fixed once an `initial point' $p_0\in\mathbb{C}^4$ of $\tau$ is fixed. Let us call this lattice space 
\[
\Xi_4(p_0):=\Big\{p\in\mathbb{C}^4\Big|p-p_0\in\mathbb{Z}^4\Big\}.
\]
Since, however, the `initial point' $p_0$ does not appear explicitly in our discussion, we simply write $\Xi_4(p_0)$ as $\Xi_4$. Moreover we often write $p-p_0$ as $p\in\Xi_4$ unless there is a confusion.

In order to study the HM eq. within the framework of the theory of category, it will be useful to study its geometrical feature on the lattice space $\Xi_4$. For this purpose we introduce a notion of 'difference form' on the lattice space in this subsection.

\subsubsection{4 dimensional `difference' forms}

Let us define \cite{SJNMP} an exterior `difference' operator $d_B$ by 
\begin{equation}
d_B\omega(p):=\sum_{j=1}^4D_j\omega(p)\wedge dp_j, \qquad ^\forall \omega(p):\Xi_4\to\mathbb{C}.
\label{d_B}
\end{equation}
We must emphasize that the form $D_j\omega(p)$ is on $\Xi_4$ if $p\in \Xi_4$, but, in contrast to the differential form, it is not at the same point $p$ but is at $p+\delta_j$. In particular, the operation of $d_B$ to $\omega(p)$ increases the value of the sum $p_1+p_2+p_3+p_4$ of the components of $p$ by 1. To describe the situation more precisely we define a subspace of $\Xi_4$ by
\[
\Xi_4^{(n)}:=\Big\{p\in\Xi_4\Big| p_1+p_2+p_3+p_4=n\in\mathbb{Z}\Big\},
\]
so that
\[
\bigcup_{n\in\mathbb{Z}}\Xi_4^{(n)}=\Xi_4.
\]
We notice that $\Xi_4^{(n)}$ is a lattice hyperplane in $\Xi_4$. In particular $\Xi_4^{(1)}$ is the hyperplane which includes the four points 
$\{\delta_1,\ \delta_2,\ \delta_3,\ \delta_4\}$. 
All other hyperplanes are parallel to $\Xi_4^{(1)}$.

\begin{figure}[h]
\begin{center}
\includegraphics[scale=0.4]{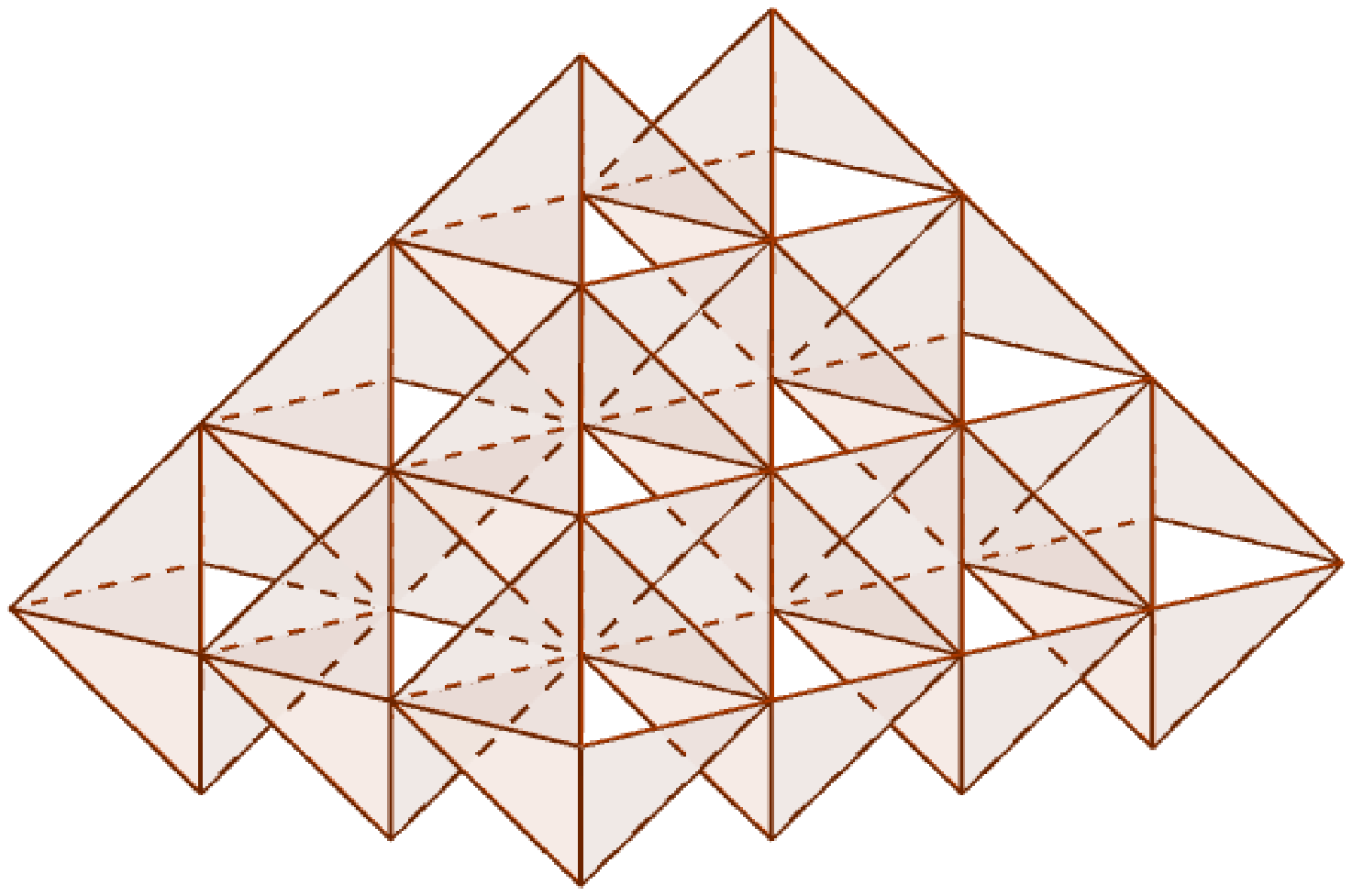}
\caption{$\Xi_4^{(n)}$}
\label{fig2}
\end{center}
\end{figure}

Each hyperplane is embedded in a three dimensional lattice space $\mathbb{Z}^3$. In fact the points of $\Xi_4^{(n)}$ occupy all corners of octahedra which fill $\mathbb{Z}^3$ together with tetrahedra, as it is illustrated in Fig.\ref{fig2}.

If $p\in\Xi_4^{(n)}$, the forms $D_j\omega(p)=\omega_j(p)$ are on $\Xi_4^{(n+1)}$ for all $j$, hence
\[
D:\quad \Xi_4^{(n)}\to \Xi_4^{(n+1)}, 
\]
with $D=(D_1,D_2,D_3, D_4)$. Since all functions in the HM eq.(\ref{HM}) are of the form $\tau_{ij}(p)=D_iD_j\tau(p)$, they are on $\Xi_4^{(n+2)}$ if $p\in\Xi_4^{(n)}$. Moreover the six functions $\tau_{ij}(p)$ in (\ref{HM}) are at the six corners of the octahedron whose center is at $p$. Hence the HM eq. determines relations among functions on $\Xi_4^{(n+2)}$. In other words solutions of the HM eq. are different if they are on different hyperplanes. We mention that this is a result of the fact that the HM eq. is a Pl\"ucker relation.\\

Let us define $\omega^{(n)}$ by
\[
\omega^{(n)}(p):\ \Xi_4^{(n)}\ \to\ \mathbb{C},\qquad n\in\mathbb{Z}
\]
when $p\in \Xi_4^{(n)}$. We then naturally obtain a graded algebra
\begin{equation}
\bigoplus_{n\in\mathbb{Z}}\ \mathbb{C}\ \omega^{(n)}
\label{4d grade}
\end{equation}
on which $d_B$ acts by
\begin{equation}
d_B^{(n)}:\quad \omega^{(n)}\to \omega^{(n+1)}.
\label{d_B}
\end{equation}

\subsubsection{3 dimensional `difference' forms}

\begin{figure}[h]
\begin{center}
\includegraphics[scale=0.6]{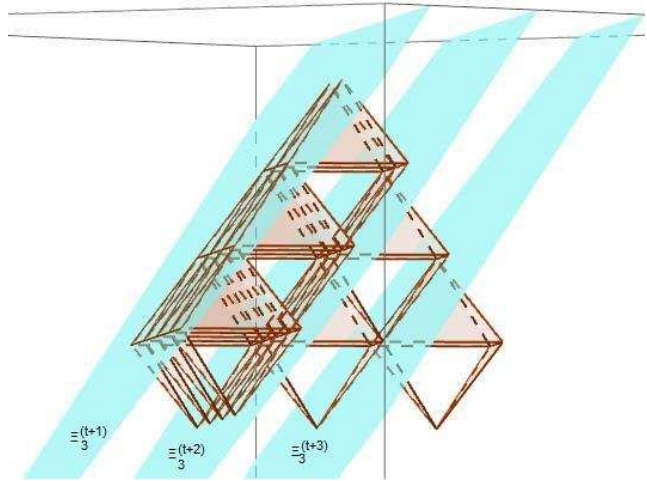}
\caption{$\Xi_3^{(t,n+2)}$}
\label{fig3}
\end{center}
\end{figure}

From Fig.\ref{fig2} we can see that the lattice space $\Xi_4^{(n+2)}$ consists of parallel planes, each filled by triangles of octahedra, as illustrated in Fig.\ref{fig3}. If we fix the direction of the planes parallel to the direction of $-p_4$, we can specify them by the values of $t:=p_1+p_2+p_3$. Such a plane is defined by 
\[
\Xi_3^{(t,n+2)}:=\Big\{(p_1,p_2,p_3)\in\mathbb{Z}^3\Big|p_1+p_2+p_3=t\in\mathbb{Z},\ 
(p_1,p_2,p_3,p_4)\in\Xi_4^{(n+2)}\Big\},
\]
\[
\bigcup_{t\in\mathbb{Z}}\Xi_3^{(t,n+2)}=\Xi_4^{(n+2)}.
\]
Since $n+2\ (=p_1+p_2+p_3+p_4)=t+p_4$ is fixed, the planes are perpendicular to the direction of $-p_4$. 

We denote a point on $\Xi_3^{(t,n+2)}$ by $\bm{p}=(p_1,p_2,p_3)\in\mathbb{Z}^3$. 
Corresponding to $d_B$ of (\ref{d_B}) we define the 3 dimensional exterior `difference' operator $d_T$ by
\[
d_T\omega(\bm{p})=\sum_{a=1}^3D_a\omega(\bm{p})\wedge dp_a,\qquad
^\forall \omega(\bm{p}): \ \Xi_3^{(t,n+2)}\to\mathbb{C},
\]
which we call the shift operator. Since, similar to the 4 dimensional case, $\omega(\bm{p})$ and $D_a\omega(\bm{p})$ are on different planes, we define graded forms of degree $t$ by $\omega[t]$ when $\bm{p}\in \Xi_3^{(t,n+2)}$. The graded algebra
\[
\bigoplus_{t\in\mathbb{Z}}\mathbb{C} \omega[t]
\]
is generated by
\begin{equation}
d_T^{[t]}:\ \  \omega[t]\to \omega[t+1].
\label{d_T}
\end{equation}

\section{Dynamical feature of the HM eq.}

Based on the geometrical structure of the HM eq. we studied in the previous section, we discuss dynamical feature of the HM eq. in this section. 

\subsection{Linear B\"acklund transformation}

In order to study dynamical feature of the HM eq. we consider a pair of 
4 dimensional `difference' 2-forms \cite{SJNMP}
\begin{eqnarray}
F:=\sum_{i,j=1}^4F_{ij}dp_i\wedge dp_j,&\quad& F_{ij}(p)=a_{ij}\tau_{ij}(p),
\label{F}\\
\tilde F:=\sum_{i,j=1}^4\tilde F_{ij}dp_i\wedge dp_j,&\quad&\tilde F_{ij}(p)={}^*a_{ij}\tau_{ij}(p)
\label{tilde F}
\end{eqnarray}
with $^*a_{ij}=\sum_{kl}\epsilon_{ijkl}a_{kl}$, where $\epsilon_{ijkl}$ is the Levi-Civita symbol. We can easily check that each of the followings  \cite{ShiS}
\begin{equation}
\det F_{ij}=0,\qquad \det \tilde F_{ij}=0
\label{det F=0}
\end{equation}
is equivalent to the HM eq. (\ref{HM}). By this reason we call $F$ of (\ref{F}) the HM 2-form in this paper.

Let us consider, for any $\tau(p)$ and $\sigma(p)$, the following forms
\begin{eqnarray*}
\tilde F\wedge {\rm d_B}\sigma
&=&\sum_{j,k,l} {}^*a_{kl}\tau_{kl}\sigma_jdp_k\wedge dp_l\wedge dp_j
=\sum_{i,j}\tilde F_{ij}\sigma_j\ ^*dp_i,\\
d_B(\tilde F\wedge {\rm d_B}\sigma)
&=&\sum_{i,j}a_{ij}{}^*\tau_{j}\sigma_{ij}dp_i\wedge {^*dp_i}=:\sum_{i,j}G_{ij}{}^*\tau_{j}\delta_i,
\end{eqnarray*}
where we used the notation ${}^*dp_i:=\epsilon_{ijkl}dp_j\wedge dp_k\wedge dp_l$ and
$\sum_{kl}\epsilon_{ijkl}{}^*a_{kl}\tau_{ikl}=a_{ij}{}^*\tau_j$. If we require 
$\tilde F\wedge {\rm d_B}\sigma=0$, it is compatible {\it iff} 
$d_B(\tilde F\wedge {\rm d_B}\sigma)=0$, or equivalently
\begin{eqnarray}
&\sum_{j}\tilde F_{ij}\sigma_j=0,&\qquad i=1,2,3,4
\label{Fg=0}\\
&\Updownarrow&\nonumber\\
&\sum_{j}G_{ij}{}^*\tau_{j}=0,&\qquad i=1,2,3,4.
\label{Gf=0}
\end{eqnarray}

Now suppose $\tau^{(0)}(p)$ is a solution of the HM eq. (\ref{HM}). If it is substituted to $\tilde F$ in (\ref{Fg=0}), we can solve the linear equation for $\sigma$, because $\tilde F$ satisfies (\ref{det F=0}). Let us denote by $\tau^{(1)}(p)$ the solution of (\ref{Fg=0}) which is just obtained dependent on $\tau^{(0)}$. Then we can substitute $\tau^{(1)}$ into $G$ of (\ref{Gf=0}). Because (\ref{Fg=0}) and (\ref{Gf=0}) are the same equation, the equation (\ref{Gf=0}) for $\tau$ has certainly solution. 
This means $\det G$ must vanish, so that, by (\ref{det F=0}) again, $\tau^{(1)}(p)$ is a solution of the HM eq.. 

As $\tau^{(1)}(p)$ is substituted the linear equation (\ref{Gf=0}) has  
solutions other than $\tau^{(0)}$ in general. Let $\tau^{(2)}$ be one of them which is independent from $\tau^{(0)}$.  $\tau^{(2)}$ is again a solution of the HM eq., so that we can substitute it to (\ref{Fg=0}) to find $\tau^{(3)}$, and so on. Here we must mention that if the initial function $\tau^{(0)}(p)$ is on $\Xi_4^{(n)}$, the solution $\tau^{(1)}(p)$ is on $\Xi_4^{(n+1)}$. 
We can repeat this procedure and obtain a sequence of the B\"acklund transformation \cite{SS1}
\begin{equation}
\tau^{(0)}\quad \maprightu{{\rm d_B^{(0)}}}\quad \tau^{(1)}
\quad \maprightu{{\rm d_B^{(1)}}}\quad \tau^{(2)}
\quad \maprightu{{\rm d_B^{(2)}}}\quad \tau^{(3)}\quad \maprightu{{\rm d_B^{(3)}}}\quad \cdots
\label{BT}
\end{equation}

If we choose another solution of the HM eq. for the initial function $\tau^{(0)}$, we should have different series of solutions. In fact, in the string theory, we can add loops of closed strings so that the world sheet increases holes as is depicted in the
Fig.\ref{Riemann}. There is also a vertex operator which substitutes a D-brane, so that the world sheet is attached to a boundary \cite{Sato-Saito}. All such variations change the topology of the state $|G\rangle$, and corresponds to different initial solution $\tau^{(0)}$ of the HM eq.. If we denote the complex 
(\ref{BT}) by ${\mathcal BT}_G$ 
corresponding to the state $|G\rangle$, we will find a family of complexes of the B\"acklund transformation.

\subsection{Dynamical evolution}

In this subsection we fix the hyperplane $\Xi_4^{(n+2)}$, hence we simply denote $\Xi_3^{(t,n+2)}$ by $\Xi_3^{[t]}$, and study behavior of a particular solution of the HM eq. For this purpose we rewrite the HM 2-form by using the operator $d_T$ of (\ref{d_T}) as
\[
F=\sum_{i,j=1}^4F_{ij}dp_i\wedge dp_j=
d_T\left(F_4(p)dp_4+\sum_{b=1}^3F_b(p)dp_b\right)
\]
and see that the six components $F_{ij}$ split into two parts
\begin{eqnarray}
d_TF_4(p)&=&\sum_{a=1}^3F_{a4}(\bm{p})dp_a,
\label{3d 1form}\\
d_T\sum_{b=1}^3F_b(p)dp_b&=&\sum_{b,c=1}^3F_{bc}(\bm{p})dp_b\wedge dp_c,
\label{3d 2 form}
\end{eqnarray}
corresponding to 3 dimensional 1-form and 2-form, respectively.

Let us denote by $\tilde S_j$ and $S_j$ the triangles in an octahedron which are perpendicular to $p_j$ and parallel each other. Then we can see that $\{F_{a4}\}$ and $\{F_{bc}\}$ form triangles
\[
\tilde S_4:=\Big(\tau_{14}, \tau_{24}, \tau_{34}\Big),\qquad
S_4:=\Big(\tau_{23}, \tau_{31}, \tau_{12}\Big),
\]
which are perpendicular to $p_4$. We denote the octahedron, which consists of these triangles, by 
${\mathcal O}=(\tilde S_4, S_4)$, as illustrated in Fig.4.

\begin{figure}[h]
\begin{center}
\input{Nordfjordpic38nocolor.tex}
\caption{}
\label{fig4}
\end{center}
\end{figure}

From Fig.\ref{fig3} we see that every octahedron is put between two nearest planes, such that two parallel triangles are on each plane. In fact the triangle $\tilde S_4$ is on $\Xi_3^{[t+1]}$ and $S_4$ is on $\Xi_3^{[t+2]}$ if $\tau(\bm{p})$ is on $\Xi_3^{[t]}$. Define ${\mathcal O}[t]=(\tilde S_4[t+1], S_4[t+2])$ with
\begin{eqnarray*}
\tilde S_4[t+1]&=&\Big(\tau_{14}[t+1],\tau_{24}[t+1],\tau_{34}[t+1]\Big),\\
S_4[t+2]&=&\Big(\tau_{23}[t+2],\tau_{31}[t+2],\tau_{12}[t+2]\Big)
\end{eqnarray*}
and let $\{{\mathcal O}\}[t]$ be all set of octahedra put between $\Xi_3^{[t+1]}$ and $\Xi_3^{(t+2)}$. Then the procedure to solve initial value problem is to determine the following sequence
\begin{equation}
\{{\mathcal O}\}\quad \stackrel{d_T}{\longrightarrow}\quad
\{{\mathcal O}\}[1]\quad \stackrel{d_T}{\longrightarrow}\quad
\{{\mathcal O}\}[2]\quad \stackrel{d_T}{\longrightarrow}\quad
\{{\mathcal O}\}[3]\quad \stackrel{d_T}{\longrightarrow}\quad\cdots
\label{O series}
\end{equation}
when information $\{{\mathcal O}\}$ at initial time $t=0$ is given.

\subsection{Deterministic rule for the flow of information}

We want to know how information on $\{{\mathcal O}\}$ transfers to other octahedra as $t$ increases. The HM eq determines a relation between $\tilde S_4$ and $S_4$. But it does not tell us how the information of $\tilde S_4$ transfers to $S_4[1]$. In order to solve an initial value problem we must know some deterministic rules which decide uniquely the local flow of information. In this subsection we set up a rule for the flow of information in an octahedron. 

Let us consider two lattice points $p$ and $p'$ on $\Xi_4$ which are separated by
\[
p'-p=m\in \mathbb{Z}^4.
\]
If they are on the same hyperplane $\Xi_4^{(n)}$, the separation $m$ must satisfy
\begin{equation}
m_1+m_2+m_3+m_4=0.
\label{m+m+m+m=0}
\end{equation}
If $p$ and $p'$ are neighbor of an octahedron, 
\begin{equation}
m=\delta_i-\delta_j,\qquad i\ne j
\label{1-1}
\end{equation}
corresponding to the edge parallel to the vector $p_i-p_j$. 
There will be many possible routs along which information can transfer between two points $p$ and $p'$ fixed arbitrary. Since an addition of a path of the type (\ref{1-1}) does not change the condition (\ref{m+m+m+m=0}), any routs can connect $p$ and $p'$ as long as they are connected by edges of octahedra. \\

When we decide the rule of transfer of information we must keep in mind the following items:
\begin{itemize}
\item 
Information at $p+\delta_i+\delta_j$ is transfered properly to $p'+\delta_k+\delta_l$ if all operators corresponding to all possible routs $\{r\}$ change $\tau_{ij}(p)$ to $\tau_{kl}(p')$ uniquely. 
\[
D_r(ij,kl)_{p,p'}: \ \ \tau_{ij}(p)\stackrel{r}{\longrightarrow} \tau_{kl}(p'),\qquad \forall r.
\]
\item Our system is deterministic if a rule of transfer of information is fixed along edges of an octahedron, and is the same for all octahedra. 
\item The system is integrable if this rule is sufficient to predict the values of $\tau$  on $\Xi_3^{[t]}$ for all $t$ when the values on $\Xi_3^{(0)}$ are given arbitrary. 
\end{itemize} 

There are 12 edges in an octahedron ${\mathcal O}$. Every object $\tau_{ij}$ at a corner has connections to its four neighbors but no direct connection to its diagonal one. 
Since we are interested in a flow of information from one corner to another of ${\mathcal O}$, we must decide direction (hence a rule) of the flow. In other words we fix the order of points in ${\mathcal O}$. A natural way is the cyclic ordering of 
suffixes, {\it i.e.}, 
\begin{equation}
1<2<3<4<1<2<3.
\label{1<2<}
\end{equation}
We notice in Fig. 4 that every pair of corners connected by an edge have a common suffix, like $(\tau_{14}, \tau_{12})$. Therefore we define the direction of transfer by
\begin{equation}
D(ij,ik): \tau_{ij}\longrightarrow \tau_{ik},
\qquad {\it iff} \quad k<i<j.
\label{order}
\end{equation}
For example, $D(34,24)$ is possible because $3<4<2$, but $D(24,12)$ is not possible because $4<2<1$ is not compatible with (\ref{1<2<}), {\it etc}. In this way we can define uniquely the directions which are necessary to connect all objects in ${\mathcal O}$. In Fig.\ref{fig5} $(a)$ the directions allowed by this rule (\ref{order}) are shown by arrows. If we project the diagram along $p_3$ we have
Fig.\ref{fig5} $(b)$. 

\begin{figure}[h]
\centering
\input{Nordfjordpic20.tex}
\caption{}
\label{fig5}
\end{figure}

The action of  $D(kl,ij)$ is to remove punctures at $z_i$ and $z_j$ from $|G\rangle$ and insert other punctures at $z_k$ and $z_l$. It will be convenient to represent this action explicitly by means of the operators:
\begin{equation}
D(ij,kl)=D_kD_lD_j^{-1}D_i^{-1}.
\label{Hom}
\end{equation}
We can easily check the rule of product,
\[
D(ij,kl)\circ D(kl,mn)=D(ij,mn).
\]
From
\[
D(ij,kl)\circ D(kl,ij)
= D(ij,ij)=id
\]
we see that $D(ij,kl)$ is an isomorphism.

Since corners connected by an edge have a common suffix, (\ref{Hom}) simplifies 
\[
D(ij,ik)=D_kD_j^{-1}
\]
whereas the morphism connecting the diagonals of an octahedron can be obtained by a product of morphisms, for instance, like
\[
D(23,24)\circ D(24,14)=D(23,14)=D_2D_3D_4^{-1}D_1^{-1}.
\]

\subsection{Transfer of information along a chain of octahedra}

In this subsection we want to see how information flows along a chain of octahedra. Let us denote by ${\mathcal O}(p)=(\tilde S_4(p),S_4(p))$ the octahedron whose center is at $p\in \Xi_4^{(n)}$. There are three octahedra which share three edges of the triangle $S_4(p)$. Since these three neighbors are on the same hyperplane $\Xi_4^{(n+2)}$ their centers must be at $p+\delta_1-\delta_4,\ p+\delta_2-\delta_4,\ p+\delta_3-\delta_4$, respectively. Let ${\mathcal O}(p)[1]$ be one of them, say ${\mathcal O}(p+\delta_3-\delta_4)$. Then these octahedra are connected as illustrated in Fig.\ref{fig6}, where we use the abbreviations:
\[
\tilde S_4=(X,Y,Z)=\Big(\tau_{14}, \tau_{24}, \tau_{34}\Big),\quad
S_4=(X',Y',Z')=\Big(\tau_{23}, \tau_{31}, \tau_{12}\Big).
\]

\begin{figure}[h]
\centering
\input{Nordfjordpic27.tex}
\caption{}
\label{fig6}
\end{figure}

For the information of ${\mathcal O}$ to transfer to its neighbor ${\mathcal O}[1]$ properly, we must impose the following conditions: 
\begin{equation}
\tau_{14}[1]=\tau_{13},\qquad \tau_{24}[1]=\tau_{23}.
\label{connection cond}
\end{equation}

Repeating this procedure we can define a chain of octahedra by
\begin{equation}
{\mathcal O}(p)[t]={\mathcal O}(p+t\delta_3-t\delta_4),\quad t\in\mathbb{Z}.
\label{chain}
\end{equation}
We recall that $\tilde S_4(p)$ is on $\Xi_3^{[t+1]}$ and $S_4(p)$ is on $\Xi_3^{(t+2)}$, if $p_1+p_2+p_3=t$. Because $d_T$ transfers information on $\Xi_3^{[t]}$ to  $\Xi_3^{[t+1]}$ we see that the information on ${\mathcal O}$ can be transfered along the chain of octahedra 
\begin{equation}
{\mathcal O}\maprightu{{\rm d}_T}{\mathcal O}[1]\maprightu{{\rm d}_T}{\mathcal O}[2]
\maprightu{{\rm d}_T}{\mathcal O}[3]\maprightu{{\rm d}_T}\cdots.
\label{O chain}
\end{equation}
if the connection conditions (\ref{connection cond}) are satisfied at every site of connection. This is certainly compatible with (\ref{O series}), and explains the local flow of information.


\section{View from the theory of category}

We are now ready to summarize our result in the previous section in terms of the category theory.

\subsection{Triangulated category}

We have pointed out in our previous paper \cite{SJNMP} a possible explanation of the flow of information by means of the triangulated category. In order to see this correspondence more in detail let us first recall the axioms of triangulated category \cite{Kashiwara-Schapira, GelfandManin}. 

\noindent
{\bf Definition}

Let ${\mathcal D}$ be an additive category, $X,Y,Z,X',Y',Z'$ be  objects and $u,v,w$ be morphisms  of ${\mathcal D}$.  The structure of a {\it triangulated category} on ${\mathcal D}$ is defined by the {\it shift functor} $T$ and the class of {\it distinguished triangles} satisfying the following axioms:\\

\begin{enumerate}
\item[Tr1]
$(1)$Any triangle of the form
$$
X\stackrel{\rm id}{\longrightarrow} X \longrightarrow 0 \longrightarrow T(X) 
$$
is in the class of distinguished triangles.\\
$(2)$ Any triangle isomorphic to a distinguished triangle is distinguished.\\
$(3)$ Any morphism $u:X\rightarrow Y$ can be completed to a distinguished triangle
$$
X\stackrel{u}\longrightarrow Y \longrightarrow C(u) \longrightarrow T(X) 
$$
by the object $C(u)$ obtained by morphism $u$.

\item[Tr2]
The triangle
$$
X\stackrel{u}\longrightarrow Y \stackrel{v}\longrightarrow Z \stackrel{w}\longrightarrow T(X) 
$$
is a distinguished triangle if and only if
$$
Y \stackrel{v}\longrightarrow Z \stackrel{w}\longrightarrow T(X) \stackrel{-T(u)}\longrightarrow T(Y)
$$
is a distinguished triangle.

\item[Tr3]
Suppose there exists a commutative diagram of distinguished triangles,
\begin{align*}
 &\hspace{0.5mm}X\hspace{0.5mm}\longrightarrow Y \hspace{0.5mm}\longrightarrow \hspace{0.5mm}Z \hspace{1mm}\longrightarrow T(X)\\
u&\downarrow \hspace{6.5mm}v\downarrow \hspace{25mm}\downarrow T(u)\\
 &X'\longrightarrow Y' \longrightarrow Z' \longrightarrow T(X').
\end{align*}
Then this diagram can be completed to a commutative diagram by a (not necessarily unique) morphism $w:Z\rightarrow Z'$.

\item[Tr4] (the octahedron axiom) 
Let $X\stackrel{u}\longrightarrow Y\stackrel{v}\longrightarrow Z$ be a triangle. Then the following commutative diagram holds:
\begin{equation}
\input{Octahedronaxiom.tex}
\label{axiom of 8}
\end{equation}
\end{enumerate}
\bigskip

Comparing this set of axioms Tr1 $\sim$ Tr4 with our arguments in \S 3.3 and \S 3.4, we naturally find the following correspondence:

\begin{equation}
\begin{array}{ccc}
objects&\longleftrightarrow& \tau\  functions,\cr
morphism &\longleftrightarrow&D(ij,ik),\cr
 T: A\to T(A)&\longleftrightarrow&d_T: A\to A[1],\cr
octahedron\ axiom&\longleftrightarrow&flow\ diagram\ Fig.\ref{fig6}.
\end{array}
\label{HM category}
\end{equation}

For example there are flows of information which start from $A\in (X,Y,Z)$ and then end at $A[1]\in\Big(X[1],Y[1],Z[1]\Big)$ as
\begin{equation}
\begin{array}{ccccccc}
X&\to& Y&\to& Z&\to &X[1],\cr
X&\to& Y&\to& Z'&\to &X[1],\cr
X&\to&Z&\to&Y'&\stackrel{id}{\to}&X[1],\cr
Y&\to&Z&\to&X'&\stackrel{id}{\to}&Y[1],\cr
Z&\to&Y'&\to&X'&\to&Z[1].\cr
\end{array}
\label{third cond}
\end{equation} 
They are all compatible with the axioms as we can check easily. Here $id$'s in (\ref{third cond}) owe to the condition (\ref{connection cond}). Via these routs information in ${\mathcal O}$ can transfer to ${\mathcal O}[1]$. Let us denote the triangulated category (\ref{HM category}) as ${\mathcal HM}_G^{(n)}$ in this paper. The suffix $(n)$ is to remind that a solution of the HM eq is given separately for each lattice space $\Xi_4^{(n)}$, while the suffix $G$ is to remind that the solution is dependent on the initial state $|G\rangle$.

\subsection{The category theory of global behavior}

From above arguments we see that the objects of the category ${\mathcal HM}_G^{(n)}$
\[
{\rm Ob}({\mathcal HM}_G^{(n)})=\{\tau^{(n)}_{ij}(p)\ \Big|\ p\in\Xi_4^{(n)}\}
\]
are nothing but a solution of the HM eq (\ref{HM}), which we denote as
\begin{equation}
{\mathcal HM}_G^{(n)}\simeq \tau_G^{(n)}
\label{HG simeq tau}
\end{equation}
when $|G\rangle$ is given.

On the other hand we learned in \S 3.1 that a series of the B\"acklund transformations forms a complex ${\mathcal BT}_G$ of solutions of the HM eq associated to the state $|G\rangle$, {\it i.e.,}
\[
d_B:\ \tau_G^{(n)}\ \to\ \tau_G^{(n+1)}.
\]
If we combine this with the result (\ref{HG simeq tau}) we obtain
\[
{\mathcal BT}_G=\Big\{\{{\mathcal HM}_G^{(n)}\}_{n\in\mathbb{Z}},\ d_B\Big\}.
\]
Therefore ${\mathcal BT}_G$ is a category of categories and $d_B$ is a functor.
\bigskip

Moreover, as we mentioned in \S 3.1, the change of the state $|G\rangle$
\[
|G\rangle\to |G'\rangle={\mathcal F}|G\rangle.
\]
will generate a different complex ${\mathcal BT}_{G'}$ of the B\"acklund transformations. Therefore we obtain another functor
\[
{\mathcal F}:\ {\mathcal BT}_G\ \to\ {\mathcal BT}_{G'}.
\]

\bigskip

We have thus found various type of categories which are related from each other. Among others our interpretation of the information flow of the HM eq. by means of the triangulated category is the most fundamental.  However the correspondence of the flow with the triangulated category seems not quite right by two reasons. 
\begin{enumerate}
\item An addition of objects is no longer an object in general, because our objects are solutions of a nonlinear equation. Although most of studies of triangulated category have been based on additive categories in mathematics, the above axioms themselves do not use the notion of additivity. Therefore nonadditive nature of our theory will not cause any problem.
\item
Second reason is that we have not yet defined null object 0, which appears in the first axiom Tr1. This is the main subject of this paper which we are going to discuss in the following section.    
\end{enumerate}


\section{Localization and singularity confinement}

It is well known that a localization of a triangulated category is also a triangulated category. We show in this section that the singularity confinement of a rational map obtained from the $\tau$ function of the HM eq. can be described in terms of the localization of the triangulated category. 

\subsection{Reduction of the lattice space}

We have studied difference geometry of 4 and 3 dimensions in \S 2. Our concern in this section is a 2 dimensional lattice space. 

If we fix $n=p_1+p_2+p_3+p_4-2$ and $t=p_1+p_2+p_3$ we are left with 2 dimensional lattice space, which we parameterize by
\[
q=p_1+p_2,\qquad j:=p_2-p_1.
\]
Like the higher dimensional lattice cases we define 2 dimensional lattice space
\[
\Xi_2^{(q,t,n+2)}:=\Big\{(p_1,p_2)\in\mathbb{Z}^2\Big| p_1+p_2=q\in\mathbb{Z},\  (p_1,p_2,p_3)\in \Xi_3^{(t,n+2)}\Big\},
\]
\[
\bigcup_{q\in\mathbb{Z}}\Xi_2^{(q,t,n+2)}=\Xi_3^{(t,n+2)}
\]
and the exterior difference operator d$_Q$ by
\[
{\rm d}_Q\tau(\bm{p})=D_1\tau(\bm{p})dp_1+ D_2\tau(\bm{p})dp_2,
\]
which displaces the lattice space
\[
D: \Xi_2^{(q,t,n+2)}\to \Xi_2^{(q+1,t,n+2)}.
\]

Because $t=q+p_3$, we can fix $p_3$ instead of $t$. In this frame of coordinate, the change of $t$ is exactly the same with the change of $q$. 
We recall that the diagram Fig.\ref{fig6} (b) was the projection Fig.\ref{fig6} (a) along $p_3$. Since $j=p_2-p_1\in\mathbb{Z}$ is still free we denote $\tau^{[t]}_j:=\tau(\bm{p})$ and consider the lattice space 
\[
\Big\{ (j,t)\in\mathbb{Z}^2\Big| \Delta t=\Delta q, p_2-p_1=j\in\mathbb{Z}, (p_1,p_2)\in\Xi_2^{(q,t,n+2)}\Big\}.
\]

In our previous section we discussed the transfer of information along the chain of octahedra. We now extend the study to consider a transfer of information of many octahedra linked along a line in the $j$ direction.

We have to mention that the connection condition (\ref{connection cond}) is already taken into account in this expression. Moreover the projection along $p_3$ enforces degeneration of $Z'[t]$ in ${\mathcal O}[t]$   and $Z[t+2]$ in ${\mathcal O}[t+2]$. Therefore the shift operation d$_T$ brings $\tilde S_4$ directly to $\tilde S_4[1]$, so that $\tau_j^{[t]}$ is determined uniquely for all $t$ and $j$ as we can see in the Fig.\ref{fig7}.
\begin{figure}[h]
\centering
\input{Nordfjordpic16.tex}
\caption{}
\label{fig7}
\end{figure}

In the theory of KP hierarchy it is known that we can either truncate the function $\tau_j^{[t]}$, or impose periodicity in the direction of $j$ at any value, with no violation of integrability.  For example we can impose
\begin{equation}
\tau_{j+d}^{[t]}=\tau_j^{[t]}
\label{j+d=j}
\end{equation}
to obtain a reduced map of $d$ dimension. In this case it is more convenient to consider 
\[
\bm{\tau}^{[t]}=(\tau^{[t]}_0,\tau^{[t]}_1, \tau^{[t]}_2, \tau^{[t]}_3, ...,\tau^{[t]}_{d-1}),
\]
instead of a triangle $\tilde S_4$ of each octahedron separately. When $d=3$, the chain of Fig.\ref{fig7} becomes a chain of triangles. 

\subsection{Localization of a triangulated category}

The theory of triangulated category tells us that, if there is a null system, the theory can be localized such that the localized theory again satisfies the axioms of triangulated category \cite{Kashiwara-Schapira, resolution}. A null system ${\mathcal N}\subset {\mathcal T}$ of the triangulated category ${\mathcal T}$ is a set of objects defined by
\begin{enumerate}
\item $0\in {\mathcal N}$.
\item $Z\in {\mathcal N}$ if $X, Y\in{\mathcal N}$ and $X\to Y\to Z\to X[1]$ is a distinguished triangle.
\item $X[1]\in{\mathcal N}$\quad {\it iff}\quad $X\in{\mathcal N}$.
\end{enumerate}

For any triangulated category ${\mathcal T}$ and a null system ${\mathcal N}\subset {\mathcal T}$, we define a multiplicative system by
\[
S({\mathcal N}):=\{ g|X\stackrel{g}{\to}Y\to Z\to X[1],\  X,Y\in {\mathcal T},\ Z\in {\mathcal N}\}.
\]
Then the localization is defined by the functor ${\mathcal T}\to {\mathcal T}/
S({\mathcal N})$. The following theorem is known in the theory of triangulated category:
\bigskip

\noindent
{\bf Theorem}

{\it ${\mathcal T}/S({\mathcal N})$ is again a triangulated category whose null object is 0 itself.}\\

Therefore, in order to discuss the localization of our system we must know a null system of our triangulated category. Our objects are solutions $\{\tau_{ij}\}$ of the HM eq. which are assigned at corners of each octahedron.  They are generically finite, because the $\tau$ functions are defined by the ratios (\ref{string amp}) of correlation functions such that the zeros of correlation functions cancel from each other. 

As we explained in \S 2.1 the correlation functions $\{\Phi_{ij}\}$ vanish by themselves when two punctures encounter on the Riemann surface. It is, however, important to notice that, when the correlation functions $\Phi(p,z;G)$ and $\Phi(p,z;0)$ vanish simultaneously, the cancellation of their zeros does not mean the value of their ratio being definite. Let $\lambda(p)$ be the ratio of the correlation functions  at the point where they vanish, {\it i.e.}
\begin{equation}
\lambda(p):=\left\{\left.{\Phi(p,z;G)\over\Phi(p,z;0)}\right|\Phi(p,z;G)=\Phi(p,z;0)=0\right\},
\label{lambda=}
\end{equation}
then 
$\lambda(p)$ is \underline{indeterminate} in general, hence can take any value. Since zero is not excluded in (\ref{lambda=}) we call the zero of $\lambda(p)$ the null object and denote by 0, {\it i.e.,}
\[
0\in\lambda(p).
\]
We now focus our attention to this subtle object in the following discussion and show how the localization of triangulated category resolves the subtlety.\\

The localization of our system will be introduced by considering rational maps of the $\tau$ functions.
To be specific we consider some reduced flow diagrams of Fig.\ref{fig7} which satisfy the condition (\ref{j+d=j}). In particular we study in detail rational maps defined by the following variables:
\begin{equation}
x_j^{[t]}=\left\{\begin{array}{cll}
\displaystyle{{\tau_{j+\epsilon +1}^{[t]}\tau_{j-\epsilon}^{[t+1]}\over \tau_{j+1}^{[t]}\tau_{j}^{[t+1]}}},&\quad j,\epsilon=1,2,3,...,d,&\quad{\rm LV},\cr
&&\cr
\displaystyle{{\tau_{j+1}^{[t]}\tau_{j-\epsilon}^{[t+1]}\over 
\tau_{j}^{[t]}\tau_{j+1-\epsilon}^{[t+1]}}},&\quad j,\epsilon=1,2,3,...,d,&\quad{\rm KdV}.\cr
\end{array}\right.
\label{rational}
\end{equation}
The maps which are obtained by the new variables LV and KdV are called the Lotka-Volterra map and the Korteweg-de Vries map, respectively \cite{HTI, HT2}. 

An important feature of the variables (\ref{rational}) is that they are invariant under the local gauge transformation of $\tau_j^{[t]}$
\begin{equation}
g:\ \tau_j^{[t]}\to \exp\left(\int^t\nu(t',j)dt'+\int^j\mu(t,j')dj'\right)\tau_j^{[t]},
\label{gauge}
\end{equation}
where $\nu(t,j)$ and $\mu(t,j)$ are arbitrary functions. For example, 
we can write the right hand side of  (\ref{rational}) as
\begin{equation}
x_j^{[t]}=\left\{\begin{array}{cll}
\displaystyle{{\Phi_{j+\epsilon +1}^{[t]}(p,z)\Phi_{j-\epsilon}^{[t+1]}(p,z)\over \Phi_{j+1}^{[t]}(p,z)\Phi_{j}^{[t+1]}(p,z)}},&\quad j,\epsilon=1,2,3,...,d,&\quad{\rm LV},\cr
&&\cr
\displaystyle{{\Phi_{j+1}^{[t]}(p,z)\Phi_{j-\epsilon}^{[t+1]}(p,z)\over 
\Phi_{j}^{[t]}(p,z)\Phi_{j+1-\epsilon}^{[t+1]}(p,z)}},&\quad j,\epsilon=1,2,3,...,d,&\quad{\rm KdV}.\cr
\end{array}\right.
\label{rational2}
\end{equation}
This follows from the fact that the denominator of $\tau_{j+k}^{[t+u]}(p)$ is given by
\[
\Phi_{j+k}^{[t+u]}(p,z,0)=(z_1-z_2)^{(p_1-k)(p_2+k)}(z_3-z_4)^{(p_3+u)(p_4-u)}
\]
so that all denominators of $\tau$ functions in (\ref{rational}) are eliminated exactly from the expression. 

We notice that we can not distinguish a change of $\lambda(p)$ in (\ref{lambda=}) with the gauge transformation
\begin{equation}
g: \lambda(p) \longrightarrow \lambda'(p).	
\label{gauge}
\end{equation}
This means that, if $\Lambda(\infty)$ is the set of all possible $\lambda(p)$, {\it i.e.,}
\begin{equation}
\Lambda(\infty):=\Big\{ \lambda(p)\Big| p\in\Xi_4^{(n+2)}\Big\}.
\label{Lambda}
\end{equation}
$\Lambda(\infty)$ is invariant under the gauge transformation.\\

Now suppose that $\tau_{j}^{[t+1]}$ (or $\tau_{j+1-\epsilon}^{[t+1]}$ in the KdV case) in the denominator of $x_j^{[t]}$ in (\ref{rational}) is the null object, hence takes the value zero. Then $x_j^{[t]}$ and also $x_{j-1}^{[t+1]}$ diverge while all components of $\bm{x}^{[t+u]}$ with $u\ge 2$ are finite, as far as other $\tau$ functions are finite. This owes to the fact that the same $\tau$ function does not propagate beyond two steps. There is no way to determine the values of $\tau^{[t+1]}$'s because the null object is invariant under the gauge (\ref{gauge}), {\it i.e.,}
\[
g: 0\longrightarrow g0=0.
\]
In other words the null object is transfered to an indeterminate object,
\begin{equation}
d_T: 0\longrightarrow \lambda(p).
\label{d_T 0 to lambda}
\end{equation}
which is an element of $\Lambda(\infty)$. It should be emphasized that $\Lambda(\infty)$ does not appear in the localized theory, because the localized variables $x_j^{[t]}$'s are gauge invariant. Thus we are strongly suggested to identify  $\Lambda(\infty)$ with the null system
${\mathcal N}$ of our map
\begin{equation}
{\mathcal N}\sim\Lambda(\infty).
\label{conjecture}
\end{equation}
Our argument in the rest of this paper will be devoted to support this conjecture. 


\subsection{Singularity confinement}

To proceed our argument further we consider the case $d=3$ and $\epsilon=1$ in (\ref{rational}) for simplicity. Then the HM eq. becomes the following rational maps,
\begin{eqnarray}
&{\rm LV\ map}&\qquad
\left\{
\begin{array}{c}
x_1^{[t+1]}=x_1^{[t]}\displaystyle{{1-x_2^{[t]}+x_2^{[t]}x_3^{[t]}\over 1-x_3^{[t]}+x_3^{[t]}x_1^{[t]}}},\\
x_2^{[t+1]}=x_2^{[t]}\displaystyle{{1-x_3^{[t]}+x_3^{[t]}x_1^{[t]}\over 1-x_1^{[t]}+x_1^{[t]}x_2^{[t]}}},\\
x_3^{[t+1]}=x_3^{[t]}\displaystyle{{1-x_1^{[t]}+x_1^{[t]}x_2^{[t]}\over 1-x_2^{[t]}+x_2^{[t]}x_3^{[t]}}},\\
\end{array}\right.
\label{LV}\\
&{\rm KdV\ map}&\qquad
\left\{
\begin{array}{c}
x_1^{[t+1]} = x_1^{[t]}\displaystyle{{1+x_1^{[t]}x_3^{[t]}+x_1^{[t]}x_2^{[t]}(x_3^{[t]})^2\over 1+x_1^{[t]}x_2^{[t]}+(x_1^{[t]})^2x_2^{[t]}x_3^{[t]}}},\\
x_2^{[t+1]} = x_2^{[t]}\displaystyle{{1+x_1^{[t]}x_2^{[t]}
+(x_1^{[t]})^2x_2^{[t]}x_3^{[t]}\over 1+x_2^{[t]}x_3^{[t]}+x_1^{[t]}(x_2^{[t]})^2x_3^{[t]}}},\\
x_3^{[t+1]} = x_3^{[t]}\displaystyle{{1+x_2^{[t]}x_3^{[t]}+x_1^{[t]}(x_2^{[t]})^2x_3^{[t]}\over 1+x_1^{[t]}x_3^{[t]}+x_1^{[t]}x_2^{[t]}(x_3^{[t]})^2}}.\\
\end{array}\right.
\label{KdV}
\end{eqnarray}

These maps have two invariants. If $\bm{x}=(x_1,x_2,x_3)$ denotes the initial value $\bm{x}^{[0]}$, the invariants are given by
\begin{eqnarray}
&{\rm LV\ map}&\quad r=x_1x_2x_3,\quad s=(1-x_1)(1-x_2)(1-x_3),
\label{LV inv}\\
&{\rm KdV\ map}&\quad r=x_1x_2x_3,\quad s=(1+x_1x_2)(1+x_2x_3)(1+x_3x_1),
\end{eqnarray}
respectively.
\bigskip

We can solve the initial value problem of the HM eq following to the algorithm:\\
\begin{enumerate}
\item[A1] Fix initial values $\bm{\tau}^{[0]}=(\tau_0^{[0]},\tau_1^{[0]},\tau_2^{[0]})$
and $\bm{\tau}^{[1]}=(\tau_0^{[1]},\tau_1^{[1]},\tau_2^{[1]})$ by hand to determine $\bm{x}=(x_1, x_1, x_3)$.

\item[A2] 
$\bm{x}^{[t]}:=(x_1^{[t]},x_2^{[t]},x_3^{[t]}),\ \ t\ge 1$ are obtained as functions of $\bm{x}$ iteratively by the maps (\ref{LV}) or (\ref{KdV}).
\item[A3] $\bm{\tau}^{[t+1]}=(\tau_0^{[t+1]},\tau_1^{[t+1]},\tau_2^{[t+1]}),\ \  t\ge 1$ are determined by (\ref{rational}) from $\bm{\tau}^{[t]}$ and $\bm{x}^{[t]}$.
\end{enumerate}
\bigskip

Needless to say this procedure of solving the HM eq. (\ref{HM}) is compatible with the flow of information through the chain of octahedra, since the rational maps (\ref{LV}) and (\ref{KdV}) are derived from the HM eq. by the transformation of dependent variables (\ref{rational}). The algorithm is certainly deterministic, since values of $\bm{\tau}^{[t]}$ for all $t\ge 2$ are determined if the initial values $\bm{\tau}^{[0]}$ and $\bm{\tau}^{[1]}$ are fixed. As we will show in the following, however, it becomes not clear how the null object appears during the procedure.\\

The singularity confinement of  the LV map (\ref{LV}) and KdV map  (\ref{KdV}) have been studied in detail \cite{SS2, IVPP, SS3}. To see what happens we review this problem from the view point of the theory of category. 

Since we are interested in studying the singularity confinement we fix the initial conditions such that $\bm{x}^{[1]}$ is divergent.
Without loss of generality this condition is satisfied by requiring for the denominator of $x_1^{[1]}$ to vanish. Let us solve the 3 dimensional LV map case following to our algorithm.

\begin{enumerate}
\item[A1]
We fix the initial values $\bm{\tau}^{[1]}$ at
\[
\bm{\tau}^{[1]}=(\lambda_0,\lambda_1,\lambda_2)
\] and, instead of fixing $\bm{\tau}^{[0]}$ by hand, we require
\begin{enumerate}
\item denominator of $x_1^{[1]}$ vanishes:
\[
1-x_3+x_3x_1=0
\]
\item invariants $r,s$ are fixed by
\[
r=x_1x_2x_3,\qquad
s=(1-x_1)(1-x_2)(1-x_3),
\]
\end{enumerate}
from which we obtain
\begin{equation}
\bm{x}
:=\left({r-s\over r+1}, r{s+1\over r-s}, {r+1\over s+1}\right),
\label{p^0}
\end{equation}
and
\[
\bm{\tau}^{[0]}=\Big((r-s)\lambda_1,\ (s+1)\lambda_2,\ (r+1)\lambda_0\Big).
\]

\item[A2]
Iteration of the map (\ref{LV}) yields the sequence of singularity confinement,
\begin{equation}
\bm{x}\to (\infty,0,1) \to (1,0,\infty)\to \bm{x}^{[3]}\to \bm{x}^{[4]}\to\cdots
\label{p^0top^1to..}
\end{equation}
where
\[
\bm{x}^{[3]}=
\left({\alpha^{(2)}\over\gamma^{(2)}},\ r{\gamma^{(2)}\over\beta^{(2)}},\ {\beta^{(2)}\over\alpha^{(2)}}\right),\quad
\bm{x}^{[4]}=\left({\alpha^{(2)}\alpha^{(3)}\over \beta^{(2)}\gamma^{(3)}},
r{\gamma^{(2)}\gamma^{(3)}\over \beta^{(2)}\beta^{(3)}}, 
{\beta^{(2)}\beta^{(3)}\over \gamma^{(2)}\alpha^{(3)}}\right),
\]
\begin{equation}
\alpha^{(2)}=r+1,\quad \beta^{(2)}=r-s,\quad \gamma^{(2)}=s+1,
\label{gamma2}
\end{equation}
\[
\alpha^{(3)}:=r^2-3rs-s-rs^2,\qquad
\beta^{(3)}:=sr^2+3rs-s^2+r,
\]
\begin{equation}
\gamma^{(3)}:=r^2+s^2+r+s-rs+1.
\label{gamma3}
\end{equation}

\item[A3] 

\begin{enumerate}
\item  From $\bm{x}^{[1]}=(\infty, 0, 1)$, we find\ \  $\tau_1^{[2]}=0\ $ and $\ \tau_2^{[2]}\lambda_2=\tau_0^{(2)}\lambda_1$.
Since an overall factor is irrelevant we obtain
\[
\bm{\tau}^{[2]}=(\lambda_2, 0, \lambda_1).
\]
\item 
From $\bm{x}^{[2]}=(1, 0, \infty)$ we find $\tau_0^{[3]}\lambda_2=\tau_1^{[3]}\lambda_1$, 
but $\tau_2^{[3]}$ is undetermined.

\item Since $\bm{x}^{[t]}$'s are finite for all $t\ge 3$, the rest of $\bm{\tau}^{[t+1]}$ are determined for all $t\ge 3$, thus we obtain, up to overall factors,
\[
\bm{\tau}^{[0]}\to
(\lambda_0,\lambda_1,\lambda_2)\to (\lambda_2,0,\lambda_1)\to (\lambda'_0,\lambda'_1,\lambda'_2)\qquad\qquad\qquad\qquad\qquad\qquad
\]
\begin{equation}
\to \Big(\lambda'_2\alpha^{(2)},\lambda'_0\gamma^{(2)}, \lambda'_1\beta^{(2)}\Big)
\to\left(\lambda'_1\alpha^{(3)}, \lambda'_2\gamma^{(3)}, \lambda'_0\beta^{(3)}\right)\to\cdots
\label{sequence}
\end{equation}
Here we defined new functions 
\[
(\lambda'_0, \lambda'_1,\lambda'_2):=(\tau_0^{[3]},\tau_1^{[3]},\tau_2^{[3]}),
\]
which are free as far as 
\begin{equation}
\lambda'_0\lambda_2=\lambda'_1\lambda_1
\label{lambda'lambda=lambda'lambda}
\end{equation}
is satisfied.
\end{enumerate}
\end{enumerate}

From this result it is clear how the singularity confinement undergoes.
The singularities of $\bm{x}^{[1]}$ and $\bm{x}^{[2]}$ in (\ref{p^0top^1to..}) come from $\tau_1^{[2]}=0$. This null object is the source of the singularities.
This information, however, can transfer only to its neighbor 
since $\tau_1^{[2]}$ does not appear beyond $\bm{x}^{[2]}$. Hence it does not transfer directly to remote objects. 

We can extend the sequence of  (\ref{sequence}) to the left, if we apply the inverse map of  (\ref{LV}) to $\bm{x}$. We find, with $\lambda_{t+3}:=\lambda_t,\ \lambda'_{t+3}:=\lambda'_t$,
\begin{equation}
\bm{\tau}^{[t+2]}=\left\{
\begin{array}{cl}
\Big(
\lambda'_{1-t}\alpha^{(t)},\ \ \lambda'_{2-t}\gamma^{[t]},\ \ \lambda'_{-t}\beta^{(t)}\Big),&
t\ge 2,\\
(\lambda'_0, \lambda'_1, \lambda'_2), &t=1,\\
(\lambda_2,\ 0,\ \lambda_1),&t=0,\\
(\lambda_0, \lambda_1, \lambda_2),& t=-1,\\
\Big(\lambda_{2-t}\beta^{(-t)},\lambda_{-t}\gamma^{[-t]},\lambda_{1-t}\alpha^{(-t)}\Big),&
t\le -2.\\
\end{array}\right.
\label{tau to tau to}
\end{equation}
Here we denote by $\gamma^{[t]}$ the product
\[
\gamma^{[t]}:=\prod_{\{\nu\}}\gamma^{(\nu)}
\]
and $\{\nu\}$ means the set of all prime numbers which divides $t$. Explicit forms of $\alpha^{(t)},\ \beta^{(t)},\ \gamma^{(t)}$ are given in Appendix and (\ref{gamma}).

\begin{figure}[h]
\begin{center}
\input{nullsystem.tex}
\caption{$\bm{\tau}^{[t]},\ t\in\mathbb{Z}$}
\label{nullsystem}
\end{center}
\end{figure}
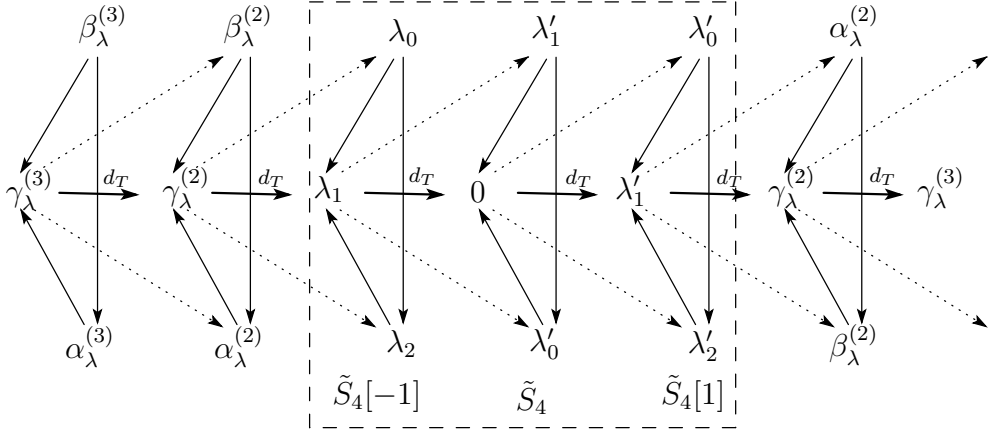

We can summarize our result of this subsection by the diagram in Fig.\ref{nullsystem}.


\noindent
Here we used the notations $\alpha_\lambda^{(t)}, \beta_\lambda^{(t)}$ and $\gamma_\lambda^{[t]}$ to represent 
the elements of $\bm{\tau}$ in (\ref{tau to tau to}).

From this analysis we learned that the existence of a null object 
introduces free functions $(\lambda'_0, \lambda'_1, \lambda'_2)$ in addition to the 
initial values $(\lambda_0, \lambda_1, \lambda_2)$. Since it is constrained by (\ref{lambda'lambda=lambda'lambda}) and the overall factor is irrelevant there is only one degree of freedom. As we have seen it comes from the gauge invariance of the null object $g 0\sim  0$.

\subsection{Projective resolution}

We are now ready to apply the localization theorem of the category theory discussed in \S 5.2. The diagram of Fig.\ref{nullsystem} shows how the effect of this gauge freedom propagates as $t$ increases. It is important to notice that 
all objects in three triangles $\tilde S_4[-1], \tilde S_4, \tilde S_4[1]$ are indeterminate. In other words
\[
\tilde S_4[-1], \tilde S_4, \tilde S_4[1]\subset \Lambda(\infty).
\]

If we accept the conjecture (\ref{conjecture}) to identify ${\mathcal N}\sim\Lambda(\infty)$, we naturally define our multiplicative system by the set of gauge transformations
\[
S({\mathcal N}):=\{\ g\ |\ g: \lambda(p)\to \lambda'(p),\ 0\to g0=0\},
\]
so that our local theory is obtained simply by gauge fixing all $\lambda_j$'s and $\lambda'_j$'s at 1.

\begin{center}
\input{localnullsystem.tex}
\end{center}
\centerline{\bf Fig. 9}
\bigskip

Let us consider a subchain of Fig.\ref{nullsystem},\\

\begin{center}
\input{reductionnullsystem.tex}
\end{center}
\bigskip

\centerline{\bf Fig. 10}
\bigskip

\noindent
which can be obtained by iteration of the map $\bm{x}^{[t]}\to \bm{x}^{[t+k]}$. Especially in the diagram\\

\begin{center}
\input{projectiveobject.tex}
\end{center}
\bigskip

\centerline{\bf Fig. 11}
\bigskip
\noindent
the morphism
$u:\ \gamma_\lambda^{[k]}\to \lambda'_0$ passes the epimorphism $\pi: \lambda'_1\to\lambda'_0$. Hence the object $\gamma_\lambda^{[k]}$ is a projective object for all $k$, and the exact sequence
\begin{equation}
P:=\cdots \to \gamma_\lambda^{[4]}\to \gamma_\lambda^{[3]}\to \gamma_\lambda^{[2]}\to\lambda_1\to \lambda'_0\to 0
\label{projective resolution}
\end{equation}
is a projective resolution of $\lambda'_0$.

This is a result of the theory of triangulated category in mathematics\cite{resolution}. It tells us that infinitely many projections by $\gamma_\lambda^{[t]}$'s constitute the object $\lambda'_0$. But it is certainly unclear at this moment if it has something to do with integrability of the map. To explain their relation we must recall some results of our previous works.

\subsection{Invariant varieties of periodic points (IVPP)}

\subsubsection{Generation of IVPPs}

Let us consider a $d$ dimensional rational map whose initial point is at $\bm{x}=(x_1,x_2,...,x_d)$. The period $t$ condition of the map is a set of $d$ equations
\begin{equation}
\bm{x}^{[t]}(\bm{x})-\bm{x}=0.
\label{periodicity cond}
\end{equation}
Then it was proved in our papers \cite{SS2, IVPP, SS3} the following theorem:
\bigskip

\noindent
{\bf IVPP theorem}

{\it All periodic points of the map form a variety for each period if the map has invariants  more than $d/2$ and there is no Julia set.}\\
  
The varieties are called invariant varieties of periodic points (IVPPs) because they are algebraic varieties which are determined by the invariants alone. Namely if we denote the variety by
\begin{equation}
v^{[t]}=\Big\{\bm{x}\Big| \gamma^{(t)}=0\Big\},
\label{v^n}
\end{equation}
the functions $\gamma^{(t)}$'s are polynomial functions of the invariants. They can be derived from (\ref{periodicity cond}) directly by means of the method of Gr\"obner bases, although it is not easy. 

In our recent papers \cite{SS3, YSW} we have shown, however, that, if we use the method of singularity confinement, we can derive them quite easily. In fact we have already encountered $\gamma^{(2)}$ in (\ref{gamma2}) and $\gamma^{(3)}$ in (\ref{gamma3}) for the 3dLV map in the previous subsection,
\[
\gamma^{(2)}=s+1,\qquad
\gamma^{(3)}=r^2+s^2+r+s-rs+1.
\]
If we continue the map we will find
\begin{eqnarray}
\gamma^{(4)}&=&(r-s)^3-s(r+1)^3\nonumber\\
\gamma^{(5)}&=&
-(r-s)^6+(r-2)(r+1)(s+1)(r-s)^4\nonumber\\
&&+(2s+r)(r+1)^2(s+1)^2(r-s)^2-s^2(r+1)^3(s+1)^3\nonumber\\
\gamma^{(6)}
&=&-3(r-s)^4-(r+1)(s+1)\big((r+1)(s+1)-3\big)(r-s)^2
\nonumber
\\
&&-3(r+1)(s+1)(r^3+s^3-r^2s^2-2r^2s-2rs^2-r^2-s^2)
\nonumber\\
{\rm etc}..&&
\label{gamma}
\end{eqnarray}

This can be understood as follows. Since $\tau_1^{[2]}=0$,  a point satisfying $\tau_1^{[t+2]}=0$ must include periodic points of period $t$. On the other hand $\tau_1^{[t+2]}\propto\gamma^{[t]}$ holds as we can see from (\ref{tau to tau to}). Hence the points on $v^{[t]}$ of (\ref{v^n}) are periodic points of period $t$. If we repeat the procedure explained in the previous subsections we derive a sequence of polynomial functions of the invariants. Thus we have found an important fact:
\bigskip

\centerline{\it $P$ of (\ref{projective resolution}) is a chain of polynomial functions whose zero sets are IVPPs.}

\subsubsection{Degeneration of IVPPs}

Now we must call attention to another remarkable feature of IVPPs. In \cite{IVPP, SS3} we have found that IVPPs of all periods intersect on a variety which we called a variety of singular points (VSP). It is a variety of singular points by two reasons. 
\begin{enumerate}
\item Every point on VSP is singular because it is a point which is occupied by periodic points of all periods simultaneously. 

\item It is an indeterminate point of the map. Namely VSP is a set of points on which the denominators and the numerators of the rational map vanish simultaneously.
\end{enumerate}

In order to explore this odd behavior on VSP we studied  deformation of integrable maps by introducing a parameter, say $a$, by hand. When $a\ne 0$, periodic points form sets of discrete points for each period, including the Julia set. We have shown that as $a$ becomes small some of these isolated points approach IVPPs, while a large part of them approach VSP and crash there altogether in the $a=0$ limit \cite{IVPP, SS3}. 

Moreover in our recent paper \cite{SSHYW} we have shown, by studying a simple map of 2 dimension, how the Julia set approach to VSP. We have found that the periodic points move along an algebraic curve for each period as $a$ changes its value. As $a$ becomes zero the points approach a singular locus of the curve for each period. This locus is common to all curves of different periods. Hence periodic points of all periods degenerate at this point. 

In all cases the VSP itself is a variety of fixed points when $a\ne 0$, while it turns to a set of indeterminate points as $a$ becomes zero. Therefore, from these observations, we learn that a large part of periodic points of a generic map approach to a set of fixed points as the map becomes integrable and the fixed points turn to VSP. \\

Now we can translate information of VSP into the language of $\tau$ functions using the formula (\ref{rational}). 
In the case of LV map (\ref{LV}) the points $\bm{x}^{[1]}=(\infty,0,1)$ and $\bm{x}^{[2]}=(1,0,\infty)$ are on VSP. From the algorithm of \S 5.3 they correspond exactly to the three triangles $\Big(\tilde S_4[-1], \tilde S_4, \tilde S_4[1]\Big)$ of the chain in Fig.\ref{nullsystem}. The latter belongs to $\Lambda(\infty)$, which is a set of indeterminate points again, but of the $\tau$ functions instead of the map functions. 

The correspondence of VSP with $\Lambda(\infty)$ is clear because the indeterminacy of both functions comes from the same source, {\it i.e.}, the zero set of correlation functions $\Phi(p,z;G)$. Moreover from this correspondence we see that $\lambda'_0$ and $\lambda_1$ in $P$ of (\ref{projective resolution}) must be objects  in which all IVPPs are degenerate.   In other words the null set is a source from which IVPPs are generated. This will provide us an interpretation of the projective resolution.

\section{Conclusion}

Before we close this paper we would like to discuss briefly about the characterization of integrability. 

The Julia set is the closure of the set of unstable periodic points of iterations of a map \cite{Devaney}. If there is a Julia set the map is nonintegrable, because the map has some chaotic orbits whose behavior no one can predict. The existence of the Julia set is  not necessary but sufficient for a map being nonintegrable.

The IVPP theorem, which we explained in \S 5.5.1, tells us that a Julia set and an IVPP of any period can not exist in one map simultaneously. This, however does not mean that the existence of an IVPP is sufficient to guarantee integrability of a rational map in general, because the theorem was proved only for maps with sufficient number of invariants. Nevertheless it will be worthwhile to study conditions for IVPPs being generated from the null object, in order to clarify the notion of integrability.\\

The rational maps, which we studied in this paper, belong to the KP hierarchy. The dynamical variables of these maps are related to the $\tau$ functions of the HM eq. by the formulas (\ref{rational}) of the form 
\begin{equation}
x_j^{[t]}={\tau_a^{[t]}\tau_b^{[t+1]}\over \tau_c^{[t]}\tau_d^{[t+1]}}.
\label{tautau/tautau}
\end{equation}
This particular form of the transformation is the key of all our arguments. 

In order to explain what this means, let us consider an arbitrary rational map $\bm{x}\to f(\bm{x})$. As we repeat the map $t$ times, we will obtain a rational function $f^{[t]}(\bm{x})$ whose degree increases exponentially as $t$ increases, unless cancellations of factors in the numerator and the denominator happen to take place. If an initial condition was such that one of the factors of the denominators vanishes the map will diverge. This particular factor appears in the map repeatedly since it has no chance to be canceled. In other words the singularity confinement does not take place in general.

In the case of the map of the KP hierarchy, on the other hand, the function $f^{[t]}(\bm{x})$ is always factorized in the form (\ref{tautau/tautau}). This fact guarantees that the same factor is not transferred beyond two steps, hence the singularity confinement becomes possible. We can understand this phenomenon from the very construction of the $\tau$ functions. Namely the zero of a $\tau$ function does not forces its neighbor to vanish, because of the Grassmann nature of the correlation function $\Phi(p,z;G)$.

Now we want to rephrase this phenomenon in terms of the category theory.
The IVPPs are generated from the null set of the triangulated category as the resolution of an object. This is a natural consequence of the localization of the $\tau$ functions. The localization itself is supported by the axioms of the triangulated category as far as there exists a null object. Therefore the integrability of the KP hierarchy is guaranteed by the axioms of distinguished triangles.



\appendix

\section{$\alpha^{(t)}$ and $\beta^{(t)}$}

\begin{eqnarray*}
\alpha^{(2)}&=&r+1\\
\alpha^{(3)}&=&-rs^2+r^2-3rs-s\\
\alpha^{(4)}&=&r^4s+3r^3s+r^2s^3+6r^2s+r^2-r^3s^2-s^4r-6rs^2-3rs^3-s^2\\
\alpha^{(5)}&=&-s^6r-3rs^5+3r^2s^5-6s^4r-6r^3s^4+3s^4r^2+r^2s^3-r^3s^3+10r^4s^3\\
&&-10rs^3-s^3+s^3r^5+21r^2s^2
+21r^4s^2+27r^3s^2+6r^2s+3r^3s-3r^4s\\
&&-6r^5s+r^5+r^2+r^4+r^3+r^6\\
\beta^{(2)}&=&r-s\\
\beta^{(3)}&=&3rs+r-s^2+r^2s\\
\beta^{(4)}&=&r^4-3r^3s+r^2s^3+6r^2s^2+r^2+r^3+3rs+6rs^2+r-s^3\\
\beta^{(5)}&=&-s^6r^2+s^5-6r^2s^5+s^5r^3-21s^4r^2-s^4r^4+3r^3s^4-10s^4r+3r^4s^3\\
&&-r^2s^3-6rs^3+27r^3s^3+s^3r^5-3r^2s^2-3rs^2-21r^4s^2-r^3s^2-6r^5s^2\\
&&-r^6s^2-rs-3r^2s-6r^3s-10r^4s+r^5,\\
etc..&&
\end{eqnarray*}

\end{document}

%% file: Riemannsurface.tex
\unitlength 1pt
\begin{picture}(203.8014, 80.9424)( 19.5129,-96.1191)
%
\special{pn 4}%
\special{sh 1}%
\special{ar 660 900 8 8 0  6.28318530717959E+0000}%
\special{sh 1}%
\special{ar 660 900 8 8 0  6.28318530717959E+0000}%
%
\special{pn 4}%
\special{sh 1}%
\special{ar 1700 1010 8 8 0  6.28318530717959E+0000}%
\special{sh 1}%
\special{ar 1700 1010 8 8 0  6.28318530717959E+0000}%
%
\special{pn 4}%
\special{sh 1}%
\special{ar 2080 910 8 8 0  6.28318530717959E+0000}%
\special{sh 1}%
\special{ar 2080 910 8 8 0  6.28318530717959E+0000}%
%
\special{pn 4}%
\special{sh 1}%
\special{ar 1310 740 8 8 0  6.28318530717959E+0000}%
\special{sh 1}%
\special{ar 1310 740 8 8 0  6.28318530717959E+0000}%
\put(48.4209,-75.1608){\makebox(0,0)[lb]{$z_1$}}%
\put(86.0013,-62.8749){\makebox(0,0)[lb]{$z_2$}}%
\put(125.0271,-73.7154){\makebox(0,0)[lb]{$z_3$}}%
\put(153.2124,-73.7154){\makebox(0,0)[lb]{$z_4$}}%
\put(172.0026,-54.2025){\makebox(0,0)[lb]{$|G\rangle$}}%
%
\special{pn 8}%
\special{ar 1680 840 1410 430  0.8176450 6.2831853}%
\special{ar 1680 840 1410 430  0.0000000 0.7735580}%
%
\special{pn 8}%
\special{pa 900 890}%
\special{pa 918 920}%
\special{pa 936 948}%
\special{pa 956 970}%
\special{pa 982 986}%
\special{pa 1012 996}%
\special{pa 1046 1002}%
\special{pa 1080 1006}%
\special{pa 1114 1008}%
\special{pa 1148 1008}%
\special{pa 1180 1002}%
\special{pa 1210 992}%
\special{pa 1236 978}%
\special{pa 1256 956}%
\special{pa 1274 928}%
\special{pa 1290 900}%
\special{pa 1300 880}%
\special{sp}%
%
\special{pn 8}%
\special{pa 1010 1000}%
\special{pa 1030 974}%
\special{pa 1050 950}%
\special{pa 1076 930}%
\special{pa 1106 918}%
\special{pa 1136 912}%
\special{pa 1170 908}%
\special{pa 1202 908}%
\special{pa 1234 908}%
\special{pa 1260 910}%
\special{sp}%
%
\special{pn 8}%
\special{pa 1530 730}%
\special{pa 1542 762}%
\special{pa 1556 790}%
\special{pa 1576 816}%
\special{pa 1600 834}%
\special{pa 1630 848}%
\special{pa 1664 854}%
\special{pa 1700 852}%
\special{pa 1732 846}%
\special{pa 1764 834}%
\special{pa 1790 818}%
\special{pa 1812 796}%
\special{pa 1832 770}%
\special{pa 1850 742}%
\special{pa 1868 714}%
\special{pa 1880 690}%
\special{sp}%
%
\special{pn 8}%
\special{pa 1580 790}%
\special{pa 1612 786}%
\special{pa 1644 780}%
\special{pa 1676 776}%
\special{pa 1708 778}%
\special{pa 1738 788}%
\special{pa 1764 806}%
\special{pa 1780 820}%
\special{sp}%
\put(29.6307,-30.3534){\makebox(0,0)[lb]{$p_1$}}%
\put(93.9510,-24.5718){\makebox(0,0)[lb]{$p_2$}}%
\put(135.1449,-39.7485){\makebox(0,0)[lb]{$p_4$}}%
\put(148.8762,-101.1780){\makebox(0,0)[lb]{$p_3$}}%
%
\special{pn 4}%
\special{sh 1}%
\special{ar 1280 1140 8 8 0  6.28318530717959E+0000}%
\special{sh 1}%
\special{ar 1280 1140 8 8 0  6.28318530717959E+0000}%
\put(95.3964,-83.8332){\makebox(0,0)[lb]{$z$}}%
%
\special{pn 8}%
\special{pa 2730 710}%
\special{pa 2756 732}%
\special{pa 2780 754}%
\special{pa 2798 778}%
\special{pa 2814 806}%
\special{pa 2820 838}%
\special{pa 2820 872}%
\special{pa 2812 906}%
\special{pa 2796 938}%
\special{pa 2772 960}%
\special{pa 2744 972}%
\special{pa 2712 970}%
\special{pa 2710 970}%
\special{sp}%
%
\special{pn 8}%
\special{pa 2790 770}%
\special{pa 2766 792}%
\special{pa 2748 816}%
\special{pa 2740 848}%
\special{pa 2744 882}%
\special{pa 2758 914}%
\special{pa 2784 930}%
\special{pa 2790 930}%
\special{sp}%
%
\special{pn 8}%
\special{pa 2520 520}%
\special{pa 2544 544}%
\special{pa 2566 568}%
\special{pa 2586 592}%
\special{pa 2606 618}%
\special{pa 2622 644}%
\special{pa 2636 674}%
\special{pa 2646 702}%
\special{pa 2654 734}%
\special{pa 2660 766}%
\special{pa 2662 798}%
\special{pa 2662 832}%
\special{pa 2660 864}%
\special{pa 2656 896}%
\special{pa 2652 928}%
\special{pa 2646 960}%
\special{pa 2638 990}%
\special{pa 2628 1022}%
\special{pa 2618 1052}%
\special{pa 2598 1112}%
\special{pa 2588 1142}%
\special{pa 2580 1160}%
\special{sp -0.045}%
%
\special{pn 8}%
\special{pa 660 900}%
\special{pa 460 450}%
\special{fp}%
%
\special{pn 8}%
\special{pa 1310 740}%
\special{pa 1280 290}%
\special{fp}%
%
\special{pn 8}%
\special{pa 1710 1020}%
\special{pa 2060 1330}%
\special{fp}%
%
\special{pn 8}%
\special{pa 2080 910}%
\special{pa 1960 580}%
\special{fp}%
%
\special{pn 8}%
\special{pa 1280 1140}%
\special{pa 920 1310}%
\special{fp}%
%
\special{pn 4}%
\special{sh 1}%
\special{ar 1310 760 14 14 0  6.28318530717959E+0000}%
\special{sh 1}%
\special{ar 1310 760 14 14 0  6.28318530717959E+0000}%
%
\special{pn 4}%
\special{sh 1}%
\special{ar 660 900 14 14 0  6.28318530717959E+0000}%
\special{sh 1}%
\special{ar 660 900 14 14 0  6.28318530717959E+0000}%
%
\special{pn 4}%
\special{sh 1}%
\special{ar 2080 910 14 14 0  6.28318530717959E+0000}%
\special{sh 1}%
\special{ar 2080 910 14 14 0  6.28318530717959E+0000}%
%
\special{pn 4}%
\special{sh 1}%
\special{ar 1700 1020 14 14 0  6.28318530717959E+0000}%
\special{sh 1}%
\special{ar 1700 1020 14 14 0  6.28318530717959E+0000}%
%
\special{pn 4}%
\special{sh 1}%
\special{ar 1280 1140 14 14 0  6.28318530717959E+0000}%
\special{sh 1}%
\special{ar 1280 1140 14 14 0  6.28318530717959E+0000}%
\end{picture}%

%% file: Nordfjordpic38nocolor.tex
\unitlength 1pt
\begin{picture}(126.1834,103.2738)(  5.7816,-114.1866)
\put(129.5801,-78.7020){\makebox(0,0)[lb]{$d_T$}}%
\put(119.6069,-48.9991){\makebox(0,0)[lb]{$S_4$}}%
%
\special{pn 8}%
\special{pa 464 406}%
\special{pa 464 858}%
\special{fp}%
%
\special{pn 8}%
\special{pa 464 406}%
\special{pa 1500 406}%
\special{fp}%
\special{pa 1500 406}%
\special{pa 464 858}%
\special{fp}%
%
\special{pn 8}%
\special{pa 464 406}%
\special{pa 256 1536}%
\special{fp}%
%
\special{pn 8}%
\special{pa 464 858}%
\special{pa 256 1536}%
\special{fp}%
%
\special{pn 8}%
\special{pa 256 1536}%
\special{pa 1292 1536}%
\special{fp}%
\special{pa 1292 1536}%
\special{pa 464 858}%
\special{fp}%
%
\special{pn 8}%
\special{pa 1500 406}%
\special{pa 1292 1084}%
\special{fp}%
%
\special{pn 8}%
\special{pa 1292 1084}%
\special{pa 1292 1536}%
\special{fp}%
%
\special{pn 8}%
\special{pa 1292 1084}%
\special{pa 816 710}%
\special{fp}%
%
\special{pn 8}%
\special{pa 1292 1084}%
\special{pa 910 1220}%
\special{fp}%
%
\special{pn 4}%
\special{pa 796 406}%
\special{pa 464 768}%
\special{fp}%
\special{pa 858 406}%
\special{pa 464 836}%
\special{fp}%
\special{pa 920 406}%
\special{pa 536 824}%
\special{fp}%
\special{pa 982 406}%
\special{pa 640 778}%
\special{fp}%
\special{pa 1044 406}%
\special{pa 744 734}%
\special{fp}%
\special{pa 1106 406}%
\special{pa 848 688}%
\special{fp}%
\special{pa 1168 406}%
\special{pa 950 644}%
\special{fp}%
\special{pa 1230 406}%
\special{pa 1054 598}%
\special{fp}%
\special{pa 1292 406}%
\special{pa 1158 554}%
\special{fp}%
\special{pa 1356 406}%
\special{pa 1260 508}%
\special{fp}%
\special{pa 1418 406}%
\special{pa 1364 462}%
\special{fp}%
\special{pa 734 406}%
\special{pa 464 700}%
\special{fp}%
\special{pa 670 406}%
\special{pa 464 632}%
\special{fp}%
\special{pa 608 406}%
\special{pa 464 566}%
\special{fp}%
\special{pa 546 406}%
\special{pa 464 496}%
\special{fp}%
%
\special{pn 4}%
\special{pa 1272 1072}%
\special{pa 734 1072}%
\special{dt 0.027}%
\special{pa 1148 970}%
\special{pa 608 970}%
\special{dt 0.027}%
\special{pa 1014 870}%
\special{pa 484 870}%
\special{dt 0.027}%
\special{pa 888 768}%
\special{pa 682 768}%
\special{dt 0.027}%
\special{pa 1024 1174}%
\special{pa 858 1174}%
\special{dt 0.027}%
%
\special{pn 4}%
\special{pa 432 970}%
\special{pa 360 970}%
\special{dt 0.027}%
\special{pa 402 1072}%
\special{pa 340 1072}%
\special{dt 0.027}%
\special{pa 370 1174}%
\special{pa 318 1174}%
\special{dt 0.027}%
\special{pa 464 870}%
\special{pa 380 870}%
\special{dt 0.027}%
\special{pa 464 768}%
\special{pa 402 768}%
\special{dt 0.027}%
\special{pa 464 666}%
\special{pa 412 666}%
\special{dt 0.027}%
\special{pa 464 566}%
\special{pa 432 566}%
\special{dt 0.027}%
\put(22.4037,-26.7399){\makebox(0,0)[lb]{$\tau_{14}$}}%
\put(103.6352,-27.3181){\makebox(0,0)[lb]{$\tau_{12}$}}%
\put(10.1178,-123.5817){\makebox(0,0)[lb]{$\tau_{34}$}}%
\put(40.0376,-68.8733){\makebox(0,0)[lb]{$\tau_{24}$}}%
\put(79.9306,-69.8851){\makebox(0,0)[lb]{$\tau_{31}$}}%
\put(92.5056,-122.8590){\makebox(0,0)[lb]{$\tau_{23}$}}%
\put(38.6644,-97.7813){\makebox(0,0)[lb]{$S_1$}}%
\put(59.9841,-79.6415){\makebox(0,0)[lb]{$S_2$}}%
\put(43.1452,-46.5419){\makebox(0,0)[lb]{$S_3$}}%
%
\special{pn 8}%
\special{pa 1568 1052}%
\special{pa 1826 1132}%
\special{fp}%
\special{sh 1}%
\special{pa 1826 1132}%
\special{pa 1768 1092}%
\special{pa 1776 1116}%
\special{pa 1756 1132}%
\special{pa 1826 1132}%
\special{fp}%
\put(64.1758,-20.3079){\makebox(0,0)[lb]{${\mathcal O}$}}%
\put(5.7816,-46.3251){\makebox(0,0)[lb]{$\tilde S_4$}}%
%
\special{pn 8}%
\special{pa 1406 778}%
\special{pa 1456 788}%
\special{pa 1502 796}%
\special{pa 1542 798}%
\special{pa 1570 794}%
\special{pa 1584 780}%
\special{pa 1580 754}%
\special{pa 1572 722}%
\special{pa 1582 696}%
\special{pa 1612 676}%
\special{pa 1634 666}%
\special{sp}%
%
\special{pn 8}%
\special{pa 402 914}%
\special{pa 356 898}%
\special{pa 314 880}%
\special{pa 280 862}%
\special{pa 258 842}%
\special{pa 250 820}%
\special{pa 262 796}%
\special{pa 292 782}%
\special{pa 316 774}%
\special{pa 320 752}%
\special{pa 308 720}%
\special{pa 284 682}%
\special{pa 266 656}%
\special{sp}%
%
\special{pn 8}%
\special{pa 686 1022}%
\special{pa 686 1524}%
\special{dt 0.045}%
\special{pa 804 1130}%
\special{pa 804 1524}%
\special{dt 0.045}%
\special{pa 922 1224}%
\special{pa 922 1524}%
\special{dt 0.045}%
\special{pa 1042 1320}%
\special{pa 1042 1524}%
\special{dt 0.045}%
\special{pa 1160 1414}%
\special{pa 1160 1524}%
\special{dt 0.045}%
\special{pa 568 926}%
\special{pa 568 1524}%
\special{dt 0.045}%
\special{pa 448 900}%
\special{pa 448 1524}%
\special{dt 0.045}%
\special{pa 330 1280}%
\special{pa 330 1524}%
\special{dt 0.045}%
%
\special{pn 8}%
\special{pa 1502 398}%
\special{pa 1292 1524}%
\special{fp}%
%
\special{pn 4}%
\special{pa 1356 1184}%
\special{pa 1292 1184}%
\special{dt 0.027}%
\special{pa 1330 1306}%
\special{pa 1292 1306}%
\special{dt 0.027}%
\special{pa 1384 1062}%
\special{pa 1292 1062}%
\special{dt 0.027}%
\special{pa 1396 940}%
\special{pa 1330 940}%
\special{dt 0.027}%
\special{pa 1422 818}%
\special{pa 1370 818}%
\special{dt 0.027}%
\end{picture}%

%% file: Nordfjordpic20.tex
\unitlength 1pt
\begin{picture}(352.6776,138.7584)( 14.4540,-157.5486)
%
\special{pn 8}%
\special{pa 3880 430}%
\special{pa 3500 1130}%
\special{fp}%
\special{pa 3500 1130}%
\special{pa 3880 1830}%
\special{fp}%
\put(270.2898,-28.9080){\makebox(0,0)[lb]{$\tau_{14}$}}%
\put(234.1548,-86.0013){\makebox(0,0)[lb]{$\tau_{34}$}}%
\put(270.2898,-145.2627){\makebox(0,0)[lb]{$\tau_{24}$}}%
\put(231.2640,-51.3117){\makebox(0,0)[lb]{$p_3-p_1$}}%
\put(231.2640,-117.0774){\makebox(0,0)[lb]{$p_3-p_2$}}%
\put(292.6935,-28.1853){\makebox(0,0)[lb]{$p_4-p_3$}}%
\put(330.2739,-145.9854){\makebox(0,0)[lb]{$\tau_{23}$}}%
\put(367.1316,-85.2786){\makebox(0,0)[lb]{$\tau_{12}$}}%
\put(341.1144,-29.6307){\makebox(0,0)[lb]{$\tau_{13}$}}%
%
\special{pn 8}%
\special{pa 680 450}%
\special{pa 1880 450}%
\special{fp}%
%
\special{pn 8}%
\special{pa 680 450}%
\special{pa 680 1050}%
\special{fp}%
%
\special{pn 8}%
\special{pa 680 1050}%
\special{pa 1880 450}%
\special{fp}%
%
\special{pn 8}%
\special{pa 680 450}%
\special{pa 380 1750}%
\special{fp}%
%
\special{pn 8}%
\special{pa 380 1750}%
\special{pa 680 1050}%
\special{fp}%
\special{pa 680 1050}%
\special{pa 1580 1750}%
\special{fp}%
\special{pa 1580 1750}%
\special{pa 380 1750}%
\special{fp}%
%
\special{pn 8}%
\special{pa 1880 450}%
\special{pa 1580 1750}%
\special{fp}%
\special{pa 1580 1750}%
\special{pa 1580 1250}%
\special{fp}%
\special{pa 1580 1250}%
\special{pa 1580 1150}%
\special{fp}%
\special{pa 1580 1150}%
\special{pa 1880 450}%
\special{fp}%
%
\special{pn 8}%
\special{pa 1580 1150}%
\special{pa 1150 830}%
\special{fp}%
%
\special{pn 8}%
\special{pa 1580 1150}%
\special{pa 1090 1370}%
\special{fp}%
\put(102.6234,-72.9927){\makebox(0,0)[lb]{$\tau_{13}$}}%
\put(135.1449,-31.0761){\makebox(0,0)[lb]{$\tau_{12}$}}%
\put(15.1767,-137.3130){\makebox(0,0)[lb]{$\tau_{34}$}}%
\put(56.3706,-81.6651){\makebox(0,0)[lb]{$\tau_{24}$}}%
\put(39.7485,-30.3534){\makebox(0,0)[lb]{$\tau_{14}$}}%
\put(112.0185,-138.0357){\makebox(0,0)[lb]{$\tau_{23}$}}%
\put(65.0430,-50.5890){\makebox(0,0)[lb]{$S_3$}}%
\put(50.5890,-113.4639){\makebox(0,0)[lb]{$S_1$}}%
\put(80.9424,-84.5559){\makebox(0,0)[lb]{$S_2$}}%
%
\special{pn 8}%
\special{pa 1740 850}%
\special{pa 1768 836}%
\special{pa 1796 824}%
\special{pa 1822 810}%
\special{pa 1850 798}%
\special{pa 1880 786}%
\special{pa 1908 776}%
\special{pa 1938 766}%
\special{pa 1970 756}%
\special{pa 2002 750}%
\special{pa 2034 744}%
\special{pa 2070 742}%
\special{pa 2106 740}%
\special{pa 2144 742}%
\special{pa 2178 748}%
\special{pa 2194 770}%
\special{pa 2186 804}%
\special{pa 2158 826}%
\special{pa 2138 846}%
\special{pa 2148 876}%
\special{pa 2174 902}%
\special{pa 2204 918}%
\special{pa 2236 924}%
\special{pa 2268 922}%
\special{pa 2280 920}%
\special{sp}%
\put(166.9437,-70.8246){\makebox(0,0)[lb]{$S_4$}}%
%
\special{pn 8}%
\special{pa 3880 430}%
\special{pa 3880 900}%
\special{pa 3510 1130}%
\special{pa 3510 1130}%
\special{pa 3880 430}%
\special{pa 3880 900}%
\special{fp}%
%
\special{pn 8}%
\special{pa 3510 1130}%
\special{pa 3880 1340}%
\special{pa 3880 1820}%
\special{pa 3880 1820}%
\special{pa 3510 1130}%
\special{pa 3880 1340}%
\special{fp}%
%
\special{pn 8}%
\special{pa 3880 1830}%
\special{pa 4270 1590}%
\special{pa 4270 1590}%
\special{pa 4680 1830}%
\special{pa 4680 1830}%
\special{pa 3880 1830}%
\special{pa 4270 1590}%
\special{fp}%
%
\special{pn 8}%
\special{pa 5060 1140}%
\special{pa 4680 1360}%
\special{pa 4680 1830}%
\special{pa 4680 1830}%
\special{pa 5060 1140}%
\special{pa 4680 1360}%
\special{fp}%
%
\special{pn 8}%
\special{pa 4680 430}%
\special{pa 5060 1130}%
\special{fp}%
%
\special{pn 8}%
\special{pa 3880 430}%
\special{pa 4680 430}%
\special{fp}%
%
\special{pn 8}%
\special{pa 4270 1590}%
\special{pa 4680 1360}%
\special{fp}%
%
\special{pn 8}%
\special{pa 3880 430}%
\special{pa 5060 1140}%
\special{fp}%
%
\special{pn 8}%
\special{pa 3880 900}%
\special{pa 3880 1340}%
\special{fp}%
%
\special{pn 13}%
\special{pa 4930 890}%
\special{pa 4680 890}%
\special{dt 0.045}%
\special{pa 4900 830}%
\special{pa 4680 830}%
\special{dt 0.045}%
\special{pa 4860 770}%
\special{pa 4680 770}%
\special{dt 0.045}%
\special{pa 4830 710}%
\special{pa 4680 710}%
\special{dt 0.045}%
\special{pa 4800 650}%
\special{pa 4680 650}%
\special{dt 0.045}%
\special{pa 4770 590}%
\special{pa 4680 590}%
\special{dt 0.045}%
\special{pa 4730 530}%
\special{pa 4680 530}%
\special{dt 0.045}%
\special{pa 4700 470}%
\special{pa 4680 470}%
\special{dt 0.045}%
\special{pa 4960 950}%
\special{pa 4750 950}%
\special{dt 0.045}%
\special{pa 4990 1010}%
\special{pa 4850 1010}%
\special{dt 0.045}%
\special{pa 5030 1070}%
\special{pa 4950 1070}%
\special{dt 0.045}%
%
\special{pn 13}%
\special{pa 4830 1550}%
\special{pa 4680 1550}%
\special{dt 0.045}%
\special{pa 4800 1610}%
\special{pa 4680 1610}%
\special{dt 0.045}%
\special{pa 4770 1670}%
\special{pa 4680 1670}%
\special{dt 0.045}%
\special{pa 4740 1730}%
\special{pa 4680 1730}%
\special{dt 0.045}%
\special{pa 4700 1790}%
\special{pa 4680 1790}%
\special{dt 0.045}%
\special{pa 4870 1490}%
\special{pa 4680 1490}%
\special{dt 0.045}%
\special{pa 4900 1430}%
\special{pa 4680 1430}%
\special{dt 0.045}%
\special{pa 4930 1370}%
\special{pa 4680 1370}%
\special{dt 0.045}%
\special{pa 4970 1310}%
\special{pa 4770 1310}%
\special{dt 0.045}%
\special{pa 5000 1250}%
\special{pa 4870 1250}%
\special{dt 0.045}%
\special{pa 5030 1190}%
\special{pa 4980 1190}%
\special{dt 0.045}%
%
\special{pn 13}%
\special{pa 3830 1310}%
\special{pa 3830 1720}%
\special{dt 0.045}%
\special{pa 3770 1280}%
\special{pa 3770 1610}%
\special{dt 0.045}%
\special{pa 3710 1240}%
\special{pa 3710 1500}%
\special{dt 0.045}%
\special{pa 3650 1210}%
\special{pa 3650 1390}%
\special{dt 0.045}%
\special{pa 3590 1180}%
\special{pa 3590 1270}%
\special{dt 0.045}%
%
\special{pn 13}%
\special{pa 4370 1650}%
\special{pa 4370 1830}%
\special{dt 0.045}%
\special{pa 4430 1680}%
\special{pa 4430 1830}%
\special{dt 0.045}%
\special{pa 4490 1720}%
\special{pa 4490 1830}%
\special{dt 0.045}%
\special{pa 4550 1750}%
\special{pa 4550 1830}%
\special{dt 0.045}%
\special{pa 4610 1790}%
\special{pa 4610 1830}%
\special{dt 0.045}%
\special{pa 4310 1610}%
\special{pa 4310 1830}%
\special{dt 0.045}%
\special{pa 4250 1600}%
\special{pa 4250 1830}%
\special{dt 0.045}%
\special{pa 4190 1640}%
\special{pa 4190 1830}%
\special{dt 0.045}%
\special{pa 4130 1680}%
\special{pa 4130 1830}%
\special{dt 0.045}%
\special{pa 4070 1710}%
\special{pa 4070 1830}%
\special{dt 0.045}%
\special{pa 4010 1750}%
\special{pa 4010 1830}%
\special{dt 0.045}%
\special{pa 3950 1790}%
\special{pa 3950 1830}%
\special{dt 0.045}%
%
\special{pn 8}%
\special{pa 4270 670}%
\special{pa 4690 430}%
\special{fp}%
%
\special{pn 13}%
\special{pa 3880 900}%
\special{pa 4270 670}%
\special{dt 0.045}%
%
\special{pn 4}%
\special{pa 4680 440}%
\special{pa 4680 920}%
\special{fp}%
%
\special{pn 13}%
\special{pa 4680 910}%
\special{pa 4680 1360}%
\special{dt 0.045}%
\special{pa 4680 1360}%
\special{pa 4680 1360}%
\special{dt 0.045}%
%
\special{pn 13}%
\special{pa 3880 1340}%
\special{pa 4270 1590}%
\special{dt 0.045}%
%
\special{pn 13}%
\special{pa 3770 640}%
\special{pa 3770 970}%
\special{dt 0.045}%
\special{pa 3710 760}%
\special{pa 3710 1010}%
\special{dt 0.045}%
\special{pa 3650 870}%
\special{pa 3650 1040}%
\special{dt 0.045}%
\special{pa 3590 980}%
\special{pa 3590 1080}%
\special{dt 0.045}%
\special{pa 3830 530}%
\special{pa 3830 930}%
\special{dt 0.045}%
%
\special{pn 13}%
\special{pa 4310 450}%
\special{pa 4310 650}%
\special{dt 0.045}%
\special{pa 4250 430}%
\special{pa 4250 650}%
\special{dt 0.045}%
\special{pa 4190 430}%
\special{pa 4190 620}%
\special{dt 0.045}%
\special{pa 4130 430}%
\special{pa 4130 580}%
\special{dt 0.045}%
\special{pa 4070 430}%
\special{pa 4070 540}%
\special{dt 0.045}%
\special{pa 4010 430}%
\special{pa 4010 510}%
\special{dt 0.045}%
\special{pa 3950 430}%
\special{pa 3950 470}%
\special{dt 0.045}%
\special{pa 4370 430}%
\special{pa 4370 610}%
\special{dt 0.045}%
\special{pa 4430 430}%
\special{pa 4430 580}%
\special{dt 0.045}%
\special{pa 4490 430}%
\special{pa 4490 540}%
\special{dt 0.045}%
\special{pa 4550 430}%
\special{pa 4550 510}%
\special{dt 0.045}%
\special{pa 4610 430}%
\special{pa 4610 480}%
\special{dt 0.045}%
%
\special{pn 8}%
\special{pa 4130 590}%
\special{pa 3880 840}%
\special{fp}%
\special{pa 4170 610}%
\special{pa 3910 870}%
\special{fp}%
\special{pa 4210 630}%
\special{pa 4050 790}%
\special{fp}%
\special{pa 4250 650}%
\special{pa 4200 700}%
\special{fp}%
\special{pa 4100 560}%
\special{pa 3880 780}%
\special{fp}%
\special{pa 4060 540}%
\special{pa 3880 720}%
\special{fp}%
\special{pa 4020 520}%
\special{pa 3880 660}%
\special{fp}%
\special{pa 3980 500}%
\special{pa 3880 600}%
\special{fp}%
\special{pa 3950 470}%
\special{pa 3880 540}%
\special{fp}%
\special{pa 3910 450}%
\special{pa 3880 480}%
\special{fp}%
%
\special{pn 4}%
\special{pa 4580 860}%
\special{pa 4020 1420}%
\special{fp}%
\special{pa 4620 880}%
\special{pa 4060 1440}%
\special{fp}%
\special{pa 4660 900}%
\special{pa 4090 1470}%
\special{fp}%
\special{pa 4670 950}%
\special{pa 4130 1490}%
\special{fp}%
\special{pa 4670 1010}%
\special{pa 4170 1510}%
\special{fp}%
\special{pa 4670 1070}%
\special{pa 4200 1540}%
\special{fp}%
\special{pa 4670 1130}%
\special{pa 4240 1560}%
\special{fp}%
\special{pa 4670 1190}%
\special{pa 4280 1580}%
\special{fp}%
\special{pa 4670 1250}%
\special{pa 4410 1510}%
\special{fp}%
\special{pa 4670 1310}%
\special{pa 4550 1430}%
\special{fp}%
\special{pa 4550 830}%
\special{pa 3980 1400}%
\special{fp}%
\special{pa 4510 810}%
\special{pa 3950 1370}%
\special{fp}%
\special{pa 4470 790}%
\special{pa 3910 1350}%
\special{fp}%
\special{pa 4430 770}%
\special{pa 3880 1320}%
\special{fp}%
\special{pa 4400 740}%
\special{pa 3880 1260}%
\special{fp}%
\special{pa 4360 720}%
\special{pa 3880 1200}%
\special{fp}%
\special{pa 4320 700}%
\special{pa 3880 1140}%
\special{fp}%
\special{pa 4280 680}%
\special{pa 3880 1080}%
\special{fp}%
\special{pa 4150 750}%
\special{pa 3880 1020}%
\special{fp}%
\special{pa 4000 840}%
\special{pa 3880 960}%
\special{fp}%
%
\special{pn 4}%
\special{pa 4920 1060}%
\special{pa 4680 1300}%
\special{fp}%
\special{pa 4960 1080}%
\special{pa 4690 1350}%
\special{fp}%
\special{pa 5000 1100}%
\special{pa 4830 1270}%
\special{fp}%
\special{pa 5030 1130}%
\special{pa 4970 1190}%
\special{fp}%
\special{pa 4880 1040}%
\special{pa 4680 1240}%
\special{fp}%
\special{pa 4850 1010}%
\special{pa 4680 1180}%
\special{fp}%
\special{pa 4810 990}%
\special{pa 4680 1120}%
\special{fp}%
\special{pa 4770 970}%
\special{pa 4680 1060}%
\special{fp}%
\special{pa 4730 950}%
\special{pa 4680 1000}%
\special{fp}%
%
\special{pn 4}%
\special{pa 4150 1530}%
\special{pa 3880 1800}%
\special{fp}%
\special{pa 4120 1500}%
\special{pa 3880 1740}%
\special{fp}%
\special{pa 4080 1480}%
\special{pa 3880 1680}%
\special{fp}%
\special{pa 4040 1460}%
\special{pa 3880 1620}%
\special{fp}%
\special{pa 4010 1430}%
\special{pa 3880 1560}%
\special{fp}%
\special{pa 3970 1410}%
\special{pa 3880 1500}%
\special{fp}%
\special{pa 3930 1390}%
\special{pa 3880 1440}%
\special{fp}%
\special{pa 3900 1360}%
\special{pa 3880 1380}%
\special{fp}%
\special{pa 4190 1550}%
\special{pa 3970 1770}%
\special{fp}%
\special{pa 4230 1570}%
\special{pa 4120 1680}%
\special{fp}%
%
\special{pn 4}%
\special{pa 1590 450}%
\special{pa 1310 730}%
\special{fp}%
\special{pa 1530 450}%
\special{pa 1230 750}%
\special{fp}%
\special{pa 1470 450}%
\special{pa 1070 850}%
\special{fp}%
\special{pa 1410 450}%
\special{pa 950 910}%
\special{fp}%
\special{pa 1350 450}%
\special{pa 830 970}%
\special{fp}%
\special{pa 1260 480}%
\special{pa 710 1030}%
\special{fp}%
\special{pa 1220 460}%
\special{pa 680 1000}%
\special{fp}%
\special{pa 1170 450}%
\special{pa 680 940}%
\special{fp}%
\special{pa 1110 450}%
\special{pa 680 880}%
\special{fp}%
\special{pa 1050 450}%
\special{pa 680 820}%
\special{fp}%
\special{pa 990 450}%
\special{pa 680 760}%
\special{fp}%
\special{pa 930 450}%
\special{pa 680 700}%
\special{fp}%
\special{pa 870 450}%
\special{pa 680 640}%
\special{fp}%
\special{pa 810 450}%
\special{pa 680 580}%
\special{fp}%
\special{pa 750 450}%
\special{pa 680 520}%
\special{fp}%
\special{pa 1650 450}%
\special{pa 1430 670}%
\special{fp}%
\special{pa 1710 450}%
\special{pa 1550 610}%
\special{fp}%
\special{pa 1770 450}%
\special{pa 1670 550}%
\special{fp}%
%
\special{pn 8}%
\special{pa 1710 1200}%
\special{pa 1580 1200}%
\special{dt 0.045}%
\special{pa 1690 1260}%
\special{pa 1580 1260}%
\special{dt 0.045}%
\special{pa 1680 1320}%
\special{pa 1580 1320}%
\special{dt 0.045}%
\special{pa 1670 1380}%
\special{pa 1590 1380}%
\special{dt 0.045}%
\special{pa 1650 1440}%
\special{pa 1580 1440}%
\special{dt 0.045}%
\special{pa 1640 1500}%
\special{pa 1580 1500}%
\special{dt 0.045}%
\special{pa 1620 1560}%
\special{pa 1580 1560}%
\special{dt 0.045}%
\special{pa 1610 1620}%
\special{pa 1580 1620}%
\special{dt 0.045}%
\special{pa 1720 1140}%
\special{pa 1580 1140}%
\special{dt 0.045}%
\special{pa 1730 1080}%
\special{pa 1610 1080}%
\special{dt 0.045}%
\special{pa 1700 1020}%
\special{pa 1640 1020}%
\special{dt 0.045}%
\special{pa 1760 960}%
\special{pa 1660 960}%
\special{dt 0.045}%
\special{pa 1780 900}%
\special{pa 1690 900}%
\special{dt 0.045}%
\special{pa 1790 840}%
\special{pa 1730 840}%
\special{dt 0.045}%
\special{pa 1800 780}%
\special{pa 1740 780}%
\special{dt 0.045}%
\special{pa 1820 720}%
\special{pa 1760 720}%
\special{dt 0.045}%
\special{pa 1830 660}%
\special{pa 1790 660}%
\special{dt 0.045}%
\special{pa 1750 1020}%
\special{pa 1710 1020}%
\special{dt 0.045}%
%
\special{pn 4}%
\special{pa 1330 1260}%
\special{pa 950 1260}%
\special{dt 0.027}%
\special{pa 1460 1200}%
\special{pa 870 1200}%
\special{dt 0.027}%
\special{pa 1560 1140}%
\special{pa 800 1140}%
\special{dt 0.027}%
\special{pa 1480 1080}%
\special{pa 720 1080}%
\special{dt 0.027}%
\special{pa 1400 1020}%
\special{pa 740 1020}%
\special{dt 0.027}%
\special{pa 1320 960}%
\special{pa 860 960}%
\special{dt 0.027}%
\special{pa 1240 900}%
\special{pa 980 900}%
\special{dt 0.027}%
\special{pa 1160 840}%
\special{pa 1100 840}%
\special{dt 0.027}%
\special{pa 1200 1320}%
\special{pa 1030 1320}%
\special{dt 0.027}%
%
\special{pn 4}%
\special{pa 640 1140}%
\special{pa 530 1140}%
\special{dt 0.027}%
\special{pa 620 1200}%
\special{pa 510 1200}%
\special{dt 0.027}%
\special{pa 590 1260}%
\special{pa 490 1260}%
\special{dt 0.027}%
\special{pa 560 1320}%
\special{pa 480 1320}%
\special{dt 0.027}%
\special{pa 530 1380}%
\special{pa 470 1380}%
\special{dt 0.027}%
\special{pa 510 1440}%
\special{pa 450 1440}%
\special{dt 0.027}%
\special{pa 490 1500}%
\special{pa 440 1500}%
\special{dt 0.027}%
\special{pa 460 1560}%
\special{pa 420 1560}%
\special{dt 0.027}%
\special{pa 670 1080}%
\special{pa 540 1080}%
\special{dt 0.027}%
\special{pa 680 1020}%
\special{pa 550 1020}%
\special{dt 0.027}%
\special{pa 680 960}%
\special{pa 560 960}%
\special{dt 0.027}%
\special{pa 680 900}%
\special{pa 580 900}%
\special{dt 0.027}%
\special{pa 680 840}%
\special{pa 590 840}%
\special{dt 0.027}%
\special{pa 670 780}%
\special{pa 600 780}%
\special{dt 0.027}%
\special{pa 680 720}%
\special{pa 620 720}%
\special{dt 0.027}%
\special{pa 680 660}%
\special{pa 630 660}%
\special{dt 0.027}%
\special{pa 680 600}%
\special{pa 650 600}%
\special{dt 0.027}%
\special{pa 680 540}%
\special{pa 660 540}%
\special{dt 0.027}%
%
\special{pn 4}%
\special{pa 780 1130}%
\special{pa 780 1750}%
\special{dt 0.027}%
\special{pa 840 1170}%
\special{pa 840 1750}%
\special{dt 0.027}%
\special{pa 900 1220}%
\special{pa 900 1750}%
\special{dt 0.027}%
\special{pa 960 1270}%
\special{pa 960 1740}%
\special{dt 0.027}%
\special{pa 1020 1310}%
\special{pa 1020 1750}%
\special{dt 0.027}%
\special{pa 1080 1360}%
\special{pa 1080 1750}%
\special{dt 0.027}%
\special{pa 1140 1410}%
\special{pa 1140 1750}%
\special{dt 0.027}%
\special{pa 1200 1450}%
\special{pa 1200 1750}%
\special{dt 0.027}%
\special{pa 1260 1500}%
\special{pa 1260 1750}%
\special{dt 0.027}%
\special{pa 1320 1550}%
\special{pa 1320 1750}%
\special{dt 0.027}%
\special{pa 1380 1590}%
\special{pa 1380 1750}%
\special{dt 0.027}%
\special{pa 1440 1640}%
\special{pa 1440 1750}%
\special{dt 0.027}%
\special{pa 1500 1690}%
\special{pa 1500 1750}%
\special{dt 0.027}%
\special{pa 720 1080}%
\special{pa 720 1750}%
\special{dt 0.027}%
\special{pa 660 1100}%
\special{pa 660 1750}%
\special{dt 0.027}%
\special{pa 600 1240}%
\special{pa 600 1750}%
\special{dt 0.027}%
\special{pa 540 1400}%
\special{pa 540 1750}%
\special{dt 0.027}%
\special{pa 480 1520}%
\special{pa 480 1750}%
\special{dt 0.027}%
\special{pa 420 1660}%
\special{pa 420 1750}%
\special{dt 0.027}%
\put(304.2567,-84.5559){\makebox(0,0)[lb]{$S_3$}}%
%
\special{pn 8}%
\special{pa 2500 1248}%
\special{pa 2820 1248}%
\special{fp}%
\special{sh 1}%
\special{pa 2820 1248}%
\special{pa 2754 1228}%
\special{pa 2768 1248}%
\special{pa 2754 1268}%
\special{pa 2820 1248}%
\special{fp}%
\put(164.7756,-103.1293){\makebox(0,0)[lb]{projection}}%
\put(75.1608,-166.9437){\makebox(0,0)[lb]{$(a)$}}%
\put(286.9119,-165.4983){\makebox(0,0)[lb]{$(b)$}}%
%
\special{pn 8}%
\special{pa 1240 450}%
\special{pa 1180 450}%
\special{fp}%
\special{sh 1}%
\special{pa 1180 450}%
\special{pa 1248 470}%
\special{pa 1234 450}%
\special{pa 1248 430}%
\special{pa 1180 450}%
\special{fp}%
%
\special{pn 8}%
\special{pa 540 1070}%
\special{pa 530 1120}%
\special{fp}%
\special{sh 1}%
\special{pa 530 1120}%
\special{pa 564 1060}%
\special{pa 540 1068}%
\special{pa 524 1052}%
\special{pa 530 1120}%
\special{fp}%
%
\special{pn 8}%
\special{pa 680 770}%
\special{pa 680 820}%
\special{fp}%
\special{sh 1}%
\special{pa 680 820}%
\special{pa 700 754}%
\special{pa 680 768}%
\special{pa 660 754}%
\special{pa 680 820}%
\special{fp}%
%
\special{pn 8}%
\special{pa 1220 780}%
\special{pa 1290 750}%
\special{fp}%
\special{sh 1}%
\special{pa 1290 750}%
\special{pa 1222 758}%
\special{pa 1242 772}%
\special{pa 1238 796}%
\special{pa 1290 750}%
\special{fp}%
%
\special{pn 8}%
\special{pa 1280 920}%
\special{pa 1220 880}%
\special{fp}%
\special{sh 1}%
\special{pa 1220 880}%
\special{pa 1264 934}%
\special{pa 1264 910}%
\special{pa 1288 900}%
\special{pa 1220 880}%
\special{fp}%
%
\special{pn 8}%
\special{pa 1250 1300}%
\special{pa 1300 1270}%
\special{fp}%
\special{sh 1}%
\special{pa 1300 1270}%
\special{pa 1234 1288}%
\special{pa 1254 1298}%
\special{pa 1254 1322}%
\special{pa 1300 1270}%
\special{fp}%
%
\special{pn 8}%
\special{pa 1580 1370}%
\special{pa 1580 1430}%
\special{fp}%
\special{sh 1}%
\special{pa 1580 1430}%
\special{pa 1600 1364}%
\special{pa 1580 1378}%
\special{pa 1560 1364}%
\special{pa 1580 1430}%
\special{fp}%
%
\special{pn 8}%
\special{pa 1740 780}%
\special{pa 1720 810}%
\special{fp}%
\special{sh 1}%
\special{pa 1720 810}%
\special{pa 1774 766}%
\special{pa 1750 766}%
\special{pa 1740 744}%
\special{pa 1720 810}%
\special{fp}%
%
\special{pn 8}%
\special{pa 1710 1170}%
\special{pa 1730 1090}%
\special{fp}%
\special{sh 1}%
\special{pa 1730 1090}%
\special{pa 1694 1150}%
\special{pa 1718 1142}%
\special{pa 1734 1160}%
\special{pa 1730 1090}%
\special{fp}%
%
\special{pn 8}%
\special{pa 570 1330}%
\special{pa 550 1370}%
\special{fp}%
\special{sh 1}%
\special{pa 550 1370}%
\special{pa 598 1320}%
\special{pa 574 1322}%
\special{pa 562 1302}%
\special{pa 550 1370}%
\special{fp}%
%
\special{pn 8}%
\special{pa 1210 1460}%
\special{pa 1140 1400}%
\special{fp}%
\special{sh 1}%
\special{pa 1140 1400}%
\special{pa 1178 1460}%
\special{pa 1180 1436}%
\special{pa 1204 1428}%
\special{pa 1140 1400}%
\special{fp}%
%
\special{pn 8}%
\special{pa 940 1750}%
\special{pa 1020 1750}%
\special{fp}%
\special{sh 1}%
\special{pa 1020 1750}%
\special{pa 954 1730}%
\special{pa 968 1750}%
\special{pa 954 1770}%
\special{pa 1020 1750}%
\special{fp}%
%
\special{pn 8}%
\special{pa 4310 430}%
\special{pa 4250 430}%
\special{fp}%
\special{sh 1}%
\special{pa 4250 430}%
\special{pa 4318 450}%
\special{pa 4304 430}%
\special{pa 4318 410}%
\special{pa 4250 430}%
\special{fp}%
%
\special{pn 8}%
\special{pa 4340 1830}%
\special{pa 4270 1830}%
\special{fp}%
\special{sh 1}%
\special{pa 4270 1830}%
\special{pa 4338 1850}%
\special{pa 4324 1830}%
\special{pa 4338 1810}%
\special{pa 4270 1830}%
\special{fp}%
%
\special{pn 8}%
\special{pa 4020 820}%
\special{pa 4080 790}%
\special{fp}%
\special{sh 1}%
\special{pa 4080 790}%
\special{pa 4012 802}%
\special{pa 4032 814}%
\special{pa 4030 838}%
\special{pa 4080 790}%
\special{fp}%
%
\special{pn 8}%
\special{pa 4910 850}%
\special{pa 4880 790}%
\special{fp}%
\special{sh 1}%
\special{pa 4880 790}%
\special{pa 4892 860}%
\special{pa 4904 838}%
\special{pa 4928 842}%
\special{pa 4880 790}%
\special{fp}%
%
\special{pn 8}%
\special{pa 4680 1080}%
\special{pa 4680 1140}%
\special{fp}%
\special{sh 1}%
\special{pa 4680 1140}%
\special{pa 4700 1074}%
\special{pa 4680 1088}%
\special{pa 4660 1074}%
\special{pa 4680 1140}%
\special{fp}%
%
\special{pn 8}%
\special{pa 4870 1510}%
\special{pa 4900 1440}%
\special{fp}%
\special{sh 1}%
\special{pa 4900 1440}%
\special{pa 4856 1494}%
\special{pa 4880 1490}%
\special{pa 4892 1510}%
\special{pa 4900 1440}%
\special{fp}%
%
\special{pn 8}%
\special{pa 3720 1530}%
\special{pa 3690 1470}%
\special{fp}%
\special{sh 1}%
\special{pa 3690 1470}%
\special{pa 3702 1540}%
\special{pa 3714 1518}%
\special{pa 3738 1522}%
\special{pa 3690 1470}%
\special{fp}%
%
\special{pn 8}%
\special{pa 3880 1060}%
\special{pa 3880 1110}%
\special{fp}%
\special{sh 1}%
\special{pa 3880 1110}%
\special{pa 3900 1044}%
\special{pa 3880 1058}%
\special{pa 3860 1044}%
\special{pa 3880 1110}%
\special{fp}%
%
\special{pn 8}%
\special{pa 4450 1500}%
\special{pa 4520 1450}%
\special{fp}%
\special{sh 1}%
\special{pa 4520 1450}%
\special{pa 4454 1472}%
\special{pa 4478 1482}%
\special{pa 4478 1506}%
\special{pa 4520 1450}%
\special{fp}%
%
\special{pn 8}%
\special{pa 4510 800}%
\special{pa 4460 770}%
\special{fp}%
\special{sh 1}%
\special{pa 4460 770}%
\special{pa 4508 822}%
\special{pa 4506 798}%
\special{pa 4528 788}%
\special{pa 4460 770}%
\special{fp}%
%
\special{pn 8}%
\special{pa 4050 1450}%
\special{pa 4130 1490}%
\special{fp}%
\special{sh 1}%
\special{pa 4130 1490}%
\special{pa 4080 1442}%
\special{pa 4082 1466}%
\special{pa 4062 1478}%
\special{pa 4130 1490}%
\special{fp}%
%
\special{pn 8}%
\special{pa 3730 730}%
\special{pa 3690 780}%
\special{fp}%
\special{sh 1}%
\special{pa 3690 780}%
\special{pa 3748 740}%
\special{pa 3724 738}%
\special{pa 3716 716}%
\special{pa 3690 780}%
\special{fp}%
\put(14.4540,-55.6479){\makebox(0,0)[lb]{$\tilde S_4$}}%
%
\special{pn 8}%
\special{pa 620 970}%
\special{pa 582 980}%
\special{pa 544 988}%
\special{pa 506 996}%
\special{pa 470 1000}%
\special{pa 436 1000}%
\special{pa 404 998}%
\special{pa 374 988}%
\special{pa 348 974}%
\special{pa 326 952}%
\special{pa 306 922}%
\special{pa 300 892}%
\special{pa 320 874}%
\special{pa 354 858}%
\special{pa 380 834}%
\special{pa 390 804}%
\special{pa 380 776}%
\special{pa 370 760}%
\special{sp}%
\end{picture}%

%% file: Nordfjordpic27.tex
\unitlength 1pt
\begin{picture}(403.2666,140.2038)( 10.1178,-161.8848)
\put(153.2124,-106.2369){\makebox(0,0)[lb]{$\scriptstyle{d_T}$}}%
\put(107.3932,-54.9252){\makebox(0,0)[lb]{$S_4$}}%
%
\special{pn 8}%
\special{pa 430 534}%
\special{pa 430 910}%
\special{fp}%
%
\special{pn 8}%
\special{pa 430 534}%
\special{pa 1348 534}%
\special{fp}%
\special{pa 1348 534}%
\special{pa 430 910}%
\special{fp}%
%
\special{pn 8}%
\special{pa 430 534}%
\special{pa 246 1476}%
\special{fp}%
%
\special{pn 8}%
\special{pa 1348 534}%
\special{pa 1164 1476}%
\special{fp}%
%
\special{pn 8}%
\special{pa 430 910}%
\special{pa 246 1476}%
\special{fp}%
%
\special{pn 8}%
\special{pa 246 1476}%
\special{pa 1164 1476}%
\special{fp}%
\special{pa 1164 1476}%
\special{pa 430 910}%
\special{fp}%
%
\special{pn 8}%
\special{pa 1348 534}%
\special{pa 1164 1100}%
\special{fp}%
%
\special{pn 8}%
\special{pa 1164 1100}%
\special{pa 1164 1476}%
\special{fp}%
%
\special{pn 8}%
\special{pa 1164 1100}%
\special{pa 742 788}%
\special{fp}%
%
\special{pn 8}%
\special{pa 1164 1100}%
\special{pa 826 1212}%
\special{fp}%
%
\special{pn 8}%
\special{pa 1164 1476}%
\special{pa 982 2040}%
\special{fp}%
%
\special{pn 8}%
\special{pa 1164 1100}%
\special{pa 1110 1428}%
\special{fp}%
%
\special{pn 8}%
\special{pa 1092 1476}%
\special{pa 982 2040}%
\special{fp}%
%
\special{pn 8}%
\special{pa 982 2040}%
\special{pa 1900 2040}%
\special{fp}%
\special{pa 1900 2040}%
\special{pa 1164 1476}%
\special{fp}%
%
\special{pn 8}%
\special{pa 2084 1100}%
\special{pa 1900 2040}%
\special{fp}%
%
\special{pn 8}%
\special{pa 1248 1100}%
\special{pa 2084 1100}%
\special{fp}%
%
\special{pn 8}%
\special{pa 1164 1476}%
\special{pa 2084 1100}%
\special{fp}%
%
\special{pn 8}%
\special{pa 2084 1100}%
\special{pa 1900 1664}%
\special{fp}%
%
\special{pn 8}%
\special{pa 1900 1664}%
\special{pa 1900 2040}%
\special{fp}%
%
\special{pn 8}%
\special{pa 1900 1664}%
\special{pa 1478 1344}%
\special{fp}%
%
\special{pn 8}%
\special{pa 1900 1664}%
\special{pa 1560 1776}%
\special{fp}%
%
\special{pn 4}%
\special{pa 1660 1100}%
\special{pa 1386 1382}%
\special{fp}%
\special{pa 1606 1100}%
\special{pa 1294 1420}%
\special{fp}%
\special{pa 1550 1100}%
\special{pa 1202 1456}%
\special{fp}%
\special{pa 1496 1100}%
\special{pa 1174 1428}%
\special{fp}%
\special{pa 1440 1100}%
\special{pa 1192 1354}%
\special{fp}%
\special{pa 1386 1100}%
\special{pa 1202 1288}%
\special{fp}%
\special{pa 1330 1100}%
\special{pa 1220 1212}%
\special{fp}%
\special{pa 1276 1100}%
\special{pa 1230 1146}%
\special{fp}%
\special{pa 1716 1100}%
\special{pa 1478 1344}%
\special{fp}%
\special{pa 1772 1100}%
\special{pa 1570 1306}%
\special{fp}%
\special{pa 1826 1100}%
\special{pa 1660 1268}%
\special{fp}%
\special{pa 1882 1100}%
\special{pa 1752 1230}%
\special{fp}%
\special{pa 1936 1100}%
\special{pa 1844 1194}%
\special{fp}%
\special{pa 1992 1100}%
\special{pa 1936 1156}%
\special{fp}%
%
\special{pn 4}%
\special{pa 1964 1730}%
\special{pa 1900 1664}%
\special{fp}%
\special{pa 1974 1682}%
\special{pa 1918 1626}%
\special{fp}%
\special{pa 1982 1636}%
\special{pa 1928 1580}%
\special{fp}%
\special{pa 1992 1588}%
\special{pa 1946 1542}%
\special{fp}%
\special{pa 2000 1542}%
\special{pa 1954 1494}%
\special{fp}%
\special{pa 2010 1494}%
\special{pa 1974 1456}%
\special{fp}%
\special{pa 2020 1448}%
\special{pa 1982 1410}%
\special{fp}%
\special{pa 2028 1400}%
\special{pa 2000 1372}%
\special{fp}%
\special{pa 2038 1354}%
\special{pa 2010 1324}%
\special{fp}%
\special{pa 1954 1776}%
\special{pa 1900 1720}%
\special{fp}%
\special{pa 1946 1824}%
\special{pa 1900 1776}%
\special{fp}%
\special{pa 1936 1870}%
\special{pa 1900 1834}%
\special{fp}%
\special{pa 1928 1918}%
\special{pa 1900 1890}%
\special{fp}%
\special{pa 1918 1964}%
\special{pa 1900 1946}%
\special{fp}%
%
\special{pn 4}%
\special{pa 1220 1194}%
\special{pa 1164 1136}%
\special{fp}%
\special{pa 1210 1240}%
\special{pa 1164 1194}%
\special{fp}%
\special{pa 1202 1288}%
\special{pa 1164 1250}%
\special{fp}%
\special{pa 1192 1334}%
\special{pa 1164 1306}%
\special{fp}%
\special{pa 1184 1382}%
\special{pa 1164 1362}%
\special{fp}%
\special{pa 1230 1146}%
\special{pa 1174 1090}%
\special{fp}%
\special{pa 1238 1100}%
\special{pa 1184 1042}%
\special{fp}%
\special{pa 1248 1052}%
\special{pa 1202 1004}%
\special{fp}%
\special{pa 1256 1004}%
\special{pa 1210 958}%
\special{fp}%
\special{pa 1266 958}%
\special{pa 1230 920}%
\special{fp}%
\special{pa 1276 910}%
\special{pa 1238 874}%
\special{fp}%
\special{pa 1284 864}%
\special{pa 1256 836}%
\special{fp}%
\special{pa 1294 816}%
\special{pa 1266 788}%
\special{fp}%
%
\special{pn 4}%
\special{pa 1836 1682}%
\special{pa 1440 1682}%
\special{dt 0.027}%
\special{pa 1808 1598}%
\special{pa 1330 1598}%
\special{dt 0.027}%
\special{pa 1698 1514}%
\special{pa 1220 1514}%
\special{dt 0.027}%
\special{pa 1588 1428}%
\special{pa 1284 1428}%
\special{dt 0.027}%
%
\special{pn 4}%
\special{pa 1146 1090}%
\special{pa 670 1090}%
\special{dt 0.027}%
\special{pa 1036 1004}%
\special{pa 558 1004}%
\special{dt 0.027}%
\special{pa 918 920}%
\special{pa 448 920}%
\special{dt 0.027}%
\special{pa 806 836}%
\special{pa 624 836}%
\special{dt 0.027}%
\special{pa 926 1174}%
\special{pa 780 1174}%
\special{dt 0.027}%
%
\special{pn 4}%
\special{pa 402 1004}%
\special{pa 338 1004}%
\special{dt 0.027}%
\special{pa 376 1090}%
\special{pa 320 1090}%
\special{dt 0.027}%
\special{pa 348 1174}%
\special{pa 302 1174}%
\special{dt 0.027}%
\special{pa 430 920}%
\special{pa 356 920}%
\special{dt 0.027}%
\special{pa 430 836}%
\special{pa 376 836}%
\special{dt 0.027}%
\special{pa 430 750}%
\special{pa 384 750}%
\special{dt 0.027}%
\special{pa 430 666}%
\special{pa 402 666}%
\special{dt 0.027}%
%
\special{pn 4}%
\special{pa 1100 1682}%
\special{pa 1054 1682}%
\special{dt 0.027}%
\special{pa 1128 1598}%
\special{pa 1064 1598}%
\special{dt 0.027}%
\special{pa 1156 1514}%
\special{pa 1082 1514}%
\special{dt 0.027}%
%
\special{pn 4}%
\special{pa 1164 1428}%
\special{pa 1110 1428}%
\special{fp}%
\special{pa 1164 1344}%
\special{pa 1128 1344}%
\special{fp}%
\special{pa 1164 1260}%
\special{pa 1138 1260}%
\special{fp}%
\put(23.1264,-36.1350){\makebox(0,0)[lb]{$X$}}%
\put(93.2283,-36.8577){\makebox(0,0)[lb]{$Z'$}}%
\put(10.1178,-115.6320){\makebox(0,0)[lb]{$Z$}}%
\put(35.7014,-71.9086){\makebox(0,0)[lb]{$Y$}}%
\put(74.4381,-75.1608){\makebox(0,0)[lb]{$Y'$}}%
\put(138.0357,-153.9351){\makebox(0,0)[lb]{$X'[1]$}}%
\put(50.5890,-154.6578){\makebox(0,0)[lb]{$Z[1]$}}%
\put(88.8921,-112.7412){\makebox(0,0)[lb]{$X'$}}%
\put(151.7670,-81.6651){\makebox(0,0)[lb]{$Z'[1]$}}%
\put(35.7014,-95.6855){\makebox(0,0)[lb]{$S_1$}}%
\put(52.3235,-78.7020){\makebox(0,0)[lb]{$S_2$}}%
\put(39.6762,-52.9016){\makebox(0,0)[lb]{$S_3$}}%
%
\special{pn 8}%
\special{pa 2080 1480}%
\special{pa 2310 1546}%
\special{fp}%
\special{sh 1}%
\special{pa 2310 1546}%
\special{pa 2252 1508}%
\special{pa 2260 1530}%
\special{pa 2240 1546}%
\special{pa 2310 1546}%
\special{fp}%
\put(58.2496,-31.0761){\makebox(0,0)[lb]{${\mathcal O}$}}%
\put(109.1277,-72.9927){\makebox(0,0)[lb]{${\mathcal O}[1]$}}%
\put(10.8405,-53.4798){\makebox(0,0)[lb]{$\tilde S_4$}}%
\put(140.2038,-131.5314){\makebox(0,0)[lb]{$S_4[1]$}}%
\put(36.1350,-130.0860){\makebox(0,0)[lb]{$\tilde S_4[1]$}}%
%
\special{pn 8}%
\special{pa 1266 844}%
\special{pa 1316 854}%
\special{pa 1362 862}%
\special{pa 1398 862}%
\special{pa 1418 854}%
\special{pa 1422 832}%
\special{pa 1412 800}%
\special{pa 1424 776}%
\special{pa 1454 756}%
\special{pa 1468 750}%
\special{sp}%
%
\special{pn 8}%
\special{pa 376 958}%
\special{pa 328 942}%
\special{pa 288 924}%
\special{pa 256 906}%
\special{pa 242 886}%
\special{pa 246 864}%
\special{pa 276 848}%
\special{pa 300 840}%
\special{pa 300 816}%
\special{pa 282 782}%
\special{pa 256 742}%
\special{sp}%
%
\special{pn 8}%
\special{pa 850 1720}%
\special{pa 900 1730}%
\special{pa 946 1738}%
\special{pa 982 1738}%
\special{pa 1002 1730}%
\special{pa 1006 1708}%
\special{pa 996 1676}%
\special{pa 1008 1652}%
\special{pa 1038 1632}%
\special{pa 1052 1626}%
\special{sp}%
%
\special{pn 8}%
\special{pa 3740 750}%
\special{pa 3440 1250}%
\special{fp}%
\special{pa 3440 1250}%
\special{pa 3740 1750}%
\special{fp}%
%
\special{pn 8}%
\special{pa 4340 750}%
\special{pa 4640 1250}%
\special{fp}%
\special{pa 4640 1250}%
\special{pa 4340 1750}%
\special{fp}%
%
\special{pn 8}%
\special{pa 4940 750}%
\special{pa 5240 1250}%
\special{fp}%
\special{pa 5240 1250}%
\special{pa 4940 1750}%
\special{fp}%
%
\special{pn 8}%
\special{pa 3740 750}%
\special{pa 4640 1250}%
\special{fp}%
%
\special{pn 8}%
\special{pa 4640 1250}%
\special{pa 3730 1750}%
\special{fp}%
%
\special{pn 8}%
\special{pa 4340 750}%
\special{pa 4340 1100}%
\special{fp}%
%
\special{pn 8}%
\special{pa 4340 1750}%
\special{pa 4340 1420}%
\special{fp}%
%
\special{pn 8}%
\special{pa 3440 1250}%
\special{pa 3740 1440}%
\special{fp}%
%
\special{pn 8}%
\special{pa 4030 1590}%
\special{pa 4340 1750}%
\special{fp}%
%
\special{pn 8}%
\special{pa 3740 750}%
\special{pa 4940 750}%
\special{fp}%
%
\special{pn 8}%
\special{pa 3740 1750}%
\special{pa 4940 1750}%
\special{fp}%
%
\special{pn 8}%
\special{pa 3740 750}%
\special{pa 3740 750}%
\special{fp}%
\special{pa 3740 750}%
\special{pa 3740 1750}%
\special{fp}%
%
\special{pn 8}%
\special{pa 4340 750}%
\special{pa 5240 1250}%
\special{fp}%
%
\special{pn 8}%
\special{pa 5240 1250}%
\special{pa 4330 1750}%
\special{fp}%
%
\special{pn 8}%
\special{pa 4940 750}%
\special{pa 4940 1100}%
\special{fp}%
%
\special{pn 8}%
\special{pa 4940 1750}%
\special{pa 4940 1420}%
\special{fp}%
%
\special{pn 8}%
\special{pa 4040 920}%
\special{pa 4340 750}%
\special{fp}%
%
\special{pn 8}%
\special{pa 3440 1250}%
\special{pa 3740 1060}%
\special{fp}%
%
\special{pn 8}%
\special{pa 4640 920}%
\special{pa 4940 750}%
\special{fp}%
%
\special{pn 8}%
\special{pa 4630 1590}%
\special{pa 4940 1750}%
\special{fp}%
%
\special{pn 8}%
\special{pa 4160 1010}%
\special{pa 3740 1430}%
\special{fp}%
\special{pa 4190 1040}%
\special{pa 3740 1490}%
\special{fp}%
\special{pa 4250 1040}%
\special{pa 3740 1550}%
\special{fp}%
\special{pa 4290 1060}%
\special{pa 3740 1610}%
\special{fp}%
\special{pa 4330 1080}%
\special{pa 3740 1670}%
\special{fp}%
\special{pa 4370 1100}%
\special{pa 3740 1730}%
\special{fp}%
\special{pa 4410 1120}%
\special{pa 3850 1680}%
\special{fp}%
\special{pa 4450 1140}%
\special{pa 3980 1610}%
\special{fp}%
\special{pa 4480 1170}%
\special{pa 4160 1490}%
\special{fp}%
\special{pa 4520 1190}%
\special{pa 4250 1460}%
\special{fp}%
\special{pa 4560 1210}%
\special{pa 4380 1390}%
\special{fp}%
\special{pa 4600 1230}%
\special{pa 4510 1320}%
\special{fp}%
\special{pa 4140 970}%
\special{pa 3740 1370}%
\special{fp}%
\special{pa 4100 950}%
\special{pa 3740 1310}%
\special{fp}%
\special{pa 4060 930}%
\special{pa 3770 1220}%
\special{fp}%
\special{pa 4020 910}%
\special{pa 3740 1190}%
\special{fp}%
\special{pa 3980 890}%
\special{pa 3740 1130}%
\special{fp}%
\special{pa 3940 870}%
\special{pa 3740 1070}%
\special{fp}%
\special{pa 3910 840}%
\special{pa 3740 1010}%
\special{fp}%
\special{pa 3870 820}%
\special{pa 3740 950}%
\special{fp}%
\special{pa 3830 800}%
\special{pa 3740 890}%
\special{fp}%
\special{pa 3790 780}%
\special{pa 3740 830}%
\special{fp}%
%
\special{pn 8}%
\special{pa 5010 1120}%
\special{pa 4450 1680}%
\special{fp}%
\special{pa 4970 1100}%
\special{pa 4380 1690}%
\special{fp}%
\special{pa 4930 1080}%
\special{pa 4540 1470}%
\special{fp}%
\special{pa 4890 1060}%
\special{pa 4560 1390}%
\special{fp}%
\special{pa 4850 1040}%
\special{pa 4640 1250}%
\special{fp}%
\special{pa 4810 1020}%
\special{pa 4620 1210}%
\special{fp}%
\special{pa 4760 1010}%
\special{pa 4600 1170}%
\special{fp}%
\special{pa 4730 980}%
\special{pa 4570 1140}%
\special{fp}%
\special{pa 4700 950}%
\special{pa 4550 1100}%
\special{fp}%
\special{pa 4660 930}%
\special{pa 4530 1060}%
\special{fp}%
\special{pa 4620 910}%
\special{pa 4510 1020}%
\special{fp}%
\special{pa 4580 890}%
\special{pa 4500 970}%
\special{fp}%
\special{pa 4540 870}%
\special{pa 4460 950}%
\special{fp}%
\special{pa 4510 840}%
\special{pa 4440 910}%
\special{fp}%
\special{pa 4470 820}%
\special{pa 4420 870}%
\special{fp}%
\special{pa 4430 800}%
\special{pa 4390 840}%
\special{fp}%
\special{pa 4390 780}%
\special{pa 4370 800}%
\special{fp}%
\special{pa 4520 1490}%
\special{pa 4490 1520}%
\special{fp}%
\special{pa 5050 1140}%
\special{pa 4580 1610}%
\special{fp}%
\special{pa 5080 1170}%
\special{pa 4780 1470}%
\special{fp}%
\special{pa 5120 1190}%
\special{pa 4850 1460}%
\special{fp}%
\special{pa 5160 1210}%
\special{pa 4980 1390}%
\special{fp}%
\special{pa 5200 1230}%
\special{pa 5110 1320}%
\special{fp}%
\special{pa 4770 1480}%
\special{pa 4710 1540}%
\special{fp}%
\put(71.5473,-171.2799){\makebox(0,0)[lb]{$(a)$}}%
\put(307.8702,-167.6664){\makebox(0,0)[lb]{$(b)$}}%
\put(265.2309,-50.5890){\makebox(0,0)[lb]{$X$}}%
\put(266.6763,-138.0357){\makebox(0,0)[lb]{$Y$}}%
\put(235.6002,-95.3964){\makebox(0,0)[lb]{$Z$}}%
\put(309.3156,-49.1436){\makebox(0,0)[lb]{$Y'$}}%
\put(309.3156,-137.3130){\makebox(0,0)[lb]{$X'$}}%
\put(351.9549,-52.7571){\makebox(0,0)[lb]{$Y'[1]$}}%
\put(350.5095,-140.2038){\makebox(0,0)[lb]{$X'[1]$}}%
\put(377.2494,-106.9596){\makebox(0,0)[lb]{$Z'[1]$}}%
\put(337.5009,-94.6737){\makebox(0,0)[lb]{$Z'$}}%
%
\special{pn 8}%
\special{pa 5360 1240}%
\special{pa 5720 1240}%
\special{fp}%
\special{sh 1}%
\special{pa 5720 1240}%
\special{pa 5654 1220}%
\special{pa 5668 1240}%
\special{pa 5654 1260}%
\special{pa 5720 1240}%
\special{fp}%
\put(393.1488,-86.7240){\makebox(0,0)[lb]{$\scriptstyle{d_T}$}}%
%
\special{pn 8}%
\special{pa 2550 1250}%
\special{pa 3110 1250}%
\special{fp}%
\special{sh 1}%
\special{pa 3110 1250}%
\special{pa 3044 1230}%
\special{pa 3058 1250}%
\special{pa 3044 1270}%
\special{pa 3110 1250}%
\special{fp}%
\put(177.7842,-88.8921){\makebox(0,0)[lb]{projection}}%
%
\special{pn 4}%
\special{pa 870 530}%
\special{pa 820 530}%
\special{fp}%
\special{sh 1}%
\special{pa 820 530}%
\special{pa 888 550}%
\special{pa 874 530}%
\special{pa 888 510}%
\special{pa 820 530}%
\special{fp}%
%
\special{pn 8}%
\special{pa 880 720}%
\special{pa 960 690}%
\special{fp}%
\special{sh 1}%
\special{pa 960 690}%
\special{pa 892 696}%
\special{pa 910 710}%
\special{pa 906 732}%
\special{pa 960 690}%
\special{fp}%
%
\special{pn 8}%
\special{pa 430 720}%
\special{pa 430 770}%
\special{fp}%
\special{sh 1}%
\special{pa 430 770}%
\special{pa 450 704}%
\special{pa 430 718}%
\special{pa 410 704}%
\special{pa 430 770}%
\special{fp}%
%
\special{pn 8}%
\special{pa 340 970}%
\special{pa 320 1040}%
\special{fp}%
\special{sh 1}%
\special{pa 320 1040}%
\special{pa 358 982}%
\special{pa 336 990}%
\special{pa 320 970}%
\special{pa 320 1040}%
\special{fp}%
%
\special{pn 8}%
\special{pa 370 1110}%
\special{pa 340 1160}%
\special{fp}%
\special{sh 1}%
\special{pa 340 1160}%
\special{pa 392 1114}%
\special{pa 368 1114}%
\special{pa 358 1094}%
\special{pa 340 1160}%
\special{fp}%
%
\special{pn 8}%
\special{pa 630 1470}%
\special{pa 670 1470}%
\special{fp}%
\special{sh 1}%
\special{pa 670 1470}%
\special{pa 604 1450}%
\special{pa 618 1470}%
\special{pa 604 1490}%
\special{pa 670 1470}%
\special{fp}%
%
\special{pn 8}%
\special{pa 920 1280}%
\special{pa 870 1250}%
\special{fp}%
\special{sh 1}%
\special{pa 870 1250}%
\special{pa 918 1302}%
\special{pa 916 1278}%
\special{pa 938 1268}%
\special{pa 870 1250}%
\special{fp}%
%
\special{pn 8}%
\special{pa 900 900}%
\special{pa 830 850}%
\special{fp}%
\special{sh 1}%
\special{pa 830 850}%
\special{pa 874 906}%
\special{pa 874 882}%
\special{pa 896 872}%
\special{pa 830 850}%
\special{fp}%
%
\special{pn 8}%
\special{pa 890 1190}%
\special{pa 970 1150}%
\special{fp}%
\special{sh 1}%
\special{pa 970 1150}%
\special{pa 902 1162}%
\special{pa 922 1174}%
\special{pa 920 1198}%
\special{pa 970 1150}%
\special{fp}%
%
\special{pn 8}%
\special{pa 1600 1090}%
\special{pa 1540 1090}%
\special{fp}%
\special{sh 1}%
\special{pa 1540 1090}%
\special{pa 1608 1110}%
\special{pa 1594 1090}%
\special{pa 1608 1070}%
\special{pa 1540 1090}%
\special{fp}%
%
\special{pn 8}%
\special{pa 1240 840}%
\special{pa 1230 880}%
\special{fp}%
\special{sh 1}%
\special{pa 1230 880}%
\special{pa 1266 820}%
\special{pa 1244 828}%
\special{pa 1228 810}%
\special{pa 1230 880}%
\special{fp}%
%
\special{pn 8}%
\special{pa 1250 1050}%
\special{pa 1260 960}%
\special{fp}%
\special{sh 1}%
\special{pa 1260 960}%
\special{pa 1234 1024}%
\special{pa 1254 1014}%
\special{pa 1274 1028}%
\special{pa 1260 960}%
\special{fp}%
%
\special{pn 8}%
\special{pa 1160 1220}%
\special{pa 1160 1280}%
\special{fp}%
\special{sh 1}%
\special{pa 1160 1280}%
\special{pa 1180 1214}%
\special{pa 1160 1228}%
\special{pa 1140 1214}%
\special{pa 1160 1280}%
\special{fp}%
%
\special{pn 8}%
\special{pa 1080 1530}%
\special{pa 1070 1580}%
\special{fp}%
\special{sh 1}%
\special{pa 1070 1580}%
\special{pa 1104 1520}%
\special{pa 1080 1528}%
\special{pa 1064 1512}%
\special{pa 1070 1580}%
\special{fp}%
%
\special{pn 8}%
\special{pa 1100 1680}%
\special{pa 1080 1740}%
\special{fp}%
\special{sh 1}%
\special{pa 1080 1740}%
\special{pa 1120 1684}%
\special{pa 1098 1690}%
\special{pa 1082 1670}%
\special{pa 1080 1740}%
\special{fp}%
%
\special{pn 8}%
\special{pa 1390 2040}%
\special{pa 1470 2040}%
\special{fp}%
\special{sh 1}%
\special{pa 1470 2040}%
\special{pa 1404 2020}%
\special{pa 1418 2040}%
\special{pa 1404 2060}%
\special{pa 1470 2040}%
\special{fp}%
%
\special{pn 8}%
\special{pa 1640 1840}%
\special{pa 1580 1790}%
\special{fp}%
\special{sh 1}%
\special{pa 1580 1790}%
\special{pa 1618 1848}%
\special{pa 1622 1824}%
\special{pa 1644 1818}%
\special{pa 1580 1790}%
\special{fp}%
%
\special{pn 8}%
\special{pa 1630 1750}%
\special{pa 1700 1720}%
\special{fp}%
\special{sh 1}%
\special{pa 1700 1720}%
\special{pa 1632 1728}%
\special{pa 1652 1742}%
\special{pa 1648 1766}%
\special{pa 1700 1720}%
\special{fp}%
%
\special{pn 8}%
\special{pa 1630 1460}%
\special{pa 1570 1410}%
\special{fp}%
\special{sh 1}%
\special{pa 1570 1410}%
\special{pa 1608 1468}%
\special{pa 1612 1444}%
\special{pa 1634 1438}%
\special{pa 1570 1410}%
\special{fp}%
%
\special{pn 8}%
\special{pa 1620 1280}%
\special{pa 1700 1240}%
\special{fp}%
\special{sh 1}%
\special{pa 1700 1240}%
\special{pa 1632 1252}%
\special{pa 1652 1264}%
\special{pa 1650 1288}%
\special{pa 1700 1240}%
\special{fp}%
%
\special{pn 8}%
\special{pa 1970 1430}%
\special{pa 1940 1500}%
\special{fp}%
\special{sh 1}%
\special{pa 1940 1500}%
\special{pa 1986 1448}%
\special{pa 1962 1452}%
\special{pa 1948 1432}%
\special{pa 1940 1500}%
\special{fp}%
%
\special{pn 8}%
\special{pa 1980 1640}%
\special{pa 1980 1560}%
\special{fp}%
\special{sh 1}%
\special{pa 1980 1560}%
\special{pa 1960 1628}%
\special{pa 1980 1614}%
\special{pa 2000 1628}%
\special{pa 1980 1560}%
\special{fp}%
%
\special{pn 8}%
\special{pa 1890 1760}%
\special{pa 1890 1820}%
\special{fp}%
\special{sh 1}%
\special{pa 1890 1820}%
\special{pa 1910 1754}%
\special{pa 1890 1768}%
\special{pa 1870 1754}%
\special{pa 1890 1820}%
\special{fp}%
%
\special{pn 8}%
\special{pa 4080 750}%
\special{pa 4010 750}%
\special{fp}%
\special{sh 1}%
\special{pa 4010 750}%
\special{pa 4078 770}%
\special{pa 4064 750}%
\special{pa 4078 730}%
\special{pa 4010 750}%
\special{fp}%
%
\special{pn 8}%
\special{pa 3610 980}%
\special{pa 3590 1010}%
\special{fp}%
\special{sh 1}%
\special{pa 3590 1010}%
\special{pa 3644 966}%
\special{pa 3620 966}%
\special{pa 3610 944}%
\special{pa 3590 1010}%
\special{fp}%
%
\special{pn 8}%
\special{pa 3610 1520}%
\special{pa 3570 1450}%
\special{fp}%
\special{sh 1}%
\special{pa 3570 1450}%
\special{pa 3586 1518}%
\special{pa 3596 1496}%
\special{pa 3620 1498}%
\special{pa 3570 1450}%
\special{fp}%
%
\special{pn 8}%
\special{pa 3740 1230}%
\special{pa 3740 1290}%
\special{fp}%
\special{sh 1}%
\special{pa 3740 1290}%
\special{pa 3760 1224}%
\special{pa 3740 1238}%
\special{pa 3720 1224}%
\special{pa 3740 1290}%
\special{fp}%
%
\special{pn 8}%
\special{pa 4150 860}%
\special{pa 4240 810}%
\special{fp}%
\special{sh 1}%
\special{pa 4240 810}%
\special{pa 4172 826}%
\special{pa 4194 836}%
\special{pa 4192 860}%
\special{pa 4240 810}%
\special{fp}%
%
\special{pn 8}%
\special{pa 4130 1640}%
\special{pa 4200 1680}%
\special{fp}%
\special{sh 1}%
\special{pa 4200 1680}%
\special{pa 4152 1630}%
\special{pa 4154 1654}%
\special{pa 4132 1664}%
\special{pa 4200 1680}%
\special{fp}%
%
\special{pn 8}%
\special{pa 4170 1520}%
\special{pa 4240 1470}%
\special{fp}%
\special{sh 1}%
\special{pa 4240 1470}%
\special{pa 4174 1492}%
\special{pa 4198 1502}%
\special{pa 4198 1526}%
\special{pa 4240 1470}%
\special{fp}%
%
\special{pn 8}%
\special{pa 4230 1020}%
\special{pa 4170 980}%
\special{fp}%
\special{sh 1}%
\special{pa 4170 980}%
\special{pa 4214 1034}%
\special{pa 4214 1010}%
\special{pa 4238 1000}%
\special{pa 4170 980}%
\special{fp}%
%
\special{pn 8}%
\special{pa 4340 1500}%
\special{pa 4340 1570}%
\special{fp}%
\special{sh 1}%
\special{pa 4340 1570}%
\special{pa 4360 1504}%
\special{pa 4340 1518}%
\special{pa 4320 1504}%
\special{pa 4340 1570}%
\special{fp}%
%
\special{pn 8}%
\special{pa 4660 750}%
\special{pa 4570 750}%
\special{fp}%
\special{sh 1}%
\special{pa 4570 750}%
\special{pa 4638 770}%
\special{pa 4624 750}%
\special{pa 4638 730}%
\special{pa 4570 750}%
\special{fp}%
%
\special{pn 8}%
\special{pa 4100 1750}%
\special{pa 4040 1750}%
\special{fp}%
\special{sh 1}%
\special{pa 4040 1750}%
\special{pa 4108 1770}%
\special{pa 4094 1750}%
\special{pa 4108 1730}%
\special{pa 4040 1750}%
\special{fp}%
%
\special{pn 8}%
\special{pa 4690 1750}%
\special{pa 4630 1750}%
\special{fp}%
\special{sh 1}%
\special{pa 4630 1750}%
\special{pa 4698 1770}%
\special{pa 4684 1750}%
\special{pa 4698 1730}%
\special{pa 4630 1750}%
\special{fp}%
%
\special{pn 8}%
\special{pa 4770 850}%
\special{pa 4840 800}%
\special{fp}%
\special{sh 1}%
\special{pa 4840 800}%
\special{pa 4774 822}%
\special{pa 4798 832}%
\special{pa 4798 856}%
\special{pa 4840 800}%
\special{fp}%
%
\special{pn 8}%
\special{pa 4740 1650}%
\special{pa 4820 1680}%
\special{fp}%
\special{sh 1}%
\special{pa 4820 1680}%
\special{pa 4766 1638}%
\special{pa 4770 1662}%
\special{pa 4752 1676}%
\special{pa 4820 1680}%
\special{fp}%
%
\special{pn 8}%
\special{pa 4940 1530}%
\special{pa 4940 1600}%
\special{fp}%
\special{sh 1}%
\special{pa 4940 1600}%
\special{pa 4960 1534}%
\special{pa 4940 1548}%
\special{pa 4920 1534}%
\special{pa 4940 1600}%
\special{fp}%
%
\special{pn 8}%
\special{pa 5080 1520}%
\special{pa 5130 1430}%
\special{fp}%
\special{sh 1}%
\special{pa 5130 1430}%
\special{pa 5080 1480}%
\special{pa 5104 1478}%
\special{pa 5116 1498}%
\special{pa 5130 1430}%
\special{fp}%
%
\special{pn 8}%
\special{pa 5100 1020}%
\special{pa 5070 950}%
\special{fp}%
\special{sh 1}%
\special{pa 5070 950}%
\special{pa 5078 1020}%
\special{pa 5092 1000}%
\special{pa 5116 1004}%
\special{pa 5070 950}%
\special{fp}%
%
\special{pn 8}%
\special{pa 4820 1010}%
\special{pa 4760 980}%
\special{fp}%
\special{sh 1}%
\special{pa 4760 980}%
\special{pa 4812 1028}%
\special{pa 4808 1004}%
\special{pa 4830 992}%
\special{pa 4760 980}%
\special{fp}%
%
\special{pn 8}%
\special{pa 4510 1030}%
\special{pa 4470 960}%
\special{fp}%
\special{sh 1}%
\special{pa 4470 960}%
\special{pa 4486 1028}%
\special{pa 4496 1006}%
\special{pa 4520 1008}%
\special{pa 4470 960}%
\special{fp}%
%
\special{pn 8}%
\special{pa 4500 1510}%
\special{pa 4530 1420}%
\special{fp}%
\special{sh 1}%
\special{pa 4530 1420}%
\special{pa 4490 1478}%
\special{pa 4514 1472}%
\special{pa 4528 1490}%
\special{pa 4530 1420}%
\special{fp}%
%
\special{pn 8}%
\special{pa 4760 1520}%
\special{pa 4840 1460}%
\special{fp}%
\special{sh 1}%
\special{pa 4840 1460}%
\special{pa 4776 1484}%
\special{pa 4798 1492}%
\special{pa 4800 1516}%
\special{pa 4840 1460}%
\special{fp}%
\put(181.3977,-101.1780){\makebox(0,0)[lb]{along $p_3$}}%
%
\special{pn 8}%
\special{pa 1030 530}%
\special{pa 810 750}%
\special{fp}%
\special{pa 970 530}%
\special{pa 710 790}%
\special{fp}%
\special{pa 880 560}%
\special{pa 610 830}%
\special{fp}%
\special{pa 840 540}%
\special{pa 500 880}%
\special{fp}%
\special{pa 790 530}%
\special{pa 430 890}%
\special{fp}%
\special{pa 730 530}%
\special{pa 430 830}%
\special{fp}%
\special{pa 670 530}%
\special{pa 430 770}%
\special{fp}%
\special{pa 610 530}%
\special{pa 430 710}%
\special{fp}%
\special{pa 550 530}%
\special{pa 430 650}%
\special{fp}%
\special{pa 490 530}%
\special{pa 430 590}%
\special{fp}%
\special{pa 910 530}%
\special{pa 880 560}%
\special{fp}%
\special{pa 1090 530}%
\special{pa 930 690}%
\special{fp}%
\special{pa 1150 530}%
\special{pa 1020 660}%
\special{fp}%
\special{pa 1210 530}%
\special{pa 1120 620}%
\special{fp}%
\special{pa 1270 530}%
\special{pa 1220 580}%
\special{fp}%
%
\special{pn 4}%
\special{pa 540 990}%
\special{pa 540 1470}%
\special{dt 0.027}%
\special{pa 630 1060}%
\special{pa 630 1450}%
\special{dt 0.027}%
\special{pa 720 1130}%
\special{pa 720 1470}%
\special{dt 0.027}%
\special{pa 810 1200}%
\special{pa 810 1470}%
\special{dt 0.027}%
\special{pa 900 1300}%
\special{pa 900 1470}%
\special{dt 0.027}%
\special{pa 990 1340}%
\special{pa 990 1470}%
\special{dt 0.027}%
\special{pa 1080 1410}%
\special{pa 1080 1470}%
\special{dt 0.027}%
\special{pa 450 930}%
\special{pa 450 1470}%
\special{dt 0.027}%
\special{pa 360 1150}%
\special{pa 360 1470}%
\special{dt 0.027}%
\special{pa 270 1390}%
\special{pa 270 1470}%
\special{dt 0.027}%
%
\special{pn 4}%
\special{pa 1350 1620}%
\special{pa 1350 2040}%
\special{dt 0.027}%
\special{pa 1440 1690}%
\special{pa 1440 2030}%
\special{dt 0.027}%
\special{pa 1530 1760}%
\special{pa 1530 2040}%
\special{dt 0.027}%
\special{pa 1620 1830}%
\special{pa 1620 2040}%
\special{dt 0.027}%
\special{pa 1710 1900}%
\special{pa 1710 2040}%
\special{dt 0.027}%
\special{pa 1800 1970}%
\special{pa 1800 2040}%
\special{dt 0.027}%
\special{pa 1260 1550}%
\special{pa 1260 2040}%
\special{dt 0.027}%
\special{pa 1170 1480}%
\special{pa 1170 2040}%
\special{dt 0.027}%
\special{pa 1080 1740}%
\special{pa 1080 2040}%
\special{dt 0.027}%
\end{picture}%

%% file: Octahedronaxiom.tex
\unitlength 1pt
\begin{picture}(156.6091,135.1449)(  7.2270,-148.1535)
\put(7.2270,-54.9252){\makebox(0,0)[lb]{$X$}}%
%
\special{pn 8}%
\special{pa 310 700}%
\special{pa 650 700}%
\special{fp}%
\special{sh 1}%
\special{pa 650 700}%
\special{pa 584 680}%
\special{pa 598 700}%
\special{pa 584 720}%
\special{pa 650 700}%
\special{fp}%
\put(41.9166,-22.4037){\makebox(0,0)[lb]{$T^{-1}(X')$}}%
%
\special{pn 8}%
\special{pa 810 350}%
\special{pa 810 560}%
\special{fp}%
\special{sh 1}%
\special{pa 810 560}%
\special{pa 830 494}%
\special{pa 810 508}%
\special{pa 790 494}%
\special{pa 810 560}%
\special{fp}%
\put(54.9252,-54.2025){\makebox(0,0)[lb]{$Y$}}%
%
\special{pn 8}%
\special{pa 1050 700}%
\special{pa 1390 700}%
\special{fp}%
\special{sh 1}%
\special{pa 1390 700}%
\special{pa 1324 680}%
\special{pa 1338 700}%
\special{pa 1324 720}%
\special{pa 1390 700}%
\special{fp}%
\put(107.6823,-53.4798){\makebox(0,0)[lb]{$Z'$}}%
%
\special{pn 8}%
\special{pa 1780 1140}%
\special{pa 2120 1140}%
\special{fp}%
\special{sh 1}%
\special{pa 2120 1140}%
\special{pa 2054 1120}%
\special{pa 2068 1140}%
\special{pa 2054 1160}%
\special{pa 2120 1140}%
\special{fp}%
\put(156.8259,-86.7240){\makebox(0,0)[lb]{$T(X)$}}%
%
\special{pn 8}%
\special{pa 810 820}%
\special{pa 810 1030}%
\special{fp}%
\special{sh 1}%
\special{pa 810 1030}%
\special{pa 830 964}%
\special{pa 810 978}%
\special{pa 790 964}%
\special{pa 810 1030}%
\special{fp}%
%
\special{pn 8}%
\special{pa 1540 810}%
\special{pa 1540 1020}%
\special{fp}%
\special{sh 1}%
\special{pa 1540 1020}%
\special{pa 1560 954}%
\special{pa 1540 968}%
\special{pa 1520 954}%
\special{pa 1540 1020}%
\special{fp}%
\put(55.6479,-87.4467){\makebox(0,0)[lb]{$Z$}}%
\put(108.4050,-86.7240){\makebox(0,0)[lb]{$Y'$}}%
\put(106.9596,-119.9682){\makebox(0,0)[lb]{$X'$}}%
%
\special{pn 8}%
\special{pa 1050 1140}%
\special{pa 1390 1140}%
\special{fp}%
\special{sh 1}%
\special{pa 1390 1140}%
\special{pa 1324 1120}%
\special{pa 1338 1140}%
\special{pa 1324 1160}%
\special{pa 1390 1140}%
\special{fp}%
%
\special{pn 8}%
\special{pa 1540 1280}%
\special{pa 1540 1490}%
\special{fp}%
\special{sh 1}%
\special{pa 1540 1490}%
\special{pa 1560 1424}%
\special{pa 1540 1438}%
\special{pa 1520 1424}%
\special{pa 1540 1490}%
\special{fp}%
%
\special{pn 8}%
\special{pa 320 810}%
\special{pa 670 1050}%
\special{fp}%
\special{sh 1}%
\special{pa 670 1050}%
\special{pa 626 996}%
\special{pa 626 1020}%
\special{pa 604 1030}%
\special{pa 670 1050}%
\special{fp}%
%
\special{pn 8}%
\special{pa 1050 1280}%
\special{pa 1400 1520}%
\special{fp}%
\special{sh 1}%
\special{pa 1400 1520}%
\special{pa 1356 1466}%
\special{pa 1356 1490}%
\special{pa 1334 1500}%
\special{pa 1400 1520}%
\special{fp}%
%
\special{pn 8}%
\special{pa 1030 350}%
\special{pa 1380 570}%
\special{fp}%
\special{sh 1}%
\special{pa 1380 570}%
\special{pa 1334 518}%
\special{pa 1336 542}%
\special{pa 1314 552}%
\special{pa 1380 570}%
\special{fp}%
%
\special{pn 8}%
\special{pa 1700 800}%
\special{pa 2120 1040}%
\special{fp}%
\special{sh 1}%
\special{pa 2120 1040}%
\special{pa 2072 990}%
\special{pa 2074 1014}%
\special{pa 2052 1024}%
\special{pa 2120 1040}%
\special{fp}%
%
\special{pn 8}%
\special{pa 1780 1590}%
\special{pa 2120 1590}%
\special{fp}%
\special{sh 1}%
\special{pa 2120 1590}%
\special{pa 2054 1570}%
\special{pa 2068 1590}%
\special{pa 2054 1610}%
\special{pa 2120 1590}%
\special{fp}%
%
\special{pn 8}%
\special{pa 2240 1260}%
\special{pa 2240 1470}%
\special{fp}%
\special{sh 1}%
\special{pa 2240 1470}%
\special{pa 2260 1404}%
\special{pa 2240 1418}%
\special{pa 2220 1404}%
\special{pa 2240 1470}%
\special{fp}%
\put(158.2713,-122.8590){\makebox(0,0)[lb]{$T(Y)$}}%
%
\special{pn 8}%
\special{pa 1780 1750}%
\special{pa 2130 1990}%
\special{fp}%
\special{sh 1}%
\special{pa 2130 1990}%
\special{pa 2086 1936}%
\special{pa 2086 1960}%
\special{pa 2064 1970}%
\special{pa 2130 1990}%
\special{fp}%
%
\special{pn 8}%
\special{pa 2240 1760}%
\special{pa 2240 1970}%
\special{fp}%
\special{sh 1}%
\special{pa 2240 1970}%
\special{pa 2260 1904}%
\special{pa 2240 1918}%
\special{pa 2220 1904}%
\special{pa 2240 1970}%
\special{fp}%
\put(157.5486,-157.5486){\makebox(0,0)[lb]{$T(Z)$}}%
\end{picture}%

%% file: Nordfjordpic16.tex
\unitlength 1pt
\begin{picture}(293.4162,239.9364)( 23.1264,-270.2898)
%
\special{pn 8}%
\special{pa 1800 600}%
\special{pa 1420 1300}%
\special{da 0.070}%
\special{pa 1420 1300}%
\special{pa 1800 2000}%
\special{da 0.070}%
%
\special{pn 8}%
\special{pa 1420 600}%
\special{pa 1390 600}%
\special{fp}%
\special{sh 1}%
\special{pa 1390 600}%
\special{pa 1458 620}%
\special{pa 1444 600}%
\special{pa 1458 580}%
\special{pa 1390 600}%
\special{fp}%
%
\special{pn 8}%
\special{pa 1570 1020}%
\special{pa 1540 1070}%
\special{fp}%
\special{sh 1}%
\special{pa 1540 1070}%
\special{pa 1592 1024}%
\special{pa 1568 1024}%
\special{pa 1558 1004}%
\special{pa 1540 1070}%
\special{fp}%
%
\special{pn 8}%
\special{pa 1560 1560}%
\special{pa 1540 1520}%
\special{fp}%
\special{sh 1}%
\special{pa 1540 1520}%
\special{pa 1552 1590}%
\special{pa 1564 1568}%
\special{pa 1588 1572}%
\special{pa 1540 1520}%
\special{fp}%
%
\special{pn 8}%
\special{pa 1000 1270}%
\special{pa 1000 1300}%
\special{fp}%
\special{sh 1}%
\special{pa 1000 1300}%
\special{pa 1020 1234}%
\special{pa 1000 1248}%
\special{pa 980 1234}%
\special{pa 1000 1300}%
\special{fp}%
%
\special{pn 8}%
\special{pa 1808 1278}%
\special{pa 1808 1308}%
\special{fp}%
\special{sh 1}%
\special{pa 1808 1308}%
\special{pa 1828 1240}%
\special{pa 1808 1254}%
\special{pa 1788 1240}%
\special{pa 1808 1308}%
\special{fp}%
%
\special{pn 8}%
\special{pa 1000 600}%
\special{pa 620 1300}%
\special{fp}%
\special{pa 620 1300}%
\special{pa 1000 2000}%
\special{fp}%
%
\special{pn 8}%
\special{pa 800 980}%
\special{pa 770 1020}%
\special{fp}%
\special{sh 1}%
\special{pa 770 1020}%
\special{pa 826 980}%
\special{pa 802 978}%
\special{pa 794 956}%
\special{pa 770 1020}%
\special{fp}%
\put(65.7657,-39.7485){\makebox(0,0)[lb]{$\tau_0^{[0]}$}}%
\put(30.3534,-99.0099){\makebox(0,0)[lb]{$\tau_1^{[0]}$}}%
%
\special{pn 8}%
\special{pa 830 1690}%
\special{pa 810 1650}%
\special{fp}%
\special{sh 1}%
\special{pa 810 1650}%
\special{pa 822 1720}%
\special{pa 834 1698}%
\special{pa 858 1702}%
\special{pa 810 1650}%
\special{fp}%
\put(24.5718,-68.6565){\makebox(0,0)[lb]{$p_3-p_1$}}%
\put(23.1264,-129.3633){\makebox(0,0)[lb]{$p_3-p_2$}}%
\put(86.0013,-41.1939){\makebox(0,0)[lb]{$p_4-p_3$}}%
%
\special{pn 8}%
\special{pa 1410 2000}%
\special{pa 1380 2000}%
\special{fp}%
\special{sh 1}%
\special{pa 1380 2000}%
\special{pa 1448 2020}%
\special{pa 1434 2000}%
\special{pa 1448 1980}%
\special{pa 1380 2000}%
\special{fp}%
%
\special{pn 8}%
\special{pa 1000 600}%
\special{pa 1800 600}%
\special{pa 1800 2000}%
\special{pa 1000 2000}%
\special{pa 1000 600}%
\special{pa 1800 600}%
\special{fp}%
\put(129.3633,-40.4712){\makebox(0,0)[lb]{$\tau_0^{[1]}$}}%
\put(186.4566,-40.4712){\makebox(0,0)[lb]{$\tau_0^{[2]}$}}%
%
\special{pn 8}%
\special{pa 4000 1990}%
\special{pa 4280 1990}%
\special{fp}%
\special{sh 1}%
\special{pa 4280 1990}%
\special{pa 4214 1970}%
\special{pa 4228 1990}%
\special{pa 4214 2010}%
\special{pa 4280 1990}%
\special{fp}%
\put(316.5426,-148.1535){\makebox(0,0)[lb]{d$_T$}}%
%
\special{pn 8}%
\special{pa 4000 2010}%
\special{pa 4000 2280}%
\special{fp}%
\special{sh 1}%
\special{pa 4000 2280}%
\special{pa 4020 2214}%
\special{pa 4000 2228}%
\special{pa 3980 2214}%
\special{pa 4000 2280}%
\special{fp}%
\put(282.5757,-181.3977){\makebox(0,0)[lb]{$j=p_2-p_1$}}%
%
\special{pn 8}%
\special{pa 2600 600}%
\special{pa 2220 1300}%
\special{da 0.070}%
\special{pa 2220 1300}%
\special{pa 2600 2000}%
\special{da 0.070}%
%
\special{pn 8}%
\special{pa 1800 600}%
\special{pa 2600 600}%
\special{pa 2600 2000}%
\special{pa 1800 2000}%
\special{pa 1800 600}%
\special{pa 2600 600}%
\special{fp}%
%
\special{pn 8}%
\special{pa 3400 600}%
\special{pa 3020 1300}%
\special{da 0.070}%
\special{pa 3020 1300}%
\special{pa 3400 2000}%
\special{da 0.070}%
%
\special{pn 8}%
\special{pa 2600 600}%
\special{pa 3400 600}%
\special{pa 3400 2000}%
\special{pa 2600 2000}%
\special{pa 2600 600}%
\special{pa 3400 600}%
\special{fp}%
%
\special{pn 8}%
\special{pa 2210 600}%
\special{pa 2180 600}%
\special{fp}%
\special{sh 1}%
\special{pa 2180 600}%
\special{pa 2248 620}%
\special{pa 2234 600}%
\special{pa 2248 580}%
\special{pa 2180 600}%
\special{fp}%
%
\special{pn 8}%
\special{pa 2990 600}%
\special{pa 2960 600}%
\special{fp}%
\special{sh 1}%
\special{pa 2960 600}%
\special{pa 3028 620}%
\special{pa 3014 600}%
\special{pa 3028 580}%
\special{pa 2960 600}%
\special{fp}%
%
\special{pn 8}%
\special{pa 2980 2000}%
\special{pa 2950 2000}%
\special{fp}%
\special{sh 1}%
\special{pa 2950 2000}%
\special{pa 3018 2020}%
\special{pa 3004 2000}%
\special{pa 3018 1980}%
\special{pa 2950 2000}%
\special{fp}%
%
\special{pn 8}%
\special{pa 2210 2000}%
\special{pa 2180 2000}%
\special{fp}%
\special{sh 1}%
\special{pa 2180 2000}%
\special{pa 2248 2020}%
\special{pa 2234 2000}%
\special{pa 2248 1980}%
\special{pa 2180 2000}%
\special{fp}%
%
\special{pn 8}%
\special{pa 2410 970}%
\special{pa 2380 1020}%
\special{fp}%
\special{sh 1}%
\special{pa 2380 1020}%
\special{pa 2432 974}%
\special{pa 2408 974}%
\special{pa 2398 954}%
\special{pa 2380 1020}%
\special{fp}%
%
\special{pn 8}%
\special{pa 3190 1000}%
\special{pa 3160 1050}%
\special{fp}%
\special{sh 1}%
\special{pa 3160 1050}%
\special{pa 3212 1004}%
\special{pa 3188 1004}%
\special{pa 3178 984}%
\special{pa 3160 1050}%
\special{fp}%
%
\special{pn 8}%
\special{pa 2410 1640}%
\special{pa 2390 1600}%
\special{fp}%
\special{sh 1}%
\special{pa 2390 1600}%
\special{pa 2402 1670}%
\special{pa 2414 1648}%
\special{pa 2438 1652}%
\special{pa 2390 1600}%
\special{fp}%
%
\special{pn 8}%
\special{pa 3220 1640}%
\special{pa 3200 1600}%
\special{fp}%
\special{sh 1}%
\special{pa 3200 1600}%
\special{pa 3212 1670}%
\special{pa 3224 1648}%
\special{pa 3248 1652}%
\special{pa 3200 1600}%
\special{fp}%
%
\special{pn 8}%
\special{pa 2600 1280}%
\special{pa 2600 1310}%
\special{fp}%
\special{sh 1}%
\special{pa 2600 1310}%
\special{pa 2620 1244}%
\special{pa 2600 1258}%
\special{pa 2580 1244}%
\special{pa 2600 1310}%
\special{fp}%
%
\special{pn 8}%
\special{pa 3400 1310}%
\special{pa 3400 1340}%
\special{fp}%
\special{sh 1}%
\special{pa 3400 1340}%
\special{pa 3420 1274}%
\special{pa 3400 1288}%
\special{pa 3380 1274}%
\special{pa 3400 1340}%
\special{fp}%
%
\special{pn 13}%
\special{pa 3400 600}%
\special{pa 3860 600}%
\special{dt 0.045}%
\special{pa 3860 600}%
\special{pa 3860 600}%
\special{dt 0.045}%
%
\special{pn 8}%
\special{pa 3400 2000}%
\special{pa 3850 2000}%
\special{dt 0.045}%
%
\special{pn 8}%
\special{pa 620 1300}%
\special{pa 1800 600}%
\special{dt 0.045}%
\special{pa 1800 600}%
\special{pa 1800 600}%
\special{dt 0.045}%
%
\special{pn 8}%
\special{pa 620 1300}%
\special{pa 1800 2000}%
\special{dt 0.045}%
%
\special{pn 8}%
\special{pa 1170 980}%
\special{pa 1240 930}%
\special{fp}%
\special{sh 1}%
\special{pa 1240 930}%
\special{pa 1174 952}%
\special{pa 1198 962}%
\special{pa 1198 986}%
\special{pa 1240 930}%
\special{fp}%
%
\special{pn 8}%
\special{pa 1190 1630}%
\special{pa 1230 1650}%
\special{fp}%
\special{sh 1}%
\special{pa 1230 1650}%
\special{pa 1180 1602}%
\special{pa 1182 1626}%
\special{pa 1162 1638}%
\special{pa 1230 1650}%
\special{fp}%
%
\special{pn 8}%
\special{pa 1420 2008}%
\special{pa 1390 2008}%
\special{fp}%
\special{sh 1}%
\special{pa 1390 2008}%
\special{pa 1458 2028}%
\special{pa 1444 2008}%
\special{pa 1458 1988}%
\special{pa 1390 2008}%
\special{fp}%
%
\special{pn 8}%
\special{pa 2210 2008}%
\special{pa 2180 2008}%
\special{fp}%
\special{sh 1}%
\special{pa 2180 2008}%
\special{pa 2248 2028}%
\special{pa 2234 2008}%
\special{pa 2248 1988}%
\special{pa 2180 2008}%
\special{fp}%
%
\special{pn 8}%
\special{pa 2990 2008}%
\special{pa 2960 2008}%
\special{fp}%
\special{sh 1}%
\special{pa 2960 2008}%
\special{pa 3028 2028}%
\special{pa 3014 2008}%
\special{pa 3028 1988}%
\special{pa 2960 2008}%
\special{fp}%
%
\special{pn 8}%
\special{pa 1800 2008}%
\special{pa 1420 2708}%
\special{da 0.070}%
\special{pa 1420 2708}%
\special{pa 1800 3408}%
\special{da 0.070}%
%
\special{pn 8}%
\special{pa 1420 2008}%
\special{pa 1390 2008}%
\special{fp}%
\special{sh 1}%
\special{pa 1390 2008}%
\special{pa 1458 2028}%
\special{pa 1444 2008}%
\special{pa 1458 1988}%
\special{pa 1390 2008}%
\special{fp}%
%
\special{pn 8}%
\special{pa 1570 2428}%
\special{pa 1540 2478}%
\special{fp}%
\special{sh 1}%
\special{pa 1540 2478}%
\special{pa 1592 2430}%
\special{pa 1568 2432}%
\special{pa 1558 2410}%
\special{pa 1540 2478}%
\special{fp}%
%
\special{pn 8}%
\special{pa 1560 2968}%
\special{pa 1540 2928}%
\special{fp}%
\special{sh 1}%
\special{pa 1540 2928}%
\special{pa 1552 2996}%
\special{pa 1564 2976}%
\special{pa 1588 2978}%
\special{pa 1540 2928}%
\special{fp}%
%
\special{pn 8}%
\special{pa 1000 2678}%
\special{pa 1000 2708}%
\special{fp}%
\special{sh 1}%
\special{pa 1000 2708}%
\special{pa 1020 2640}%
\special{pa 1000 2654}%
\special{pa 980 2640}%
\special{pa 1000 2708}%
\special{fp}%
%
\special{pn 8}%
\special{pa 1808 2684}%
\special{pa 1808 2714}%
\special{fp}%
\special{sh 1}%
\special{pa 1808 2714}%
\special{pa 1828 2648}%
\special{pa 1808 2662}%
\special{pa 1788 2648}%
\special{pa 1808 2714}%
\special{fp}%
%
\special{pn 8}%
\special{pa 1000 2008}%
\special{pa 620 2708}%
\special{fp}%
\special{pa 620 2708}%
\special{pa 1000 3408}%
\special{fp}%
%
\special{pn 8}%
\special{pa 800 2388}%
\special{pa 770 2428}%
\special{fp}%
\special{sh 1}%
\special{pa 770 2428}%
\special{pa 826 2386}%
\special{pa 802 2384}%
\special{pa 794 2362}%
\special{pa 770 2428}%
\special{fp}%
%
\special{pn 8}%
\special{pa 830 3098}%
\special{pa 810 3058}%
\special{fp}%
\special{sh 1}%
\special{pa 810 3058}%
\special{pa 822 3126}%
\special{pa 834 3106}%
\special{pa 858 3108}%
\special{pa 810 3058}%
\special{fp}%
%
\special{pn 8}%
\special{pa 1410 3408}%
\special{pa 1380 3408}%
\special{fp}%
\special{sh 1}%
\special{pa 1380 3408}%
\special{pa 1448 3428}%
\special{pa 1434 3408}%
\special{pa 1448 3388}%
\special{pa 1380 3408}%
\special{fp}%
%
\special{pn 8}%
\special{pa 1000 2008}%
\special{pa 1800 2008}%
\special{pa 1800 3408}%
\special{pa 1000 3408}%
\special{pa 1000 2008}%
\special{pa 1800 2008}%
\special{fp}%
%
\special{pn 8}%
\special{pa 2600 2008}%
\special{pa 2220 2708}%
\special{da 0.070}%
\special{pa 2220 2708}%
\special{pa 2600 3408}%
\special{da 0.070}%
%
\special{pn 8}%
\special{pa 1800 2008}%
\special{pa 2600 2008}%
\special{pa 2600 3408}%
\special{pa 1800 3408}%
\special{pa 1800 2008}%
\special{pa 2600 2008}%
\special{fp}%
%
\special{pn 8}%
\special{pa 3400 2008}%
\special{pa 3020 2708}%
\special{da 0.070}%
\special{pa 3020 2708}%
\special{pa 3400 3408}%
\special{da 0.070}%
%
\special{pn 8}%
\special{pa 2600 2008}%
\special{pa 3400 2008}%
\special{pa 3400 3408}%
\special{pa 2600 3408}%
\special{pa 2600 2008}%
\special{pa 3400 2008}%
\special{fp}%
%
\special{pn 8}%
\special{pa 2210 2008}%
\special{pa 2180 2008}%
\special{fp}%
\special{sh 1}%
\special{pa 2180 2008}%
\special{pa 2248 2028}%
\special{pa 2234 2008}%
\special{pa 2248 1988}%
\special{pa 2180 2008}%
\special{fp}%
%
\special{pn 8}%
\special{pa 2990 2008}%
\special{pa 2960 2008}%
\special{fp}%
\special{sh 1}%
\special{pa 2960 2008}%
\special{pa 3028 2028}%
\special{pa 3014 2008}%
\special{pa 3028 1988}%
\special{pa 2960 2008}%
\special{fp}%
%
\special{pn 8}%
\special{pa 2980 3408}%
\special{pa 2950 3408}%
\special{fp}%
\special{sh 1}%
\special{pa 2950 3408}%
\special{pa 3018 3428}%
\special{pa 3004 3408}%
\special{pa 3018 3388}%
\special{pa 2950 3408}%
\special{fp}%
%
\special{pn 8}%
\special{pa 2210 3408}%
\special{pa 2180 3408}%
\special{fp}%
\special{sh 1}%
\special{pa 2180 3408}%
\special{pa 2248 3428}%
\special{pa 2234 3408}%
\special{pa 2248 3388}%
\special{pa 2180 3408}%
\special{fp}%
%
\special{pn 8}%
\special{pa 2410 2378}%
\special{pa 2380 2428}%
\special{fp}%
\special{sh 1}%
\special{pa 2380 2428}%
\special{pa 2432 2380}%
\special{pa 2408 2382}%
\special{pa 2398 2360}%
\special{pa 2380 2428}%
\special{fp}%
%
\special{pn 8}%
\special{pa 3190 2408}%
\special{pa 3160 2458}%
\special{fp}%
\special{sh 1}%
\special{pa 3160 2458}%
\special{pa 3212 2410}%
\special{pa 3188 2412}%
\special{pa 3178 2390}%
\special{pa 3160 2458}%
\special{fp}%
%
\special{pn 8}%
\special{pa 2410 3048}%
\special{pa 2390 3008}%
\special{fp}%
\special{sh 1}%
\special{pa 2390 3008}%
\special{pa 2402 3076}%
\special{pa 2414 3056}%
\special{pa 2438 3058}%
\special{pa 2390 3008}%
\special{fp}%
%
\special{pn 8}%
\special{pa 3220 3048}%
\special{pa 3200 3008}%
\special{fp}%
\special{sh 1}%
\special{pa 3200 3008}%
\special{pa 3212 3076}%
\special{pa 3224 3056}%
\special{pa 3248 3058}%
\special{pa 3200 3008}%
\special{fp}%
%
\special{pn 8}%
\special{pa 2600 2688}%
\special{pa 2600 2718}%
\special{fp}%
\special{sh 1}%
\special{pa 2600 2718}%
\special{pa 2620 2650}%
\special{pa 2600 2664}%
\special{pa 2580 2650}%
\special{pa 2600 2718}%
\special{fp}%
%
\special{pn 8}%
\special{pa 3400 2718}%
\special{pa 3400 2748}%
\special{fp}%
\special{sh 1}%
\special{pa 3400 2748}%
\special{pa 3420 2680}%
\special{pa 3400 2694}%
\special{pa 3380 2680}%
\special{pa 3400 2748}%
\special{fp}%
%
\special{pn 13}%
\special{pa 3400 2008}%
\special{pa 3860 2008}%
\special{dt 0.045}%
\special{pa 3860 2008}%
\special{pa 3860 2008}%
\special{dt 0.045}%
%
\special{pn 13}%
\special{pa 3400 3408}%
\special{pa 3850 3408}%
\special{dt 0.045}%
%
\special{pn 8}%
\special{pa 620 2708}%
\special{pa 1800 2008}%
\special{dt 0.045}%
\special{pa 1800 2008}%
\special{pa 1800 2008}%
\special{dt 0.045}%
%
\special{pn 8}%
\special{pa 620 2708}%
\special{pa 1800 3408}%
\special{dt 0.045}%
%
\special{pn 8}%
\special{pa 1170 2388}%
\special{pa 1240 2338}%
\special{fp}%
\special{sh 1}%
\special{pa 1240 2338}%
\special{pa 1174 2360}%
\special{pa 1198 2368}%
\special{pa 1198 2392}%
\special{pa 1240 2338}%
\special{fp}%
%
\special{pn 8}%
\special{pa 1190 3038}%
\special{pa 1230 3058}%
\special{fp}%
\special{sh 1}%
\special{pa 1230 3058}%
\special{pa 1180 3010}%
\special{pa 1182 3034}%
\special{pa 1162 3046}%
\special{pa 1230 3058}%
\special{fp}%
%
\special{pn 8}%
\special{pa 1420 3414}%
\special{pa 1390 3414}%
\special{fp}%
\special{sh 1}%
\special{pa 1390 3414}%
\special{pa 1458 3434}%
\special{pa 1444 3414}%
\special{pa 1458 3394}%
\special{pa 1390 3414}%
\special{fp}%
%
\special{pn 8}%
\special{pa 2210 3414}%
\special{pa 2180 3414}%
\special{fp}%
\special{sh 1}%
\special{pa 2180 3414}%
\special{pa 2248 3434}%
\special{pa 2234 3414}%
\special{pa 2248 3394}%
\special{pa 2180 3414}%
\special{fp}%
%
\special{pn 8}%
\special{pa 2990 3414}%
\special{pa 2960 3414}%
\special{fp}%
\special{sh 1}%
\special{pa 2960 3414}%
\special{pa 3028 3434}%
\special{pa 3014 3414}%
\special{pa 3028 3394}%
\special{pa 2960 3414}%
\special{fp}%
\put(50.5890,-150.3216){\makebox(0,0)[lb]{$\tau_2^{[0]}$}}%
\put(30.3534,-200.1879){\makebox(0,0)[lb]{$\tau_3^{[0]}$}}%
\put(74.4381,-244.2726){\makebox(0,0)[lb]{$\tau_4^{[0]}$}}%
\put(89.6148,-99.0099){\makebox(0,0)[lb]{$\tau_1^{[1]}$}}%
\put(132.9768,-141.6492){\makebox(0,0)[lb]{$\tau_2^{[1]}$}}%
\put(86.7240,-200.1879){\makebox(0,0)[lb]{$\tau_3^{[1]}$}}%
\put(132.9768,-243.5499){\makebox(0,0)[lb]{$\tau_4^{[1]}$}}%
\put(190.7928,-244.2726){\makebox(0,0)[lb]{$\tau_4^{[2]}$}}%
\put(248.6088,-244.9953){\makebox(0,0)[lb]{$\tau_4^{[3]}$}}%
\put(246.4407,-40.4712){\makebox(0,0)[lb]{$\tau_0^{[3]}$}}%
\put(141.6492,-99.0099){\makebox(0,0)[lb]{$\tau_1^{[2]}$}}%
\put(200.1879,-99.0099){\makebox(0,0)[lb]{$\tau_1^{[3]}$}}%
\put(191.5155,-141.6492){\makebox(0,0)[lb]{$\tau_2^{[2]}$}}%
\put(249.3315,-142.3719){\makebox(0,0)[lb]{$\tau_2^{[3]}$}}%
\put(146.7081,-200.1879){\makebox(0,0)[lb]{$\tau_3^{[2]}$}}%
\put(201.6333,-200.1879){\makebox(0,0)[lb]{$\tau_3^{[3]}$}}%
%
\special{pn 13}%
\special{pa 1000 3400}%
\special{pa 1000 3740}%
\special{dt 0.045}%
%
\special{pn 13}%
\special{pa 1800 3400}%
\special{pa 1800 3720}%
\special{dt 0.045}%
%
\special{pn 13}%
\special{pa 2600 3400}%
\special{pa 2600 3720}%
\special{dt 0.045}%
%
\special{pn 13}%
\special{pa 3400 3400}%
\special{pa 3400 3740}%
\special{dt 0.045}%
%
\special{pn 8}%
\special{pa 1400 1300}%
\special{pa 2580 600}%
\special{dt 0.045}%
\special{pa 2580 600}%
\special{pa 2580 600}%
\special{dt 0.045}%
%
\special{pn 8}%
\special{pa 2200 1300}%
\special{pa 3380 600}%
\special{dt 0.045}%
\special{pa 3380 600}%
\special{pa 3380 600}%
\special{dt 0.045}%
%
\special{pn 8}%
\special{pa 1400 2700}%
\special{pa 2580 2000}%
\special{dt 0.045}%
\special{pa 2580 2000}%
\special{pa 2580 2000}%
\special{dt 0.045}%
%
\special{pn 8}%
\special{pa 2200 2700}%
\special{pa 3380 2000}%
\special{dt 0.045}%
\special{pa 3380 2000}%
\special{pa 3380 2000}%
\special{dt 0.045}%
%
\special{pn 8}%
\special{pa 1460 1300}%
\special{pa 2640 2000}%
\special{dt 0.045}%
%
\special{pn 8}%
\special{pa 2220 1300}%
\special{pa 3400 2000}%
\special{dt 0.045}%
%
\special{pn 8}%
\special{pa 1420 2700}%
\special{pa 2600 3400}%
\special{dt 0.045}%
%
\special{pn 8}%
\special{pa 2220 2700}%
\special{pa 3400 3400}%
\special{dt 0.045}%
%
\special{pn 8}%
\special{pa 1970 970}%
\special{pa 2040 920}%
\special{fp}%
\special{sh 1}%
\special{pa 2040 920}%
\special{pa 1974 942}%
\special{pa 1998 952}%
\special{pa 1998 976}%
\special{pa 2040 920}%
\special{fp}%
%
\special{pn 8}%
\special{pa 2750 970}%
\special{pa 2820 920}%
\special{fp}%
\special{sh 1}%
\special{pa 2820 920}%
\special{pa 2754 942}%
\special{pa 2778 952}%
\special{pa 2778 976}%
\special{pa 2820 920}%
\special{fp}%
%
\special{pn 8}%
\special{pa 1990 2360}%
\special{pa 2060 2310}%
\special{fp}%
\special{sh 1}%
\special{pa 2060 2310}%
\special{pa 1994 2332}%
\special{pa 2018 2342}%
\special{pa 2018 2366}%
\special{pa 2060 2310}%
\special{fp}%
%
\special{pn 8}%
\special{pa 2810 2350}%
\special{pa 2880 2300}%
\special{fp}%
\special{sh 1}%
\special{pa 2880 2300}%
\special{pa 2814 2322}%
\special{pa 2838 2332}%
\special{pa 2838 2356}%
\special{pa 2880 2300}%
\special{fp}%
%
\special{pn 8}%
\special{pa 2020 3060}%
\special{pa 2060 3080}%
\special{fp}%
\special{sh 1}%
\special{pa 2060 3080}%
\special{pa 2010 3032}%
\special{pa 2012 3056}%
\special{pa 1992 3068}%
\special{pa 2060 3080}%
\special{fp}%
%
\special{pn 8}%
\special{pa 2810 3050}%
\special{pa 2850 3070}%
\special{fp}%
\special{sh 1}%
\special{pa 2850 3070}%
\special{pa 2800 3022}%
\special{pa 2802 3046}%
\special{pa 2782 3058}%
\special{pa 2850 3070}%
\special{fp}%
%
\special{pn 8}%
\special{pa 2780 1640}%
\special{pa 2820 1660}%
\special{fp}%
\special{sh 1}%
\special{pa 2820 1660}%
\special{pa 2770 1612}%
\special{pa 2772 1636}%
\special{pa 2752 1648}%
\special{pa 2820 1660}%
\special{fp}%
%
\special{pn 8}%
\special{pa 2010 1630}%
\special{pa 2050 1650}%
\special{fp}%
\special{sh 1}%
\special{pa 2050 1650}%
\special{pa 2000 1602}%
\special{pa 2002 1626}%
\special{pa 1982 1638}%
\special{pa 2050 1650}%
\special{fp}%
\end{picture}%

%% file: nullsystem.tex
\unitlength 1pt
\begin{picture}(370.7451,161.1621)( 23.1264,-183.5658)
%
\special{pn 8}%
\special{pa 800 600}%
\special{pa 800 1990}%
\special{fp}%
\special{sh 1}%
\special{pa 800 1990}%
\special{pa 820 1924}%
\special{pa 800 1938}%
\special{pa 780 1924}%
\special{pa 800 1990}%
\special{fp}%
%
\special{pn 8}%
\special{pa 750 610}%
\special{pa 400 1200}%
\special{fp}%
\special{sh 1}%
\special{pa 400 1200}%
\special{pa 452 1154}%
\special{pa 428 1154}%
\special{pa 418 1132}%
\special{pa 400 1200}%
\special{fp}%
%
\special{pn 8}%
\special{pa 730 2000}%
\special{pa 400 1400}%
\special{fp}%
\special{sh 1}%
\special{pa 400 1400}%
\special{pa 416 1468}%
\special{pa 426 1448}%
\special{pa 450 1450}%
\special{pa 400 1400}%
\special{fp}%
%
\special{pn 8}%
\special{pa 480 1200}%
\special{pa 1450 600}%
\special{dt 0.045}%
\special{sh 1}%
\special{pa 1450 600}%
\special{pa 1384 618}%
\special{pa 1406 628}%
\special{pa 1404 652}%
\special{pa 1450 600}%
\special{fp}%
\special{pa 1450 600}%
\special{pa 1450 600}%
\special{dt 0.045}%
%
\special{pn 8}%
\special{pa 470 1400}%
\special{pa 1450 2010}%
\special{dt 0.045}%
\special{sh 1}%
\special{pa 1450 2010}%
\special{pa 1404 1958}%
\special{pa 1406 1982}%
\special{pa 1384 1992}%
\special{pa 1450 2010}%
\special{fp}%
%
\special{pn 13}%
\special{pa 600 1310}%
\special{pa 1000 1320}%
\special{fp}%
\special{sh 1}%
\special{pa 1000 1320}%
\special{pa 934 1298}%
\special{pa 948 1320}%
\special{pa 934 1338}%
\special{pa 1000 1320}%
\special{fp}%
%
\special{pn 8}%
\special{pa 1600 600}%
\special{pa 1600 1990}%
\special{fp}%
\special{sh 1}%
\special{pa 1600 1990}%
\special{pa 1620 1924}%
\special{pa 1600 1938}%
\special{pa 1580 1924}%
\special{pa 1600 1990}%
\special{fp}%
%
\special{pn 8}%
\special{pa 1550 610}%
\special{pa 1200 1200}%
\special{fp}%
\special{sh 1}%
\special{pa 1200 1200}%
\special{pa 1252 1154}%
\special{pa 1228 1154}%
\special{pa 1218 1132}%
\special{pa 1200 1200}%
\special{fp}%
%
\special{pn 8}%
\special{pa 1530 2000}%
\special{pa 1200 1400}%
\special{fp}%
\special{sh 1}%
\special{pa 1200 1400}%
\special{pa 1216 1468}%
\special{pa 1226 1448}%
\special{pa 1250 1450}%
\special{pa 1200 1400}%
\special{fp}%
%
\special{pn 8}%
\special{pa 1280 1200}%
\special{pa 2250 600}%
\special{dt 0.045}%
\special{sh 1}%
\special{pa 2250 600}%
\special{pa 2184 618}%
\special{pa 2206 628}%
\special{pa 2204 652}%
\special{pa 2250 600}%
\special{fp}%
\special{pa 2250 600}%
\special{pa 2250 600}%
\special{dt 0.045}%
%
\special{pn 8}%
\special{pa 1270 1400}%
\special{pa 2250 2010}%
\special{dt 0.045}%
\special{sh 1}%
\special{pa 2250 2010}%
\special{pa 2204 1958}%
\special{pa 2206 1982}%
\special{pa 2184 1992}%
\special{pa 2250 2010}%
\special{fp}%
%
\special{pn 13}%
\special{pa 1400 1310}%
\special{pa 1800 1320}%
\special{fp}%
\special{sh 1}%
\special{pa 1800 1320}%
\special{pa 1734 1298}%
\special{pa 1748 1320}%
\special{pa 1734 1338}%
\special{pa 1800 1320}%
\special{fp}%
%
\special{pn 8}%
\special{pa 2400 600}%
\special{pa 2400 1990}%
\special{fp}%
\special{sh 1}%
\special{pa 2400 1990}%
\special{pa 2420 1924}%
\special{pa 2400 1938}%
\special{pa 2380 1924}%
\special{pa 2400 1990}%
\special{fp}%
%
\special{pn 8}%
\special{pa 2350 610}%
\special{pa 2000 1200}%
\special{fp}%
\special{sh 1}%
\special{pa 2000 1200}%
\special{pa 2052 1154}%
\special{pa 2028 1154}%
\special{pa 2018 1132}%
\special{pa 2000 1200}%
\special{fp}%
%
\special{pn 8}%
\special{pa 2330 2000}%
\special{pa 2000 1400}%
\special{fp}%
\special{sh 1}%
\special{pa 2000 1400}%
\special{pa 2016 1468}%
\special{pa 2026 1448}%
\special{pa 2050 1450}%
\special{pa 2000 1400}%
\special{fp}%
%
\special{pn 8}%
\special{pa 2080 1200}%
\special{pa 3050 600}%
\special{dt 0.045}%
\special{sh 1}%
\special{pa 3050 600}%
\special{pa 2984 618}%
\special{pa 3006 628}%
\special{pa 3004 652}%
\special{pa 3050 600}%
\special{fp}%
\special{pa 3050 600}%
\special{pa 3050 600}%
\special{dt 0.045}%
%
\special{pn 8}%
\special{pa 2070 1400}%
\special{pa 3050 2010}%
\special{dt 0.045}%
\special{sh 1}%
\special{pa 3050 2010}%
\special{pa 3004 1958}%
\special{pa 3006 1982}%
\special{pa 2984 1992}%
\special{pa 3050 2010}%
\special{fp}%
%
\special{pn 13}%
\special{pa 2200 1310}%
\special{pa 2600 1320}%
\special{fp}%
\special{sh 1}%
\special{pa 2600 1320}%
\special{pa 2534 1298}%
\special{pa 2548 1320}%
\special{pa 2534 1338}%
\special{pa 2600 1320}%
\special{fp}%
%
\special{pn 8}%
\special{pa 3200 600}%
\special{pa 3200 1990}%
\special{fp}%
\special{sh 1}%
\special{pa 3200 1990}%
\special{pa 3220 1924}%
\special{pa 3200 1938}%
\special{pa 3180 1924}%
\special{pa 3200 1990}%
\special{fp}%
%
\special{pn 8}%
\special{pa 3150 610}%
\special{pa 2800 1200}%
\special{fp}%
\special{sh 1}%
\special{pa 2800 1200}%
\special{pa 2852 1154}%
\special{pa 2828 1154}%
\special{pa 2818 1132}%
\special{pa 2800 1200}%
\special{fp}%
%
\special{pn 8}%
\special{pa 3130 2000}%
\special{pa 2800 1400}%
\special{fp}%
\special{sh 1}%
\special{pa 2800 1400}%
\special{pa 2816 1468}%
\special{pa 2826 1448}%
\special{pa 2850 1450}%
\special{pa 2800 1400}%
\special{fp}%
%
\special{pn 8}%
\special{pa 2880 1200}%
\special{pa 3850 600}%
\special{dt 0.045}%
\special{sh 1}%
\special{pa 3850 600}%
\special{pa 3784 618}%
\special{pa 3806 628}%
\special{pa 3804 652}%
\special{pa 3850 600}%
\special{fp}%
\special{pa 3850 600}%
\special{pa 3850 600}%
\special{dt 0.045}%
%
\special{pn 8}%
\special{pa 2870 1400}%
\special{pa 3850 2010}%
\special{dt 0.045}%
\special{sh 1}%
\special{pa 3850 2010}%
\special{pa 3804 1958}%
\special{pa 3806 1982}%
\special{pa 3784 1992}%
\special{pa 3850 2010}%
\special{fp}%
%
\special{pn 13}%
\special{pa 3000 1310}%
\special{pa 3400 1320}%
\special{fp}%
\special{sh 1}%
\special{pa 3400 1320}%
\special{pa 3334 1298}%
\special{pa 3348 1320}%
\special{pa 3334 1338}%
\special{pa 3400 1320}%
\special{fp}%
%
\special{pn 8}%
\special{pa 4000 600}%
\special{pa 4000 1990}%
\special{fp}%
\special{sh 1}%
\special{pa 4000 1990}%
\special{pa 4020 1924}%
\special{pa 4000 1938}%
\special{pa 3980 1924}%
\special{pa 4000 1990}%
\special{fp}%
%
\special{pn 8}%
\special{pa 3950 610}%
\special{pa 3600 1200}%
\special{fp}%
\special{sh 1}%
\special{pa 3600 1200}%
\special{pa 3652 1154}%
\special{pa 3628 1154}%
\special{pa 3618 1132}%
\special{pa 3600 1200}%
\special{fp}%
%
\special{pn 8}%
\special{pa 3930 2000}%
\special{pa 3600 1400}%
\special{fp}%
\special{sh 1}%
\special{pa 3600 1400}%
\special{pa 3616 1468}%
\special{pa 3626 1448}%
\special{pa 3650 1450}%
\special{pa 3600 1400}%
\special{fp}%
%
\special{pn 8}%
\special{pa 3680 1200}%
\special{pa 4650 600}%
\special{dt 0.045}%
\special{sh 1}%
\special{pa 4650 600}%
\special{pa 4584 618}%
\special{pa 4606 628}%
\special{pa 4604 652}%
\special{pa 4650 600}%
\special{fp}%
\special{pa 4650 600}%
\special{pa 4650 600}%
\special{dt 0.045}%
%
\special{pn 8}%
\special{pa 3670 1400}%
\special{pa 4650 2010}%
\special{dt 0.045}%
\special{sh 1}%
\special{pa 4650 2010}%
\special{pa 4604 1958}%
\special{pa 4606 1982}%
\special{pa 4584 1992}%
\special{pa 4650 2010}%
\special{fp}%
%
\special{pn 13}%
\special{pa 3800 1310}%
\special{pa 4200 1320}%
\special{fp}%
\special{sh 1}%
\special{pa 4200 1320}%
\special{pa 4134 1298}%
\special{pa 4148 1320}%
\special{pa 4134 1338}%
\special{pa 4200 1320}%
\special{fp}%
%
\special{pn 8}%
\special{pa 4800 600}%
\special{pa 4800 1990}%
\special{fp}%
\special{sh 1}%
\special{pa 4800 1990}%
\special{pa 4820 1924}%
\special{pa 4800 1938}%
\special{pa 4780 1924}%
\special{pa 4800 1990}%
\special{fp}%
%
\special{pn 8}%
\special{pa 4750 610}%
\special{pa 4400 1200}%
\special{fp}%
\special{sh 1}%
\special{pa 4400 1200}%
\special{pa 4452 1154}%
\special{pa 4428 1154}%
\special{pa 4418 1132}%
\special{pa 4400 1200}%
\special{fp}%
%
\special{pn 8}%
\special{pa 4730 2000}%
\special{pa 4400 1400}%
\special{fp}%
\special{sh 1}%
\special{pa 4400 1400}%
\special{pa 4416 1468}%
\special{pa 4426 1448}%
\special{pa 4450 1450}%
\special{pa 4400 1400}%
\special{fp}%
%
\special{pn 8}%
\special{pa 4480 1200}%
\special{pa 5450 600}%
\special{dt 0.045}%
\special{sh 1}%
\special{pa 5450 600}%
\special{pa 5384 618}%
\special{pa 5406 628}%
\special{pa 5404 652}%
\special{pa 5450 600}%
\special{fp}%
\special{pa 5450 600}%
\special{pa 5450 600}%
\special{dt 0.045}%
%
\special{pn 8}%
\special{pa 4470 1400}%
\special{pa 5450 2010}%
\special{dt 0.045}%
\special{sh 1}%
\special{pa 5450 2010}%
\special{pa 5404 1958}%
\special{pa 5406 1982}%
\special{pa 5384 1992}%
\special{pa 5450 2010}%
\special{fp}%
%
\special{pn 13}%
\special{pa 4600 1310}%
\special{pa 5000 1320}%
\special{fp}%
\special{sh 1}%
\special{pa 5000 1320}%
\special{pa 4934 1298}%
\special{pa 4948 1320}%
\special{pa 4934 1338}%
\special{pa 5000 1320}%
\special{fp}%
\put(59.9841,-92.5056){\makebox(0,0)[lb]{$\scriptstyle{d_T}$}}%
\put(118.5228,-92.5056){\makebox(0,0)[lb]{$\scriptstyle{d_T}$}}%
\put(175.6161,-91.7829){\makebox(0,0)[lb]{$\scriptstyle{d_T}$}}%
\put(234.8775,-92.5056){\makebox(0,0)[lb]{$\scriptstyle{d_T}$}}%
\put(291.9708,-92.5056){\makebox(0,0)[lb]{$\scriptstyle{d_T}$}}%
\put(349.7868,-91.7829){\makebox(0,0)[lb]{$\scriptstyle{d_T}$}}%
\put(82.3878,-100.4553){\makebox(0,0)[lb]{$\gamma_\lambda^{(2)}$}}%
\put(311.4837,-100.4553){\makebox(0,0)[lb]{$\gamma_\lambda^{(2)}$}}%
\put(23.1264,-101.1780){\makebox(0,0)[lb]{$\gamma_\lambda^{(3)}$}}%
\put(198.7425,-98.2872){\makebox(0,0)[lb]{0}}%
\put(139.4811,-98.2872){\makebox(0,0)[lb]{$\lambda_1$}}%
\put(253.6677,-99.7326){\makebox(0,0)[lb]{$\lambda'_1$}}%
\put(367.8543,-100.4553){\makebox(0,0)[lb]{$\gamma_\lambda^{(3)}$}}%
\put(221.8689,-39.0258){\makebox(0,0)[lb]{$\lambda'_1$}}%
\put(281.1303,-39.0258){\makebox(0,0)[lb]{$\lambda'_0$}}%
\put(334.6101,-39.0258){\makebox(0,0)[lb]{$\alpha_\lambda^{(2)}$}}%
\put(334.6101,-159.7167){\makebox(0,0)[lb]{$\beta_\lambda^{(2)}$}}%
\put(281.1303,-158.2713){\makebox(0,0)[lb]{$\lambda'_2$}}%
\put(221.1462,-157.5486){\makebox(0,0)[lb]{$\lambda'_0$}}%
\put(166.9437,-156.8259){\makebox(0,0)[lb]{$\lambda_2$}}%
\put(101.9007,-160.4394){\makebox(0,0)[lb]{$\alpha_\lambda^{(2)}$}}%
\put(45.5301,-160.4394){\makebox(0,0)[lb]{$\alpha_\lambda^{(3)}$}}%
\put(167.6664,-39.0258){\makebox(0,0)[lb]{$\lambda_0$}}%
\put(105.5142,-39.0258){\makebox(0,0)[lb]{$\beta_\lambda^{(2)}$}}%
\put(50.5890,-39.0258){\makebox(0,0)[lb]{$\beta_\lambda^{(3)}$}}%
\put(216.0873,-178.5069){\makebox(0,0)[lb]{$\tilde S_4$}}%
\put(271.0125,-177.7842){\makebox(0,0)[lb]{$\tilde S_4[1]$}}%
\put(146.7081,-177.7842){\makebox(0,0)[lb]{$\tilde S_4[-1]$}}%
%
\special{pn 8}%
\special{pa 1910 310}%
\special{pa 4140 310}%
\special{pa 4140 2540}%
\special{pa 1910 2540}%
\special{pa 1910 310}%
\special{da 0.070}%
\end{picture}%

%% file: localnullsystem.tex
\unitlength 1pt
\begin{picture}(372.9132,119.2455)( 20.9583,-148.1535)
%
\special{pn 8}%
\special{pa 800 600}%
\special{pa 800 1990}%
\special{fp}%
\special{sh 1}%
\special{pa 800 1990}%
\special{pa 820 1924}%
\special{pa 800 1938}%
\special{pa 780 1924}%
\special{pa 800 1990}%
\special{fp}%
%
\special{pn 8}%
\special{pa 750 610}%
\special{pa 400 1200}%
\special{fp}%
\special{sh 1}%
\special{pa 400 1200}%
\special{pa 452 1154}%
\special{pa 428 1154}%
\special{pa 418 1132}%
\special{pa 400 1200}%
\special{fp}%
%
\special{pn 8}%
\special{pa 730 2000}%
\special{pa 400 1400}%
\special{fp}%
\special{sh 1}%
\special{pa 400 1400}%
\special{pa 416 1468}%
\special{pa 426 1448}%
\special{pa 450 1450}%
\special{pa 400 1400}%
\special{fp}%
%
\special{pn 8}%
\special{pa 480 1200}%
\special{pa 1450 600}%
\special{dt 0.045}%
\special{sh 1}%
\special{pa 1450 600}%
\special{pa 1384 618}%
\special{pa 1406 628}%
\special{pa 1404 652}%
\special{pa 1450 600}%
\special{fp}%
\special{pa 1450 600}%
\special{pa 1450 600}%
\special{dt 0.045}%
%
\special{pn 8}%
\special{pa 470 1400}%
\special{pa 1450 2010}%
\special{dt 0.045}%
\special{sh 1}%
\special{pa 1450 2010}%
\special{pa 1404 1958}%
\special{pa 1406 1982}%
\special{pa 1384 1992}%
\special{pa 1450 2010}%
\special{fp}%
%
\special{pn 13}%
\special{pa 600 1310}%
\special{pa 1000 1320}%
\special{fp}%
\special{sh 1}%
\special{pa 1000 1320}%
\special{pa 934 1298}%
\special{pa 948 1320}%
\special{pa 934 1338}%
\special{pa 1000 1320}%
\special{fp}%
%
\special{pn 8}%
\special{pa 1600 600}%
\special{pa 1600 1990}%
\special{fp}%
\special{sh 1}%
\special{pa 1600 1990}%
\special{pa 1620 1924}%
\special{pa 1600 1938}%
\special{pa 1580 1924}%
\special{pa 1600 1990}%
\special{fp}%
%
\special{pn 8}%
\special{pa 1550 610}%
\special{pa 1200 1200}%
\special{fp}%
\special{sh 1}%
\special{pa 1200 1200}%
\special{pa 1252 1154}%
\special{pa 1228 1154}%
\special{pa 1218 1132}%
\special{pa 1200 1200}%
\special{fp}%
%
\special{pn 8}%
\special{pa 1530 2000}%
\special{pa 1200 1400}%
\special{fp}%
\special{sh 1}%
\special{pa 1200 1400}%
\special{pa 1216 1468}%
\special{pa 1226 1448}%
\special{pa 1250 1450}%
\special{pa 1200 1400}%
\special{fp}%
%
\special{pn 13}%
\special{pa 1400 1310}%
\special{pa 1800 1320}%
\special{fp}%
\special{sh 1}%
\special{pa 1800 1320}%
\special{pa 1734 1298}%
\special{pa 1748 1320}%
\special{pa 1734 1338}%
\special{pa 1800 1320}%
\special{fp}%
%
\special{pn 13}%
\special{pa 2200 1310}%
\special{pa 2600 1320}%
\special{fp}%
\special{sh 1}%
\special{pa 2600 1320}%
\special{pa 2534 1298}%
\special{pa 2548 1320}%
\special{pa 2534 1338}%
\special{pa 2600 1320}%
\special{fp}%
%
\special{pn 13}%
\special{pa 3000 1310}%
\special{pa 3400 1320}%
\special{fp}%
\special{sh 1}%
\special{pa 3400 1320}%
\special{pa 3334 1298}%
\special{pa 3348 1320}%
\special{pa 3334 1338}%
\special{pa 3400 1320}%
\special{fp}%
%
\special{pn 13}%
\special{pa 3800 1310}%
\special{pa 4200 1320}%
\special{fp}%
\special{sh 1}%
\special{pa 4200 1320}%
\special{pa 4134 1298}%
\special{pa 4148 1320}%
\special{pa 4134 1338}%
\special{pa 4200 1320}%
\special{fp}%
%
\special{pn 8}%
\special{pa 4800 600}%
\special{pa 4800 1990}%
\special{fp}%
\special{sh 1}%
\special{pa 4800 1990}%
\special{pa 4820 1924}%
\special{pa 4800 1938}%
\special{pa 4780 1924}%
\special{pa 4800 1990}%
\special{fp}%
%
\special{pn 8}%
\special{pa 4750 610}%
\special{pa 4400 1200}%
\special{fp}%
\special{sh 1}%
\special{pa 4400 1200}%
\special{pa 4452 1154}%
\special{pa 4428 1154}%
\special{pa 4418 1132}%
\special{pa 4400 1200}%
\special{fp}%
%
\special{pn 8}%
\special{pa 4730 2000}%
\special{pa 4400 1400}%
\special{fp}%
\special{sh 1}%
\special{pa 4400 1400}%
\special{pa 4416 1468}%
\special{pa 4426 1448}%
\special{pa 4450 1450}%
\special{pa 4400 1400}%
\special{fp}%
%
\special{pn 8}%
\special{pa 4480 1200}%
\special{pa 5450 600}%
\special{dt 0.045}%
\special{sh 1}%
\special{pa 5450 600}%
\special{pa 5384 618}%
\special{pa 5406 628}%
\special{pa 5404 652}%
\special{pa 5450 600}%
\special{fp}%
\special{pa 5450 600}%
\special{pa 5450 600}%
\special{dt 0.045}%
%
\special{pn 8}%
\special{pa 4470 1400}%
\special{pa 5450 2010}%
\special{dt 0.045}%
\special{sh 1}%
\special{pa 5450 2010}%
\special{pa 5404 1958}%
\special{pa 5406 1982}%
\special{pa 5384 1992}%
\special{pa 5450 2010}%
\special{fp}%
%
\special{pn 13}%
\special{pa 4600 1310}%
\special{pa 5000 1320}%
\special{fp}%
\special{sh 1}%
\special{pa 5000 1320}%
\special{pa 4934 1298}%
\special{pa 4948 1320}%
\special{pa 4934 1338}%
\special{pa 5000 1320}%
\special{fp}%
\put(59.9841,-92.5056){\makebox(0,0)[lb]{$\scriptstyle{d_T}$}}%
\put(118.5228,-92.5056){\makebox(0,0)[lb]{$\scriptstyle{d_T}$}}%
\put(281.1303,-93.2283){\makebox(0,0)[lb]{$\scriptstyle{d_T}$}}%
\put(349.7868,-91.7829){\makebox(0,0)[lb]{$\scriptstyle{d_T}$}}%
\put(79.4970,-99.7326){\makebox(0,0)[lb]{$\gamma^{(2)}$}}%
\put(310.7610,-101.1780){\makebox(0,0)[lb]{$\gamma^{(2)}$}}%
\put(20.9583,-100.4553){\makebox(0,0)[lb]{$\gamma^{(3)}$}}%
\put(198.7425,-98.2872){\makebox(0,0)[lb]{0}}%
\put(367.8543,-100.4553){\makebox(0,0)[lb]{$\gamma^{(3)}$}}%
\put(336.0555,-38.3031){\makebox(0,0)[lb]{$\alpha^{(2)}$}}%
\put(340.3917,-157.5486){\makebox(0,0)[lb]{$\beta^{(2)}$}}%
\put(105.5142,-157.5486){\makebox(0,0)[lb]{$\alpha^{(2)}$}}%
\put(47.6982,-156.8259){\makebox(0,0)[lb]{$\alpha^{(3)}$}}%
\put(101.1780,-39.0258){\makebox(0,0)[lb]{$\beta^{(2)}$}}%
\put(49.8663,-39.0258){\makebox(0,0)[lb]{$\beta^{(3)}$}}%
\put(145.2627,-99.7326){\makebox(0,0)[lb]{1}}%
\put(260.8947,-99.7326){\makebox(0,0)[lb]{1}}%
%
\special{pn 8}%
\special{pa 1900 1100}%
\special{pa 4230 1100}%
\special{pa 4230 1500}%
\special{pa 1900 1500}%
\special{pa 1900 1100}%
\special{da 0.070}%
\put(224.7597,-91.7829){\makebox(0,0)[lb]{$\scriptstyle{d_T}$}}%
\put(167.6664,-91.7829){\makebox(0,0)[lb]{$\scriptstyle{d_T}$}}%
\end{picture}%

%% file: reductionnullsystem.tex
\unitlength 1pt
\begin{picture}(246.9466,119.9682)( 20.2356,-147.4308)
%
\special{pn 8}%
\special{pa 1270 580}%
\special{pa 1270 1970}%
\special{fp}%
\special{sh 1}%
\special{pa 1270 1970}%
\special{pa 1290 1904}%
\special{pa 1270 1918}%
\special{pa 1250 1904}%
\special{pa 1270 1970}%
\special{fp}%
%
\special{pn 8}%
\special{pa 1220 590}%
\special{pa 870 1180}%
\special{fp}%
\special{sh 1}%
\special{pa 870 1180}%
\special{pa 922 1134}%
\special{pa 898 1134}%
\special{pa 888 1112}%
\special{pa 870 1180}%
\special{fp}%
%
\special{pn 8}%
\special{pa 1200 1980}%
\special{pa 870 1380}%
\special{fp}%
\special{sh 1}%
\special{pa 870 1380}%
\special{pa 886 1448}%
\special{pa 896 1428}%
\special{pa 920 1430}%
\special{pa 870 1380}%
\special{fp}%
%
\special{pn 8}%
\special{pa 940 1380}%
\special{pa 1920 1990}%
\special{da 0.070}%
\special{sh 1}%
\special{pa 1920 1990}%
\special{pa 1874 1938}%
\special{pa 1876 1962}%
\special{pa 1854 1972}%
\special{pa 1920 1990}%
\special{fp}%
%
\special{pn 8}%
\special{pa 2070 580}%
\special{pa 2070 1970}%
\special{fp}%
\special{sh 1}%
\special{pa 2070 1970}%
\special{pa 2090 1904}%
\special{pa 2070 1918}%
\special{pa 2050 1904}%
\special{pa 2070 1970}%
\special{fp}%
%
\special{pn 8}%
\special{pa 2020 590}%
\special{pa 1670 1180}%
\special{fp}%
\special{sh 1}%
\special{pa 1670 1180}%
\special{pa 1722 1134}%
\special{pa 1698 1134}%
\special{pa 1688 1112}%
\special{pa 1670 1180}%
\special{fp}%
%
\special{pn 8}%
\special{pa 2000 1980}%
\special{pa 1670 1380}%
\special{fp}%
\special{sh 1}%
\special{pa 1670 1380}%
\special{pa 1686 1448}%
\special{pa 1696 1428}%
\special{pa 1720 1430}%
\special{pa 1670 1380}%
\special{fp}%
%
\special{pn 8}%
\special{pa 1750 1180}%
\special{pa 2720 580}%
\special{da 0.070}%
\special{sh 1}%
\special{pa 2720 580}%
\special{pa 2654 598}%
\special{pa 2676 608}%
\special{pa 2674 632}%
\special{pa 2720 580}%
\special{fp}%
\special{pa 2720 580}%
\special{pa 2720 580}%
\special{da 0.070}%
%
\special{pn 8}%
\special{pa 1740 1380}%
\special{pa 2720 1990}%
\special{da 0.070}%
\special{sh 1}%
\special{pa 2720 1990}%
\special{pa 2674 1938}%
\special{pa 2676 1962}%
\special{pa 2654 1972}%
\special{pa 2720 1990}%
\special{fp}%
%
\special{pn 8}%
\special{pa 2870 580}%
\special{pa 2870 1970}%
\special{fp}%
\special{sh 1}%
\special{pa 2870 1970}%
\special{pa 2890 1904}%
\special{pa 2870 1918}%
\special{pa 2850 1904}%
\special{pa 2870 1970}%
\special{fp}%
%
\special{pn 8}%
\special{pa 2820 590}%
\special{pa 2470 1180}%
\special{fp}%
\special{sh 1}%
\special{pa 2470 1180}%
\special{pa 2522 1134}%
\special{pa 2498 1134}%
\special{pa 2488 1112}%
\special{pa 2470 1180}%
\special{fp}%
%
\special{pn 8}%
\special{pa 2800 1980}%
\special{pa 2470 1380}%
\special{fp}%
\special{sh 1}%
\special{pa 2470 1380}%
\special{pa 2486 1448}%
\special{pa 2496 1428}%
\special{pa 2520 1430}%
\special{pa 2470 1380}%
\special{fp}%
%
\special{pn 8}%
\special{pa 2550 1180}%
\special{pa 3520 580}%
\special{da 0.070}%
\special{sh 1}%
\special{pa 3520 580}%
\special{pa 3454 598}%
\special{pa 3476 608}%
\special{pa 3474 632}%
\special{pa 3520 580}%
\special{fp}%
\special{pa 3520 580}%
\special{pa 3520 580}%
\special{da 0.070}%
%
\special{pn 8}%
\special{pa 2540 1380}%
\special{pa 3520 1990}%
\special{da 0.070}%
\special{sh 1}%
\special{pa 3520 1990}%
\special{pa 3474 1938}%
\special{pa 3476 1962}%
\special{pa 3454 1972}%
\special{pa 3520 1990}%
\special{fp}%
%
\special{pn 8}%
\special{pa 3670 580}%
\special{pa 3670 1970}%
\special{fp}%
\special{sh 1}%
\special{pa 3670 1970}%
\special{pa 3690 1904}%
\special{pa 3670 1918}%
\special{pa 3650 1904}%
\special{pa 3670 1970}%
\special{fp}%
%
\special{pn 8}%
\special{pa 3620 590}%
\special{pa 3270 1180}%
\special{fp}%
\special{sh 1}%
\special{pa 3270 1180}%
\special{pa 3322 1134}%
\special{pa 3298 1134}%
\special{pa 3288 1112}%
\special{pa 3270 1180}%
\special{fp}%
%
\special{pn 8}%
\special{pa 3600 1980}%
\special{pa 3270 1380}%
\special{fp}%
\special{sh 1}%
\special{pa 3270 1380}%
\special{pa 3286 1448}%
\special{pa 3296 1428}%
\special{pa 3320 1430}%
\special{pa 3270 1380}%
\special{fp}%
\put(232.7094,-96.8418){\makebox(0,0)[lb]{0}}%
\put(255.1131,-36.8577){\makebox(0,0)[lb]{$\lambda'_1$}}%
\put(255.1131,-156.8259){\makebox(0,0)[lb]{$\lambda'_0$}}%
%
\special{pn 8}%
\special{pa 970 1180}%
\special{pa 1940 580}%
\special{da 0.070}%
\special{sh 1}%
\special{pa 1940 580}%
\special{pa 1874 598}%
\special{pa 1896 608}%
\special{pa 1894 632}%
\special{pa 1940 580}%
\special{fp}%
\special{pa 1940 580}%
\special{pa 1940 580}%
\special{da 0.070}%
\put(176.3388,-98.2872){\makebox(0,0)[lb]{$\gamma_\lambda^{[k]}$}}%
\put(119.2455,-98.2872){\makebox(0,0)[lb]{$\gamma_\lambda^{[2k]}$}}%
\put(61.4295,-99.0099){\makebox(0,0)[lb]{$\gamma_\lambda^{[3k]}$}}%
\put(20.2356,-98.2872){\makebox(0,0)[lb]{$\cdots$}}%
\put(81.6651,-40.4712){\makebox(0,0)[lb]{$\beta_\lambda^{(3k)}$}}%
\put(140.9265,-39.0258){\makebox(0,0)[lb]{$\beta_\lambda^{(2k)}$}}%
\put(198.7425,-39.0258){\makebox(0,0)[lb]{$\beta_\lambda^{(k)}$}}%
\put(85.2786,-155.3805){\makebox(0,0)[lb]{$\alpha_\lambda^{(3k)}$}}%
\put(140.2038,-156.1032){\makebox(0,0)[lb]{$\alpha_\lambda^{(2k)}$}}%
\put(198.7425,-156.8259){\makebox(0,0)[lb]{$\alpha_\lambda^{(k)}$}}%
%
\special{pn 8}%
\special{pa 720 800}%
\special{pa 1160 570}%
\special{da 0.070}%
\special{sh 1}%
\special{pa 1160 570}%
\special{pa 1092 584}%
\special{pa 1114 596}%
\special{pa 1110 620}%
\special{pa 1160 570}%
\special{fp}%
%
\special{pn 8}%
\special{pa 740 1830}%
\special{pa 1120 2000}%
\special{da 0.070}%
\special{sh 1}%
\special{pa 1120 2000}%
\special{pa 1068 1956}%
\special{pa 1072 1978}%
\special{pa 1052 1992}%
\special{pa 1120 2000}%
\special{fp}%
\end{picture}%

%% file: projectiveobject.tex
\unitlength 1pt
\begin{picture}(134.4222, 59.0446)( 31.7988,-82.8937)
%
\special{pn 8}%
\special{pa 870 540}%
\special{pa 560 980}%
\special{da 0.070}%
\special{sh 1}%
\special{pa 560 980}%
\special{pa 616 938}%
\special{pa 592 936}%
\special{pa 582 914}%
\special{pa 560 980}%
\special{fp}%
\special{pa 560 980}%
\special{pa 560 980}%
\special{da 0.070}%
%
\special{pn 8}%
\special{pa 1060 540}%
\special{pa 1390 980}%
\special{da 0.070}%
\special{sh 1}%
\special{pa 1390 980}%
\special{pa 1366 916}%
\special{pa 1358 938}%
\special{pa 1334 940}%
\special{pa 1390 980}%
\special{fp}%
\put(100.4553,-86.7240){\makebox(0,0)[lb]{$\lambda'_0$}}%
\put(166.2210,-85.2786){\makebox(0,0)[lb]{0}}%
\put(88.1694,-51.3117){\makebox(0,0)[lb]{$\scriptstyle{u}$}}%
\put(66.4884,-77.3289){\makebox(0,0)[lb]{$\scriptstyle{\pi}$}}%
\put(66.4884,-33.2442){\makebox(0,0)[lb]{$\gamma_\lambda^{[k]}$}}%
\put(31.7988,-86.0013){\makebox(0,0)[lb]{$\lambda'_1$}}%
%
\special{pn 8}%
\special{pa 680 1120}%
\special{pa 1310 1120}%
\special{fp}%
\special{sh 1}%
\special{pa 1310 1120}%
\special{pa 1244 1100}%
\special{pa 1258 1120}%
\special{pa 1244 1140}%
\special{pa 1310 1120}%
\special{fp}%
%
\special{pn 8}%
\special{pa 1620 1120}%
\special{pa 2250 1120}%
\special{fp}%
\special{sh 1}%
\special{pa 2250 1120}%
\special{pa 2184 1100}%
\special{pa 2198 1120}%
\special{pa 2184 1140}%
\special{pa 2250 1120}%
\special{fp}%
\end{picture}%